\documentclass[11pt,reqno]{amsart}
\usepackage{amssymb, amsthm, amsmath}

\def\lanbox{\hbox{$\, \vrule height 0.25cm width 0.25cm depth 0.01cm 
\,$}}
\renewcommand{\Box}{\lanbox}

\def\uprho{\raise1pt\hbox{$\rho$}}

\def\R{{\mathbb R}}

\def\mfr#1/#2{\hbox{${\frac{#1}{#2}}$}}

\newcommand{\brtf}{\bar\rho}
\newcommand{\Egp}{E^{\rm GP}}

\def\R{{\mathbb R}}

\def\E{{\mathcal E}}

\def\x{{\vec x}}
\def\p{{\vec p}}
\def\z{{\vec z}}

\def\y{{\vec y}}
\def\k{{\vec k}}
\def\0{{\vec 0}}

\def\const{{\rm const.\,}}

\def\mfr#1/#2{\hbox{$\frac{{#1}}{{#2}}$}}
\def\uprho{\raise1pt\hbox{$\rho$}}
\def\upchi{\raise1pt\hbox{$\chi$}}
\def\dlambda{\lower1pt\hbox{$\lambda$}}
\newcommand{\xij}{|\x_i-\x_j|}
\newcommand{\rtf}{\rho^{\rm TF}}

\newcommand{\mtf}{\mu^{\rm TF}}

\newcommand{\hn}{{\mathord{\widehat{n}}}}
\newcommand{\an}{{\mathord{a}}^{\phantom{*}}}
\newcommand{\bn}{{\mathord{b}}^{\phantom{*}}}
\newcommand{\cA}{{\mathord{\mathcal A}}}
\newcommand{\cB}{{\mathord{\mathcal B}}}
\newcommand{\vp}{\mathord{\hbox{\boldmath $\varphi$}}}
\newcommand{\Z}{\mathbb{Z}}

\newcommand{\beq}{\begin{equation}}
\newcommand{\eeq}{\end{equation}}

\newtheorem{thm}{Theorem}[section]
\newtheorem{lem}[thm]{Lemma}
\newtheorem{corollary}[thm]{Corollary}

\theoremstyle{definition}

\theoremstyle{remark}

\numberwithin{equation}{section}

\newcommand{\eps}{\varepsilon}
\newcommand{\K}{{\mathcal K}}
\newcommand{\X}{{\vec X}}
\newcommand{\Tr}{{\rm Tr}\, }
\newcommand{\half}{\mbox{$\frac{1}{2}$}}

\newcommand{\pgp}{\phi^{\rm GP}}
\newcommand{\al}{\alpha}
\newcommand{\rmax}{\rho_{\al,{\rm max}}}
\newcommand{\rmin}{\rho_{\al,{\rm min}}}
\newcommand{\g}{g}
\newcommand{\dn}{{\mathord{d}}^{\phantom{*}}}
\newcommand{\cn}{{\mathord{c}}^{\phantom{*}}}
\newcommand{\cP}{{\mathord{\mathcal P}}}

\newcommand{\cK}{{\mathord{\mathcal K}}}
\renewcommand{\u}{{\vec u}}
\begin{document}

\title{THE QUANTUM-MECHANICAL MANY-BODY PROBLEM: THE BOSE GAS
}\thanks{The work was supported in part by US NSF grants
PHY 0139984-A01 (EHL), PHY 0353181  (RS) and DMS-0111298 (JPS); by
EU grant HPRN-CT-2002-00277 (JPS and JY); by FWF grant P17176-N02 (JY);
by MaPhySto -- A Network in Mathematical Physics and
Stochastics funded by The Danish National Research Foundation (JPS),
and by grants from the Danish research council (JPS)
}

\author[E.H.~Lieb]{ELLIOTT H.\ LIEB}
\address [E.H.~Lieb] {Departments  of Mathematics and Physics, 
Princeton
University, Jadwin Hall,  P.O. Box 708, Princeton, NJ  08544, USA}
\email{lieb@math.princeton.edu}

\author[R.~Seiringer]{ROBERT SEIRINGER} 
\address[R.~Seiringer]{Department of Physics, 
Princeton
University, Jadwin Hall,  P.O. Box 708, Princeton, NJ  08544, USA}
\email{rseiring@math.princeton.edu}   

\author[J.P.~Solovej]{JAN PHILIP SOLOVEJ}
\address[J.P.~Solovej]{School of Mathematics, Institute for Advanced
  Study, 1 Einstein Dr., Princeton, N.J. 08540. {\rm On leave from}
  Dept. of Math., University of Copenhagen, Universitetsparken 5,
  DK-2100 Co\-penhagen, Denmark} \email {solovej@math.ku.dk}

\author[J.~Yngvason]{JAKOB~YNGVASON}
\address[J.~Yngvason]{Institut f\"ur Theoretische Physik, 
Universit\"at
Wien, Boltzmanngasse 5, A-1090 Vienna, Austria}
\email{yngvason@thor.thp.univie.ac.at}

\thanks{\copyright 2004 by the authors. Reproduction of this article, 
in its
entirety, by any means, is permitted for non-commercial purposes.}

%\date{June 7, 2004}

\begin{abstract}
  Now that the low temperature properties of quantum-\-mech\-anical
  many-body systems (bosons) at low density, $\rho$, can be examined
  experimentally it is appropriate to revisit some of the formulas
  deduced by many authors 4--5 decades ago, and to explore new regimes
  not treated before. For systems with repulsive (i.e.\ positive)
  interaction potentials the experimental low temperature state and
  the ground state are effectively synonymous -- and this fact is used
  in all modeling.  In such cases, the leading term in the
  energy/particle is $2\pi\hbar^2 a \rho/m$ where $a$ is the
  scattering length of the two-body potential.  Owing to the delicate
  and peculiar nature of bosonic correlations (such as the strange
  $N^{7/5}$ law for charged bosons), four decades of research failed
  to establish this plausible formula rigorously. The only previous
  lower bound for the energy was found by Dyson in 1957, but it was 14
  times too small. The correct asymptotic formula has been obtained by
  us and this work will be presented. The reason behind the
  mathematical difficulties will be emphasized. A different formula,
  postulated as late as 1971 by Schick, holds in two dimensions and
  this, too, will be shown to be correct.  With the aid of the
  methodology developed to prove the lower bound for the homogeneous
  gas, several other problems have been successfully addressed. One is
  the proof by us that the Gross-Pitaevskii equation correctly
  describes the ground state in the `traps' actually used in the
  experiments. For this system it is also possible to prove complete
  Bose condensation and superfluidity, as we have shown.  On the
  frontier of experimental developments is the possibility that a
  dilute gas in an elongated trap will behave like a one-dimensional
  system; we have proved this mathematically.  Another topic is a
  proof that Foldy's 1961 theory of a high density Bose gas of charged
  particles correctly describes its ground state energy; using this we
  can also prove the $N^{7/5}$ formula for the ground state energy of
  the two-component charged Bose gas proposed by Dyson in 1967.  All
  of this is quite recent work and it is hoped that the mathematical
  methodology might be useful, ultimately, to solve more complex
  problems connected with these interesting systems.
\end{abstract} 

\maketitle

\section*{Foreword}

At the conference ``Perspectives in Analysis" at the KTH, Stockholm,
June 23, 2003, one of us (E.H.L.) contributed a talk with the title
above.  This talk covered material by all the authors listed above.
This contribution is a much expanded version of the talk and of
\cite{L3}.  It is based on, but supersedes, the article \cite{LSSY}.

\tableofcontents

\section{Introduction}\label{intro}

Schr\"odinger's equation of 1926 defined a new mechanics whose
Hamiltonian is based on classical mechanics, but whose
consequences are sometimes non-intuitive from the classical point
of view. One of the most extreme cases is the behavior of the
ground (= lowest energy) state of a many-body system of particles.
Since the ground state function $\Psi(\x_1,...,\x_N)$ is
automatically symmetric in the coordinates $\x_j\in \R^3$ of the
$N$ particles, we are dealing necessarily with {\it `bosons'.} If
we imposed the Pauli exclusion~principle (antisymmetry) instead,
appropriate for electrons,  the outcome would look much more
natural and, oddly, more classical. Indeed, the Pauli principle is
{\it essential} for understanding the stability of the ordinary
matter  that surrounds us.

Recent experiments have confirmed some of the bizarre properties of
bosons close to their ground state, but the theoretical ideas go back
to the 1940's -- 1960's. The first sophisticated analysis of a gas or
liquid of {\it interacting} bosons is due to Bogolubov in 1947. His
approximate theory as amplified by others, is supposed to be exact in
certain limiting cases, and some of those cases have now been verified
rigorously (for the ground state energy) --- 3 or 4 decades after they
were proposed.

The discussion will center around five main topics.

\begin{itemize}
\item [1.]  The dilute, homogeneous Bose gas with repulsive 
interaction (2D and 3D).
\item [2.]  Repulsive bosons in a trap (as used in recent experiments)
  and the \lq\lq Gross-Pitaevskii\rq\rq\ equation.
\item [3.]  Bose condensation and superfluidity for dilute trapped 
gases.
\item [4.]  One-dimensional behavior of three-dimensional gases in 
elongated traps.
\item [5.]  Foldy's \lq\lq jellium\rq\rq\ model of charged particles
  in a neutralizing background and the extension to the two-component
  gas.
\end{itemize}

Note that for potentials that tend to zero at infinity `repulsive' and
`positive' are synonymous --- in the quantum mechanical literature at
least.  In classical mechanics, in contrast, a potential that is
positive but not monotonically decreasing is not called repulsive.

The discussion below of topic~1 is based on \cite{LY1998} and
\cite{LY2d}, and of topic~2 on \cite{LSY1999} and \cite{LSY2d}. See
also
\cite{LYbham,LSYdoeb,S4,LSYnn}. 

The original references for
topic~3  are 
\cite{LS02} and \cite{LSYsuper}, but for transparency 
we also include here a section on the special case when
the trap is a rectangular box. This case already contains the 
salient points, but avoids several complications due the inhomogeneity of
the gas in a general trap.
An essential technical tool for topic 3 is a generalized Poincar\'e 
inequality, which is 
discussed in a separate section.

Topic 4 is a summary of the work in \cite{LSY}.
 
The discussion of topic 5 is based on \cite{LS} and \cite{LSo02}.

Topic 1 (3-dimensions) was the starting point and contains essential
ideas. It is explained here in some detail and is taken, with minor
modifications (and corrections), from \cite{LYbham}. In terms of
technical complexity, however, the fifth topic is the most involved
and can not be treated here in full detail.

The interaction potential between pairs of particles in the Jellium
model in topic 5 is the repulsive, {\it long-range} Coulomb potential,
while in topics 1--4 it is assumed to be repulsive and {\it short
  range}. For alkali atoms in the recent experiments on Bose Einstein
condensation the interaction potential has a repulsive hard core, but
also a quite deep attractive contribution of van der Waals type and
there are many two body bound states \cite{PS}. The Bose condensate
seen in the experiments is thus not the true ground state (which would
be a solid) but a metastable state. Nevertheless, it is usual to model
this metastable state as the ground state of a system with a repulsive
two body potential having the same scattering length as the true
potential, and this is what we shall do. In this paper all interaction
potentials will be positive.

\section{The Dilute Bose Gas in 3D} \label{sect3d}

We consider the Hamiltonian for $N$ bosons of mass
$m$ enclosed in a cubic box $\Lambda$ of side length $L$ and 
interacting by a
spherically symmetric pair potential
$v(|\x_i - \x_j|)$:
\begin{equation}\label{ham}
H_{N} = - \mu\sum_{i=1}^{N} \Delta_i +
\sum_{1 \leq i < j \leq N} v(|\x_i - \x_j|).
\end{equation}
Here  $\x_i\in\R^3$, $i=1,\dots,N$ are the positions of the
particles, $\Delta_i$ the Laplacian with respect to $\x_{i}$, and
we have denoted ${\hbar^2}/{ 2m}$ by $\mu$ for short. (By choosing
suitable units $\mu$ could, of course, be eliminated, but we want
to keep track of the dependence of the energy on  Planck's
constant and the mass.) The Hamiltonian (\ref{ham}) operates on
{\it symmetric} wave functions in $L^2(\Lambda^{N}, d\x_1\cdots
d\x_N)$ as is appropriate for bosons. The interaction potential
will be assumed to be {\it nonnegative} and to decrease faster
than $1/r^3$ at infinity.

We are interested in the ground state energy $E_{0}(N,L)$ of 
(\ref{ham}) in
the
{\it thermodynamic limit} when $N$ and $L$ tend to infinity with the
density $\rho=N/L^3$ fixed. The energy per particle in this limit is
\begin{equation}\label{eq:thmlimit} 
e_{0}(\rho)=\lim_{L\to\infty}E_{0}(\rho L^3,L)/(\rho
L^3).\end{equation}
Our results about $e_{0}(\rho)$ are based on estimates on
$E_{0}(N,L)$
for finite $N$ and $L$, which are important, e.g., for the 
considerations of
inhomogeneous systems in \cite{LSY1999}.
To define  $E_{0}(N,L)$ precisely one
must specify the boundary conditions. These should not matter for the
thermodynamic limit.
To be on the safe side we use Neumann boundary conditions for the
lower bound, and Dirichlet boundary conditions for the upper bound
since these lead, respectively, to the lowest and the highest 
energies.

For experiments with dilute gases the {\it low density asymptotics} of
$e_{0}(\rho)$ is of importance. Low density means here that the mean
interparticle distance, $\rho^{-1/3}$ is much larger than the
{\it scattering length} $a$ of the potential,
%%%% ADDITION
which is defined as follows. The zero energy scattering
Schr\"odinger equation
\begin{equation}\label{3dscatteq}
-2\mu \Delta \psi + v(r) \psi =0
\end{equation}
has a solution  of the form, asymptotically as $|\x|=r\to \infty$
(or for all $r>R_0$ if $v(r)=0$ for $ r>R_0$),
\begin{equation}\label{3dscattlength}
\psi_0(\x) = 1-a/|\x|
\end{equation}
(The factor $2$ in (\ref{3dscatteq}) comes from the reduced mass of 
the
two particle problem.) Writing $\psi_0(\x)=u_0(|\x|)/|\x|$ this is the
same as
%%%%%END ADDITION
\begin{equation}a=\lim_{r\to\infty}r-\frac{u_{0}(r)}{u_{0}'(r)},\end{equation}

where $u_{0}$ solves the zero energy (radial) scattering equation,
\begin{equation}\label{scatteq}
-2\mu u_{0}^{\prime\prime}(r)+ v(r) u_{0}(r)=0
\end{equation}
with $u_{0}(0)=0$. 

An important special case is the hard core potential $v(r)= \infty$ if
$r<a$ and  $v(r)= 0$ otherwise. Then the scattering length $a$ and the
radius $a$ are the same.

Our main result is a
rigorous proof of the formula
\begin{equation} e_{0}(\rho)\approx4\pi\mu\rho a\end{equation}
for $\rho a^3\ll 1$, more precisely of
\begin{thm}[\textbf{Low density limit of the ground state
energy}]\label{3dhomthm}
\begin{equation}\label{basic}
\lim_{\rho a^3\to 0}\frac {e_{0}(\rho)}{4\pi\mu\rho a}=1.
\end{equation}
\end{thm}
This formula is independent of the boundary conditions used for the
definition of $e_{0}(\rho)$ . It holds for every positive radially 
symmetric pair 
potential such that $\int_R^\infty v(r)r^2 dr<\infty$ for some $R$, 
which 
guarantees a finite scattering length, cf.\ Appendix A in \cite{LY2d}.

The genesis of an understanding of $e_{0}(\rho)$ was the pioneering
work \cite{BO} of Bogolubov, and in the 50's and early 60's several
derivations of (\ref{basic}) were presented \cite{Lee-Huang-YangEtc},
\cite{Lieb63}, even including higher order terms:
\begin{equation}\frac{e_{0}(\rho)}{4\pi\mu\rho a}=
1+\mfr{128}/{15\sqrt \pi}(\rho a^3)^{1/2}
+8\left(\mfr{4\pi}/{3}-\sqrt 3\right)(\rho a^3)\log (\rho a^3)
+O(\rho a^3)
\end{equation}
These early developments are reviewed in \cite{EL2}. They all rely
on some special assumptions about the ground state that have never 
been
proved, or on the selection of special terms from a perturbation 
series
which likely diverges. The only rigorous estimates of this period were
established by Dyson, who derived the following bounds in 1957 for a
gas of hard spheres \cite{dyson}:
\begin{equation} \frac1{10\sqrt 2} \leq
    \frac{e_{0}(\rho)}{ 4\pi\mu\rho a}\leq\frac{1+2 Y^{1/3}}{
(1-Y^{1/3})^2}
\end{equation}
with $Y=4\pi\rho a^3/3$. While the upper bound has the asymptotically
correct form, the lower bound is off the mark by a factor of about 
1/14.
But for about 40 years this was the best lower bound available!

Under the assumption that (\ref{basic}) is a correct asymptotic
formula for the energy, we see at once that understanding it
physically, much less proving it, is not a simple matter.
Initially, the problem presents us with two lengths, $a \ll
\rho^{-1/3}$ at low density. However, (\ref{basic}) presents us
with another length generated by the solution to the problem. This
length is the de Broglie wavelength, or  `uncertainty principle' 
length 
(sometimes called `healing length')
\begin{equation}\label{ellc}\ell_c\sim (\rho a)^{-1/2} . 
\end{equation}
 The reason for saying that
$\ell_c$ is the de Broglie wavelength is that in the hard core
case all the energy is kinetic (the hard core just imposes a $\psi
=0$ boundary condition whenever the distance between two particles
is less than $a$). By the uncertainty principle, the kinetic
energy is proportional to an inverse length squared, namely
$\ell_c$. We then have the relation (since $\rho a ^3$ is small)
\begin{equation}\label{scales} a \ll \rho^{-1/3}\ll \ell_c 
\end{equation} 
which implies, physically, that
it is impossible to localize the particles relative to each other
(even though $\rho$ is small). Bosons in their ground state are
therefore `smeared out' over distances large compared to the mean
particle distance and their individuality is entirely lost. They
cannot be localized with respect to each other without changing
the kinetic energy enormously.

Fermions, on the other hand, prefer to sit in
`private rooms', i.e., $\ell_{c}$ is never bigger than $\rho^{-1/3}$
by a fixed factor.
In this respect the quantum nature of bosons is much more pronounced
than for fermions.

Since (\ref{basic}) is a basic result about the Bose gas it is clearly
important to derive it rigorously and in reasonable generality, in
particular for more general cases than hard spheres.  The question
immediately arises for which interaction potentials one may expect it
to be true. A notable fact is that it {\it is not true for all} $v$ with
$a>0$, since there are two body potentials with positive scattering
length that allow many body bound states. (There are even such
potentials without two body bound states but with three body bound
states \cite{BA}.) For such potentials \eqref{basic} is clearly false.
Our proof, presented in the
sequel,  works for nonnegative $v$, but we conjecture that 
(\ref{basic})
holds if $a>0$ and $v$ has no $N$-body bound states for any $N$. The
lower bound is, of course, the hardest part, but the upper bound is 
not
altogether trivial either.

%%%%%%%%
Before we start with the estimates a simple computation and some
heuristics may be helpful to make
(\ref{basic}) plausible and motivate the formal proofs.

With  $\psi_{0}$ the zero energy scattering solution,
partial integration, using \eqref{3dscatteq} and 
\eqref{3dscattlength}, 
gives, for $R\geq R_0$, 
\begin{equation}\label{partint}
\int_{|\x|\leq R}\{2\mu|\nabla \psi_{0}|^2+v|\psi_{0}|^2\}d\x=
8\pi\mu a\left(1-\frac aR\right)\to 8\pi\mu a\quad\mbox{\rm for 
$R\to\infty$}. 
\end{equation}
Moreover, for positive interaction potentials the scattering solution
minimizes
the quadratic form in (\ref{partint}) for each $R\geq R_0$ with
the boundary condition $\psi_0(|\x|=R)=(1-a/R)$. Hence the energy
$E_{0}(2,L)$ of two
particles in a large box, i.e., $L\gg
a$, is approximately $8\pi\mu a/L^3$. If the gas is sufficiently
dilute it is not unreasonable to expect that the energy is essentially
a sum of all such two particle contributions. Since there are
$N(N-1)/2$ pairs, we are thus lead to $E_{0}(N,L)\approx 4\pi\mu a
N(N-1)/L^3$, which gives (\ref{basic}) in the thermodynamic limit.

This simple heuristics is far from a rigorous proof, however,
especially for the lower bound. In fact, it is  rather remarkable that
the same asymptotic formula holds both for `soft' interaction
potentials, where perturbation theory can be expected to be a good
approximation, and potentials like hard spheres where this is not so.
In the former case the ground state is approximately the constant
function and the energy is {\it mostly potential}:
According to perturbation theory
$E_{0}(N,L)\approx  N(N-1)/(2 L^3)\int v(|\x|)d\x$. In particular it 
is {\it
independent of} $\mu$, i.e. of Planck's constant and mass. Since,
however, $\int v(|\x|)d\x$ is the first Born approximation to $8\pi\mu
a$ (note that $a$ depends on $\mu$!), this is not in conflict with
(\ref{basic}).
For `hard' potentials on the other hand, the ground state is {\it
highly correlated}, i.e., it is far from being a product of single
particle states. The energy is here {\it mostly kinetic}, because the
wave function is very small where the potential is large. These two
quite different regimes, the potential energy dominated one and the
kinetic energy dominated one, cannot be distinguished by the low
density asymptotics of the energy. Whether they behave
differently with respect to other phenomena, e.g., Bose-Einstein
condensation, is not known at present.

Bogolubov's analysis \cite{BO} presupposes the existence of 
Bose-Einstein
condensation. Nevertheless, it is correct (for the energy) for the
one-dimensional delta-function Bose gas \cite{LL}, despite the fact
that there is (presumably) no condensation in that case \cite{PiSt}. 
It turns
out that BE condensation is not really needed in order to understand
the energy.  As we shall see,  `global' condensation can be replaced
by a `local' condensation on boxes whose size is independent of
$L$. It is this crucial understanding that enables us to prove Theorem
1.1 without having to decide about BE condensation.

An important idea of Dyson was to transform the hard sphere
potential into a soft potential at the cost of sacrificing the
kinetic energy, i.e., effectively to move from one
regime to the other. We shall make use of this idea in our proof
of the lower bound below. But first we discuss the simpler upper
bound, which relies on other ideas from Dyson's beautiful paper 
\cite{dyson}.

%%%%%%%%%%%
\subsection{Upper Bound}\label{upsec}

The following generalization of Dyson's upper bound holds
\cite{LSY1999}, \cite{S1999}:
\begin{thm}[\textbf{Upper bound}]\label{ub} Let  
$\rho_{1}=(N-1)/L^3$ and
$b=(4\pi\rho_{1}/3)^{-1/3}$. For non-negative potentials $v$ and $b>a$
the ground state energy of (\ref{ham}) with periodic boundary 
conditions
satisfies
\begin{equation}\label{upperbound}
E_{0}(N,L)/N\leq 4\pi \mu \rho_{1}a\frac{1-\frac{a}
{b}+\left(\frac{a}
{b}\right)^2+\frac12\left(\frac{a}
{b}\right)^3}{\left(1-\frac{a}
{b}\right)^8}.
\end{equation}
Thus in the thermodynamic limit (and for all boundary conditions)
\begin{equation}
\frac{e_{0}(\rho)}{4\pi\mu\rho a}\leq\frac{1-Y^{1/3}+Y^{2/3}-\mfr1/2Y}
{(1-Y^{1/3})^8},
\end{equation}
provided $Y=4\pi\rho a^3/3<1$.
\end{thm}

\noindent {\it Remark.} The bound (\ref{upperbound}) holds for 
potentials
with infinite range, provided $b>a$. For potentials of finite range
$R_{0}$ it can be improved for $b>R_{0}$ to
\begin{equation}\label{upperbound2}
E_{0}(N,L)/N\leq 4\pi \mu \rho_{1}a\frac{1-\left(\frac{a}
{b}\right)^2+\frac12\left(\frac{a}
{b}\right)^3}{\left(1-\frac{a}
{b}\right)^4}.
\end{equation}

\begin{proof}
We first remark that the expectation value of (\ref{ham}) with any
trial wave function gives an upper bound to the bosonic ground state
energy, even if the trial function is not symmetric under permutations
of the variables.  The reason is that an absolute ground state of the
elliptic differential operator (\ref{ham}) (i.e., a ground state
without symmetry requirement) is a nonnegative function which can
be symmetrized without changing the energy because (\ref{ham}) is
symmetric under permutations.  In other words, the absolute ground
state energy is the same as the bosonic ground state energy.

Following \cite{dyson} we choose a trial function of
the following form
\begin{equation}\label{wave}
\Psi(\x_1,\dots,\x_N)=F_1(\x_{1}) \cdot F_2(\x_{1},\x_{2}) \cdots
F_N(\x_{1},\dots,\x_{N}).
\end{equation}
More specifically, $F_{1}\equiv 1$ and $F_{i}$ depends only on the
distance of $\x_{i}$ to its nearest neighbor among the points
$\x_1,\dots ,\x_{i-1}$ (taking the periodic boundary into
account):
\begin{equation}\label{form}
F_i(\x_1,\dots,\x_i)=f(t_i), \quad t_i=\min\left(\xij,j=1,\dots,
i-1\right),
\end{equation}
with a function $f$ satisfying
\begin{equation}0\leq
f\leq 1, \quad f'\geq 0.
\end{equation}
The intuition behind the ansatz (\ref{wave}) is that the particles
are inserted into the system one at the time, taking into account
the particles previously inserted. While such a wave function
cannot reproduce all correlations present in the true ground
state, it turns out to capture the leading term in the energy for
dilute gases. The form (\ref{wave}) is  computationally easier to
handle than an ansatz of the type $\prod_{i<j}f(|\x_{i}-\x_{j}|)$,
which might appear more natural in view of the heuristic remarks
after Eq.\ \eqref{partint}.

The function $f$ is chosen to be
\begin{equation}\label{deff}
f(r)=\begin{cases}
f_0(r)/f_{0}(b)&\text{for $0\leq r\leq b$},\\
1&\textrm{for $r>b$},
\end{cases}
\end{equation}
with $f_{0}(r)=u_{0}(r)/r$ the zero energy scattering solution 
defined by 
\eqref{scatteq}. The estimates (\ref{upperbound}) and
(\ref{upperbound2}) are
obtained by somewhat lengthy computations similar as in
\cite{dyson}, but making use of
(\ref{partint}). For details we refer to \cite{LSY1999} and 
\cite{S1999}.
\end{proof}

\subsection{Lower Bound}\label{subsect22}
%%%%%%%%%%%%%%%%%%%%%%
It was explained previously in this section why the lower bound
for the bosonic ground state energy of (\ref{ham}) is not easy to  
obtain.
The three different length scales \eqref{scales} for bosons will play 
a role in the
proof below.
\begin{itemize}
\item The scattering length $a$.
\item The mean particle distance $\rho^{-1/3}$.
\item The `uncertainty principle length' $\ell_{c}$, defined by
$\mu\ell_{c}^{-2}=e_{0}(\rho)$, i.e., $\ell_{c}\sim (\rho a)^{-1/2}$.
\end{itemize}

Our lower bound for $e_{0}(\rho)$ is as follows.
\begin{thm}[\textbf{Lower bound in the thermodynamic 
limit}]\label{lbth}
For a  positive potential $v$ with finite range and $Y$ small enough
\begin{equation}\label{lowerbound}\frac{e_{0}(\rho)}{4\pi\mu\rho 
a}\geq
(1-C\,
Y^{1/17})
\end{equation}
with $C$ a constant. If $v$ does not have finite range, 
but decreases faster than  
$1/r^{3}$ (more precisely, $\int_R^\infty v(r)r^2 dr<\infty$ for some 
$R$) 
then an analogous
bound to (\ref{lowerbound})
holds, but with $CY^{1/17}$ replaced by $o(1)$ as $Y\to 0$.
\end{thm}
It should be noted right away that the error term $-C\, Y^{1/17}$ in
(\ref{lowerbound}) is of no fundamental significance and is
not believed to reflect the true state of affairs. Presumably, it
does not even have the right sign. We mention in passing that
$C$ can be taken to be
$8.9$ \cite{S1999}.

As mentioned at the beginning of this section after
Eq.\ \eqref{eq:thmlimit}, a lower bound on $E_{0}(N,L)$ for finite $N$
and $L$ is of importance for applications to inhomogeneous gases, and
in fact we derive (\ref{lowerbound}) from such a bound. We state it in
the following way:
\begin{thm}[\textbf{Lower bound in a finite box}] \label{lbthm2}
    For a  positive potential $v$ with finite range there is
a $\delta>0$ such that the ground state energy of (\ref{ham}) with 
Neumann
boundary conditions satisfies
\begin{equation}\label{lowerbound2}E_{0}(N,L)/N\geq 4\pi\mu\rho
a \left(1-C\,
Y^{1/17}\right)
\end{equation}
for all $N$ and $L$ with $Y<\delta$ and $L/a>C'Y^{-6/17}$. Here $C$
and $C'$ are positive  constants, independent of $N$ and $L$. (Note 
that the
condition on $L/a$ requires in particular that $N$ must be large
enough, $N>\hbox{\rm (const.)}Y^{-1/17}$.)  As in Theorem \ref{lbth}
such a bound, but possibly with a different error term holds also for
potentials $v$ of infinite range that decrease sufficiently fast at
infinity.
\end{thm}

The first step in the proof of Theorem \ref{lbthm2} is a 
generalization of
a lemma of Dyson, which allows us to replace $v$ by a `soft' 
potential,
at the cost of sacrificing kinetic energy and increasing the
effective range.

\begin{lem}\label{dysonl} Let $v(r)\geq 0$ with finite range 
$R_{0}$. Let
$U(r)\geq 0$
be any function satisfying $\int U(r)r^2dr\leq 1$ and $U(r)=0$ for 
$r<R_{0}$.
Let
${\mathcal B}\subset \R^3$ be star shaped with respect to $0$ (e.g.\
convex with $0\in{\mathcal B}$). Then for all differentiable
functions $\psi$
\begin{equation}\label{dysonlemma}
    \int_{\mathcal B}\left[\mu|\nabla\psi|^2+\mfr1/2
v|\psi|^2\right]
\geq \mu a \int_{\mathcal B} U|\psi|^2.\end{equation}
\end{lem}

\begin{proof}  
Actually, (\ref{dysonlemma}) holds with $\mu |\nabla \psi
(\x)|^2$
replaced by the (smaller) radial kinetic energy,
 $\mu |\partial \psi (\x)/ \partial r|^2$, and  it  suffices to
prove
the analog of (\ref{dysonlemma}) for the integral along each radial
line with fixed angular variables. Along such a line we write
$\psi(\x) = u(r)/r$ with $u(0)=0$. We consider first the special case
when $U$ is a delta-function at some radius $R\geq
R_0$,
i.e., \begin{equation}\label{deltaU}U(r)=\frac{1}{
R^2}\delta(r-R).\end{equation}
For such $U$ the analog of (\ref{dysonlemma}) along the radial line is
\begin{equation}\label{radial}\int_{0}^{R_{1}}
    \{\mu[u'(r)-(u(r)/r)]^2+\mfr1/2v(r)|u(r)]^2\}dr\geq
    \begin{cases}
        0&\text{if $R_{1}<R$}\\
            \mu a|u(R)|^2/R^2&\text{if $R\leq R_{1}$}
\end{cases}
\end{equation}
where $R_{1}$ is the length of the radial line segment in ${\mathcal
B}$.
The case $R_{1}<R$ is trivial,
because $\mu|\partial \psi/\partial r|^2+\mfr1/2 v|\psi|^2\geq 0$.
(Note that positivity of $v$ is used here.) If $R\leq R_{1}$ we
consider the integral on the left side of (\ref{radial}) from 0 
to $R$
instead of $R_{1}$ and
minimize it under the boundary condition that $u(0)=0$
and $u(R)$ is a fixed constant. Since everything is homogeneous in 
$u$ we may
normalize this value to $u(R)=R-a$.
This minimization problem leads to the zero energy
scattering
equation (\ref{scatteq}). Since $v$ is positive, the
solution is a true minimum and not just a
stationary point.

Because $v(r)=0$ for $r>R_{0}$ the solution, $u_{0}$, satisfies 
$u_{0}(r)=r-a$
for $r>R_{0}$.  By partial integration,
\begin{equation}\int_{0}^{R}\{\mu[u'_{0}(r)-(u_{0}(r)/r)]^2+
    \mfr1/2v(r)|u_{0}(r)]^2\}dr=\mu a|R-a|/R\geq \mu
a|R-a|^2/R^2.
    \end
{equation}
But $|R-a|^2/R^2$ is precisely
the right side of (\ref{radial}) if $u$ satisfies the normalization 
condition.

This derivation of (\ref{dysonlemma}) for the special case 
(\ref{deltaU})
implies the
general case, because every $U$ can be written as a
superposition of  $\delta$-functions,
$U(r)=\int R^{-2}\delta(r-R)\,U(R)R^2 dR$, and $\int U(R)R^2 dR\leq 1$
by assumption.
\end{proof}

By dividing $\Lambda$ for given points $\x_{1},\dots,\x_{N}$ into 
Voronoi
cells  ${\mathcal B}_{i}$ that contain all points
closer to $\x_{i}$ than to $\x_{j}$ with $j\neq i$ (these
cells are star shaped w.r.t. $\x_{i}$, indeed convex), the
following corollary of Lemma \ref{dysonl} can be derived in the same
way as the corresponding  Eq.\ (28) in \cite{dyson}.

\begin{corollary}\label{2.6} For any $U$ as in Lemma \ref{dysonl}
\begin{equation}\label{corollary}H_{N}\geq \mu a W\end{equation}
with $W$ the multiplication operator 
\begin{equation}\label{W}W(\x_{1},\dots,\x_{N})=\sum_{i=1}^{N}U(t_{i}),
\end{equation}
where $t_{i}$ is the distance of $\x_{i}$ to its {\it nearest
neighbor} among the other points $\x_{j}$, $j=1,\dots, N$, i.e.,
\begin{equation}\label{2.29}t_{i}(\x_{1},\dots,\x_{N})=\min_{j,\,j\neq
i}|\x_{i}-\x_{j}|.\end{equation}
\end{corollary}
\noindent
(Note that $t_{i}$ has here a slightly different meaning than in
(\ref{form}), where it denoted the distance to the nearest neighbor
among the $\x_{j}$ with $j\leq i-1$.)

Dyson considers in \cite{dyson} a one parameter family of $U$'s that
is essentially the same as the following choice, which is convenient 
for the
present purpose:
\begin{equation}\label{softened}
U_{R}(r)=\begin{cases}3(R^3-R_{0}^3)^{-1}&\text{for
$R_{0}<r<R$ }\\
0&\text{otherwise.}
\end{cases}
\end{equation}
We denote the corresponding interaction (\ref{W}) by $W_R$. For the 
hard core
gas one obtains
\begin{equation}\label{infimum}E(N,L)\geq 
\sup_{R}\inf_{(\x_{1},\dots,\x_{N})}
\mu a
W_R(\x_{1},\dots,\x_{N})\end{equation}
where the infimum is over $(\x_{1},\dots,x_{N})\in\Lambda^{N}$ with
$|\x_{i}-\x_{j}|\geq R_{0}=a$,
because of the hard core. At fixed $R$ simple geometry gives
\begin{equation}\label{fixedR}\inf_{(\x_{1},\dots,\x_{N})}
W_R(\x_{1},\dots,\x_{N})\geq \left(\frac{A}{R^3}-\frac{B}{ \rho
R^6}\right)\end{equation}
with certain constants $A$ and $B$. An evaluation of these constants
gives Dyson's bound
\begin{equation}E(N,L)/N\geq \frac{1}{10\sqrt 2} 4\pi\mu \rho
a.\end{equation}

The main reason this method does not give a better bound is that
$R$ must be chosen quite big, namely of the order of the mean
particle distance $\rho^{-1/3}$, in order to guarantee that the
spheres of radius $R$ around the $N$ points overlap. Otherwise the
infimum of $W_R$ will be zero. But large $R$ means that $W_R$ is
small. It should also be noted that this method does not work for
potentials other than hard spheres: If $|\x_{i}-\x_{j}|$ is
allowed to be less than $R_{0}$, then the right side of
(\ref{infimum}) is zero because $U(r)=0$ for $r<R_{0}$.

For these reasons we take another route. We still use  Lemma
\ref{dysonl} to get into the soft potential regime, but we do {\it
not} sacrifice {\it all} the kinetic energy as in
(\ref{corollary}). Instead we write, for $\varepsilon>0$
\begin{equation}
    H_{N}=\varepsilon H_{N}+(1-\varepsilon)H_{N}\geq \varepsilon
T_{N}+(1-\varepsilon)H_{N}
\end{equation}
with $T_{N}=-\sum_{i}\Delta_{i}$ and use (\ref{corollary}) only for 
the
part $(1-\varepsilon)H_{N}$. This gives
\begin{equation}\label{halfway}H_{N}\geq \varepsilon 
T_{N}+(1-\varepsilon)\mu
a W_R.\end{equation} We consider the operator on the right side
from the viewpoint of first order perturbation theory, with
$\varepsilon T_{N}$ as the unperturbed  part, denoted $H_{0}$.

The ground state of $H_{0}$ in a box of side length $L$ is
$\Psi_{0}(\x_{1},\dots,\x_{N})\equiv L^{-3N/2}$ and we denote
expectation values in this state by $\langle\cdot\rangle_{0}$.
A  computation, cf.\ Eq.\ (21) in \cite{LY1998} (see also Eqs.\ 
\eqref{2dfirstorder}--\eqref{firstorder2}), gives
\begin{eqnarray}\label{firstorder}4\pi\rho\left(1-\mfr1/N\right)&\geq&
\langle W_R\rangle_{0}/N  \nonumber   \\ &\geq& 4\pi\rho
\left(1-\mfr1/N\right)\left(1-\mfr{2R}/L\right)^3
\left(1+4\pi\rho(R^3-R_{0}^3)/3\right)^{-1}.\nonumber\\
\end{eqnarray}
The rationale behind the various factors is as follows: $(1-\mfr1/N)$
comes from the fact that the number of pairs is $N(N-1)/2$ and not
$N^2/2$, $(1-{2R}/L)^3$ takes into account the fact that the particles
do not interact beyond the boundary of $\Lambda$, and the last factor
measures the probability to find another particle within the
interaction range of the potential $U_R$ for a given particle.

The estimates (\ref{firstorder}) on the first order term look at first
sight quite promising, for if we let $L\to \infty$, $N\to \infty$ with
$\rho=N/L^3$ fixed, and subsequently take $R\to 0$, then $\langle
W_R\rangle_{0}/N$ converges to $4\pi\rho$, which is just what is
desired.  But the first order result (\ref{firstorder}) is not a
rigorous bound on $E_0(N,L)$, we need {\it error estimates}, and these
will depend on $\varepsilon$, $R$ and $L$.

We now recall {\it Temple's inequality} \cite{TE} for the expectation
values of an operator $H=H_{0}+V$ in the ground state
$\langle\cdot\rangle_{0}$ of $H_{0}$. It is a simple
consequence of the operator inequality
\begin{equation}(H-E_{0})(H-E_{1})\geq 0\end{equation}
for the two lowest eigenvalues, $E_{0}<E_{1}$, of
$H$ and reads
\begin{equation}\label{temple}
E_{0}\geq \langle 
H\rangle_{0}-\frac{\langle
H^2\rangle_{0}-\langle
H\rangle_{0}^2}{E_{1}-\langle H\rangle_{0}}
\end{equation}
provided $E_{1}-\langle H\rangle_{0}>0$.
Furthermore, if $V\geq 0$ we may use $E_{1}\geq E_{1}^{(0)}$= second 
lowest
eigenvalue of $H_{0}$ and replace $E_{1}$ in (\ref{temple}) by 
$E_{1}^{(0)}$.

%BEGIN CHANGE
{F}rom (\ref{firstorder}) and (\ref{temple}) we get the estimate
\begin{equation}\label{estimate2}\frac{E_{0}(N,L)}{ N}\geq 4\pi \mu 
a\rho
\left(1-{\mathcal
E}(\rho,L,R,\varepsilon)\right)\end{equation}
with
\begin{eqnarray}\nonumber
&& 1-{\mathcal
E}(\rho,L,R,\varepsilon) \\ && =(1-\varepsilon)\left(1-\mfr1/{\rho
L^3}\right)\left(1-\mfr{2R}/L\right)^3
\left(1+\mfr{4\pi}/3\rho(R^3-R_{0}^3)\right)^{-1}\nonumber\\
&&\quad \times \left(1-\frac{\mu a\big(\langle
W_R^2\rangle_0-\langle W_R\rangle_0^2\big)}{\langle
W_R\rangle_0\big(E_{1}^{(0)}-\mu a\langle W_R\rangle_0\big)}\right). 
\label{error}
\end{eqnarray}
%END CHANGE
To evaluate this further one may use the estimates (\ref{firstorder}) 
and the
bound
\begin{equation}\label{square}
\langle W_R^2\rangle_0\leq 3\frac N{R^3-R_0^3}\langle W_R\rangle_0
\end{equation}
which follows from $U_R^2=3({R^3-R_0^3})^{-1}U_R$ together with the
Schwarz inequality. A glance at the form of the error term reveals,
however, that it is {\it not} possible here to take the thermodynamic
limit $L\to\infty$ with $\rho$ fixed: We have
$E_{1}^{(0)}=\varepsilon\pi^2\mu/L^2$ (this is the kinetic energy of a
{\it single} particle in the first excited state in the box), and the
factor $E_{1}^{(0)}-\mu a\langle W_R\rangle_0$ in the denominator in
(\ref{error}) is, up to unimportant constants and lower order terms,
$\sim (\varepsilon L^{-2}-a\rho^2L^3)$. Hence the denominator
eventually becomes negative and Temple's inequality looses its
validity if $L$ is large enough.

As a way out of this dilemma we divide the big box $\Lambda$ into 
cubic {\it
cells} of side length $\ell$ that is kept {\it fixed} as $L\to 
\infty$.  The
number of cells, $L^3/\ell^3$, on the other hand, increases with 
$L$.  The $N$
particles are distributed among these cells, and we use 
(\ref{error}), with
$L$
replaced by $\ell$, $N$ by the particle number, $n$, in a cell and 
$\rho$ by
$n/\ell^3$, to estimate the energy in each cell with {\it Neumann} 
conditions
on the boundary. 
% This boundary condition leads to lower energy than any other
% boundary condition.  
For each distribution of the particles we add the
contributions from the cells, neglecting interactions across 
boundaries.
Since
$v\geq 0$ by assumption, this can only lower the energy.  Finally, we 
minimize
over all possible choices of the particle numbers for the various 
cells
adding up to $N$.  The energy obtained in this way is a lower bound to
$E_0(N,L)$,
because we are effectively allowing discontinuous test functions for 
the
quadratic form given by $H_N$.

In mathematical terms, the cell method leads to
\begin{equation}\label{sum}
E_0(N,L)/N\geq(\rho\ell^3)^{-1}\inf \sum_{n\geq 0}c_nE_0(n,\ell)
\end{equation}
where the infimum is over all choices of coefficients $c_n\geq 0$ 
(relative
number of cells containing exactly $n$ particles), satisfying the 
constraints
\begin{equation}\label{constraints}
\sum_{n\geq 0}c_n=1,\qquad \sum_{n\geq 0}c_n n=\rho\ell^3.
\end{equation}

The minimization problem for the distributions of the particles among 
the
cells would be easy if we knew that the ground state energy 
$E_0(n,\ell)$ (or
a
good
lower bound to it) were convex in $n$.  Then we could immediately 
conclude
that
it is best to have the particles as evenly distributed among the 
boxes as
possible, i.e., $c_n$ would be zero except for the $n$ equal to the
integer closest to
$\rho\ell^3$. This would give
\begin{equation}\label{estimate3}\frac{E_{0}(N,L)}{ N}\geq 4\pi \mu 
a\rho
\left(1-{\mathcal E}(\rho,\ell,R,\varepsilon)\right)\end{equation} 
i.e.,
replacement of $L$ in (\ref{estimate2}) by $\ell$, which is 
independent of
$L$.
The blow up of ${\mathcal E}$ for $L\to\infty$ would thus be avoided.

Since convexity of $E_0(n,\ell)$ is not known (except in the 
thermodynamic
limit)
we must resort to other means to show that $n=O(\rho\ell^3)$ in all
boxes. The rescue
comes from {\it superadditivity} of $E_{0}(n,\ell)$, i.e., the 
property
\begin{equation}\label{superadd}
 E_0(n+n',\ell)\geq E_0(n,\ell)+E_0(n',\ell)
\end{equation}
which follows immediately from $v\geq 0$ by dropping the interactions 
between
the $n$ particles and the $n'$ particles. The bound (\ref{superadd}) 
implies
in
particular that for any $n,p\in{\mathbb N}$ with $n\geq p$
\begin{equation}\label{superadd1}
E_{0}(n,\ell)\geq [n/p]\,E_{0}(p,\ell)\geq \frac n{2p}E_{0}(p,\ell)
\end{equation}
since the largest integer $[n/p]$ smaller than $n/p$ is in any case 
$\geq
n/(2p)$.

The way (\ref{superadd1}) is used is as follows:
Replacing $L$ by $\ell$, $N$ by $n$ and $\rho$ by $n/\ell^3$ in
(\ref{estimate2})  we have for fixed $R$ and $\varepsilon$
\begin{equation}\label{estimate4}
E_{0}(n,\ell)\geq\frac{ 4\pi \mu a}{\ell^3}n(n-1)K(n,\ell)
\end{equation}
with a certain function $K(n,\ell)$ determined by (\ref{error}).
We shall see that $K$ is monoton\-ously decreasing in $n$, so that
if $p\in{\mathbb N}$  and $n\leq p$ then
\begin{equation}\label{n<p}
E_{0}(n,\ell)\geq\frac{ 4\pi \mu a}{\ell^3}n(n-1)K(p,\ell).
\end{equation}
We now split the sum in (\ref{sum}) into two parts.
For $n<p$ we use (\ref{n<p}), and for $n\geq p$ we use 
(\ref{superadd1})
together with (\ref{n<p}) for $n=p$. The task is thus to minimize
\begin{equation}\label{task}
\sum_{n<p}c_n n(n-1)+\mfr1/2\sum_{n\geq p}c_nn(p-1)
\end{equation}
subject to the constraints ({\ref{constraints}).
Putting
\begin{equation}
k:=\rho\ell^3 \quad\text{and}\quad t:=\sum_{n<p}c_n n\leq k
\end{equation}
we have $\sum_{n\geq p}c_n n=k-t$, and since
$n(n-1)$ is convex in $n$ and vanishes for $n=0$, and 
$\sum_{n<p}c_n\leq 1$, the expression
(\ref{task})
is
\begin{equation}
\geq t(t-1)+\mfr1/2(k-t)(p-1).
\end{equation}
We have to minimize this for $1\leq t\leq k$. If $p\geq 4k$ the 
minimum is
taken
at $t=k$ and is equal to $k(k-1)$. Altogether we have thus shown that
%BEGIN CHANGE
\begin{equation}\label{estimate1}
\frac{E_{0}(N,L)}{ N}\geq 4\pi \mu a\rho\left(1-\frac1{\rho\ell^3} 
\right)
K(4\rho\ell^3,\ell).
\end{equation}
%END CHANGE

What remains is to take a closer look at $K(4\rho\ell^3,\ell)$,
which depends on the parameters $\varepsilon$ and $R$ besides
$\ell$, and choose the parameters in an optimal way.
%BEGIN CHANGE
{F}rom (\ref{error}) and (\ref{square}) we obtain
\begin{eqnarray}\label{Kformula}
K(n,\ell)&=&(1-\varepsilon) \left(1-\mfr{2R}/\ell\right)^3
\left(1+\mfr{4\pi}/3(R^3-R_{0}^3)\right)^{-1}
\nonumber
\\ &\times&\left(1-\frac3\pi
\frac{an}{(R^3-R_{0}^3)(\pi\varepsilon\ell^{-2}-4a\ell^{-3}n(n-1))}\right).
\end{eqnarray}
The estimate (\ref{estimate4}) with this $K$ is valid as long as the
denominator in the last factor
%END CHANGE
in (\ref{Kformula}) is $\geq 0$, and in order to have a formula
for
all $n$ we can take 0 as a
trivial lower bound in other cases or when (\ref{estimate4}) is
negative. As required
for (\ref{n<p}), $K$ is monotonously decreasing in $n$. We now insert
$n=4\rho\ell^3$ and obtain
%BEGIN CHANGE
\begin{eqnarray}\label{Kformula2}
K(4\rho\ell^3,\ell)&\geq&(1-\varepsilon)\left(1-\mfr{2R}/\ell\right)^3
\left(1+({\rm const.})Y(\ell/a)^3 (R^3-R_{0}^3)/\ell^3\right)^{-1}
\nonumber
\\ &\times&\left(1-
\frac{\ell^3}{(R^3-R_{0}^3)}\frac{({\rm const.})Y}
{(\varepsilon(a/\ell)^{2}-({\rm const.})Y^2(\ell/a)^3)}\right)
\end{eqnarray}
with $Y=4\pi\rho a^3/3$ as before. Also, the factor
\begin{equation}
\left(1-\frac1{\rho\ell^3} \right)=(1-({\rm 
const.})Y^{-1}(a/\ell)^{3})
\end{equation}
%END CHANGE
in (\ref{estimate1})
(which is the ratio between
$n(n-1)$ and $n^2$) must not be forgotten. We now make the ansatz
\begin{equation}\label{ans}
\varepsilon\sim Y^\alpha,\quad a/\ell\sim Y^{\beta},\quad
(R^3-R_{0}^3)/\ell^3\sim Y^{\gamma}
\end{equation}
with exponents $\alpha$, $\beta$ and $\gamma$ that we choose
in an optimal way. The conditions to be met are as follows:
%BEGIN CHANGE
\begin{itemize}
\item $\varepsilon(a/\ell)^{2}-({\rm const.})Y^2(\ell/a)^3>0$. This
holds for all small enough $Y$, provided
$\alpha+5\beta<2$ which follows from the conditions below.
\item $\alpha>0$ in order that $\varepsilon\to 0$ for $Y\to 0$.
\item $3\beta-1>0$ in order that  $Y^{-1}(a/\ell)^{3}\to 0$ for for 
$Y\to
0$.
\item $1-3\beta+\gamma>0$ in order that
$Y(\ell/a)^{3}(R^3-R_{0}^3)/\ell^3\to 0$ for for $Y\to 0$.
%END CHANGE
\item $1-\alpha-2\beta-\gamma>0$ to control the last factor in
(\ref{Kformula2}).
\end{itemize}
Taking
\begin{equation}\label{exponents}
\alpha=1/17,\quad \beta=6/17,\quad \gamma=3/17
\end{equation}
all these conditions are satisfied, and
%BEGIN CHANGE
\begin{equation}
\alpha= 3\beta-1=1-3\beta+\gamma=1-\alpha-2\beta-\gamma=1/17.
\end{equation}
It is also clear that
$2R/\ell\sim Y^{\gamma/3}=Y^{1/17}$, up to higher order terms.
%END CHANGE
This completes the proof of Theorems \ref{lbth} and \ref{lbthm2}, for
the case of potentials with finite range. By optimizing the
proportionality constants in (\ref{ans}) one can show that $C=8.9$ is
possible in Theorem \ref{lbth} \cite{S1999}. The extension to
potentials of infinite range but finite scattering length is obtained
by approximation by finite range potentials, controlling the change of
the scattering length as the cut-off is removed. See Appendix A in
\cite{LY2d} and Appendix B in \cite{LSY1999} for details. We remark
that a slower decrease of the potential than $1/r^3$ implies infinite
scattering length.
\hfill$\Box$\bigskip

The exponents (\ref{exponents}) mean in particular that
\begin{equation}a\ll R\ll \rho^{-1/3}\ll \ell \ll(\rho
a)^{-1/2},\end{equation}
whereas Dyson's method required $R\sim \rho^{-1/3}$ as already 
explained.
The condition $\rho^{-1/3}\ll \ell$ is required in order to have many
particles in each box and thus $n(n-1)\approx n^2$. The condition
$\ell \ll(\rho a)^{-1/2}$ is necessary for a spectral 
gap $\gg e_{0}(\rho)$ in Temple's inequality. It is also clear that
this choice of $\ell$  would lead to a far too big
energy and no bound for $ e_{0}(\rho)$ if we had chosen Dirichlet 
instead of
Neumann boundary
conditions for the cells. But with the latter the method works!

\section{The Dilute Bose Gas in 2D} \label{sect2d}

In contrast to the three-dimensional theory, the two-dimensional Bose
gas began to receive attention only relatively late.  The first
derivation of the correct asymptotic formula was, to our knowledge,
done by Schick \cite{schick} for a gas of hard discs. He found
\begin{equation}
    e(\rho)  \approx 4\pi \mu \rho |\ln(\rho a^2) |^{-1}.
\label{2den}
\end{equation}
This was accomplished by an infinite summation of `perturbation 
series'
diagrams. Subsequently, a corrected modification of \cite{schick} was
given in \cite{hines}. Positive temperature extensions were given in
\cite{popov} and in \cite{fishho}. All this work involved an analysis 
in
momentum space, with the exception
of a method due to one of us that works directly in configuration 
space
\cite{Lieb63}.  Ovchinnikov \cite{Ovch} derived \eqref{2den} by using,
basically, the method in \cite{Lieb63}. These derivations require
several unproven assumptions and are not rigorous. 

In two dimensions the scattering length $a$ is defined using the zero
energy scattering equation (\ref{3dscatteq}) but instead of
$\psi(r)\approx 1-a/r$ we now impose the asymptotic condition
$\psi(r)\approx \ln(r/a)$.
This is explained  in the appendix to \cite{LY2d}.

Note that in two dimensions the ground state energy  
could not possibly be $e_0(\rho)  \approx 4\pi \mu
\rho a$ as in three dimensions because that would be dimensionally 
wrong. 
Since $e_0(\rho) $ should essentially be proportional to $\rho$,
there is apparently no room for an $a$ dependence --- which is
ridiculous! It turns out that this dependence comes about in the
$\ln(\rho a^2)$ factor.

One of the intriguing facts about \eqref{2den} is that the energy for 
$N$
particles is {\it not equal} to $N(N-1)/2$  times the energy for two
particles in the low density limit --- as is the case in
three dimensions.  The latter quantity,  $E_0(2,L)$, is, 
asymptotically
for large $L$, equal to $8\pi \mu L^{-2} \left[ \ln(L^2/a^2)
\right]^{-1}$.  (This is seen in an analogous way as \eqref{partint}. 
The 
three-dimensional boundary condition $\psi_0(|\x|=R)=1-a/R$ is
replaced by $\psi_0(|\x|=R)=\ln (R/a)$ and moreover it has to be taken
into account that with this normalization $\|\psi_0\|^2={\rm
(volume)}(\ln (R/a))^2$ (to leading order), instead of just the volume
in the three-dimensional case.)  Thus, if the $N(N-1)/2$ rule were to
apply, \eqref{2den} would have to be replaced by the much smaller
quantity $4\pi \mu
\rho\left[ \ln(L^2/a^2) \right]^{-1}$. In other words, $L$, which 
tends
to $\infty$ in the thermodynamic limit, has to be replaced by the mean
particle separation, $\rho^{-1/2}$ in the logarithmic factor. Various
poetic formulations of this curious fact have been given, but the fact
remains that the non-linearity is something that does not  occur in 
more
than two dimensions and its precise nature is hardly obvious,
physically. This anomaly is the main reason that the two-dimensional 
case is
not a trivial extension of the three-dimensional one.

Eq.\ \eqref{2den} was proved in \cite{LY2d} for nonnegative, finite
range two-body potentials by finding upper and lower bounds of the
correct form, using similar ideas as in the previous section for the
three-dimensional case. We discuss below the modifications that have 
to
be made in the present two-dimensional case.  The restriction to
finite range can be relaxed as in three dimensions, but the
restriction to nonnegative $v$ cannot be removed in the current state
of our methodology.  The upper bounds will have relative remainder
terms O($|\ln(\rho a^2)|^{-1}$) while the lower bound will have
remainder O($|\ln(\rho a^2)|^{-1/5}$).  It is claimed in \cite{hines}
that the relative error for a hard core gas is negative and O$(\ln
|\ln(\rho a^2)||\ln(\rho a^2)|^{-1})$, which is consistent with our
bounds.

The upper bound is derived in complete analogy with the three
dimensional case. The function $f_0$ in the variational ansatz 
\eqref{deff} is in
two dimensions also the zero energy scattering solution --- but for 
2D, of course. The result is 
\begin{equation}\label{upperbound3}
E_{0}(N,L)/N\leq \frac{2\pi\mu\rho}{\ln(b/a)-\pi\rho 
b^{2}}\left(1+{\rm 
O}([\ln(b/a)]^{{-1}})\right).
\end{equation}
The minimum over $b$ of the leading term is obtained for 
$b=(2\pi\rho)^{{-1/2}}$. Inserting this in \eqref{upperbound3} we 
thus 
obtain
\begin{equation}\label{upperbound1}
E_{0}(N,L)/N\leq \frac{4\pi\mu\rho}{|\ln(\rho a^{2})|}\left(1+{\rm 
O}(|\ln(\rho a^{2})|^{{-1}})\right).
\end{equation}

To prove the lower bound the essential new step is to modify Dyson's 
lemma
for 2D. The 2D version of
Lemma \ref{dysonl} is:

\begin{lem}\label{dyson2d} 
Let $v(r)\geq0$ and $v(r)=0$ for $r>R_0$.
 Let $U(r)\geq 0$ be any function satisfying
\begin{equation}\label{1dyson}
 \int_0^\infty U(r)\ln(r/a)rdr \leq 1~~~~~{\rm and}~~~~~ U(r)=0 
~~~{\rm
for}~
 r<R_0.
\end{equation}
Let ${\mathcal B}\subset \R^2$ be star-shaped  with respect
to $0$ (e.g.\ convex
with $0\in{\mathcal B}$).
Then, for all functions $\psi$ in the Sobolev space 
$H^1(\mathcal{B})$,
\begin{equation}
\int_{\mathcal B} \left(\mu|\nabla \psi (\x)|^2 + \half v(|\x|)
|\psi (\x)|^2\right)~d\x  
\geq  \mu  \int_{\mathcal B} U(|\x|)
|\psi (\x)|^2 ~d\x.
\label{dysonlem2d}
\end{equation}
\end{lem}

\begin{proof}
In polar coordinates, $r,\theta$, one has 
$|\nabla \psi|^2 \geq |\partial \psi /\partial r|^2$. Therefore, it 
suffices to 
prove that for each angle $\theta \in [0,2\pi)$, and with
$\psi (r,\theta)$ denoted simply by $f(r)$,
\begin{equation} \label{radial2}
\int_0^{R(\theta)}\left( \mu |\partial f(r) /\partial r|^2 + 
\half v(r)|f(r)|^2 \right)rdr \geq  
 \mu  \int_0^{R(\theta)}  U(r)|f(r)|^2 ~rdr,
\end{equation}  
where $R(\theta)$ denotes the distance of the origin to the boundary
of $\mathcal{B}$ along the ray $\theta$. 

If $R(\theta) \leq R_0$ then \eqref{radial2} is trivial because the 
right side is zero while the left side is evidently nonnegative. 
(Here,
$v\geq0$ is used.)

If $R(\theta) > R_0$ for some given
value of $\theta$, consider the disc $\mathcal{D}(\theta)=
\{\x\in \mathbb{R}^2 \ :\ 0\leq |\x|\leq R(\theta) \}$ centered at the
origin in $\mathbb{R}^2$ and of radius $R(\theta) $.
Our function $f$ defines a spherically
symmetric function, $\x\mapsto f(\vert \x\vert)$ on
$\mathcal{D}(\theta)$, and \eqref{radial2} is 
equivalent to 
\begin{equation}\label{disc}
\int_{{\mathcal D}(\theta)} \left(\mu|\nabla f (|\x|)|^2 + 
\frac{1}{2}v(|\x|)
|f(|\x|)|^2\right)d\x\geq 
\mu\int_{{\mathcal D}(\theta)} U(|\x|)|f(|\x|)|^2 d\x.
\end{equation}

Now choose some $R\in (R_0,\ R(\theta))$ and note that the left side 
of
\eqref{disc} is not smaller than the same quantity with 
${\mathcal D}(\theta)$ replaced by the smaller disc ${\mathcal D}_R=
\{\x\in \mathbb{R}^2 \ :\ 0\leq |\x|\leq R \}$. (Again, $v\geq 0$ is 
used.)
We now minimize this integral over ${\mathcal D}_R$, fixing $f(R)$. 
This minimization problem leads to the zero energy scattering 
equation. 
Plugging in the solution and integrating by parts leads to
\begin{equation} \label{pointwise}
2\pi \int_0^{R(\theta)}\left( \mu |\partial f(r) /\partial r|^2 + 
\frac{1}{2}v(r)|f(r)|^2 \right)rdr \geq    \frac{2\pi \mu}{\ln 
(R/a)}|f(R)|^2 .
\end{equation}
The proof is completed
by multiplying  both sides of \eqref{pointwise} by $U(R)R\ln(R/a)$ 
and 
integrating with respect to $R$ from $R_0$ to $R(\theta)$.
\end{proof}

As in Corollary \ref{2.6}, Lemma \ref{dyson2d} can be used to bound 
the 
many body Hamiltonian $H_N$ from below, as follows:
\begin{corollary}
For any $U$ as in Lemma \ref{dyson2d} and any $0<\varepsilon<1$
\begin{equation}\label{epsilonbd}
H_N \geq \varepsilon T_N +(1-\varepsilon)\mu W
\end{equation}
with $T_N=-\mu\sum_{i=1}^{N}\Delta_{i}$ and 
\begin{equation}
\label{W2}W(\x_{1},\dots,\x_{N})=\sum_{i=1}^{N}U\left(\min_{j,\,j\neq
i}|\x_{i}-\x_{j}|.\right).
\end{equation}
\end{corollary} 

For $U$ we choose the following functions, parameterized by $R>R_{0}$:
\begin{equation}U_{R}(r)=\begin{cases}\nu(R)^{-1}&\text{for
$R_{0}<r<R$ }\\
0&\text{otherwise}
\end{cases}
\end{equation}
with $\nu(R)$ chosen so that
\begin{equation}
\int _{R_{0}}^{R}U_{R}(r)\ln(r/a)r\,dr=1
\end{equation}
for all $R>R_{0},$
i.e.,
\begin{equation}\label{nu}
\nu(R)=\int_{R_{0}}^{R}\ln(r/a)r\,dr=\mfr1/4 \left\{R^{2}
\left(\ln(R^{2}/a^{2})-1\right)-R_{0}^{2}
\left(\ln(R_{0}^{2}/a^{2})-1\right)\right\}.
\end{equation}
The nearest neighbor 
interaction \eqref{W2} corresponding to $U_{R}$ will be denoted 
$W_{R}$.

As in Subsection \ref{subsect22} we shall need estimates  on the 
expectation
value, $\langle W_R\rangle_{0}$,  of $W_{R}$ in the ground state of
$\varepsilon T_N$ of \eqref{epsilonbd} with
Neumann boundary conditions. This is just the average value of $W_{R}$
in a hypercube in $\R^{2N}$.  Besides the
normalization factor $\nu(R)$, the computation involves 
the volume (area) of the support of
$U_{R}$, which is
 \begin{equation} A(R)=\pi(R^{2}-R_{0}^{2}).
\end{equation}

In contrast to the three-dimensional situation the normalization 
factor
$\nu(R)$ is not just a constant ($R$ independent) multiple of $A(R)$;
 the factor $\ln(r/a)$ in \eqref{1dyson} accounts for the more
complicated expressions in the two-dimensional case.   Taking into
account that $U_{R}$ is proportional to the characteristic function 
of 
a disc of radius $R$ with a hole of radius $R_{0}$, the following 
inequalities for $n$ particles in a box of side length 
$\ell$ are obtained by the same geometric reasoning as lead to 
\eqref{firstorder}, cf.\  
\cite{LY1998}:
\begin{eqnarray}\label{2dfirstorder}
\langle 
W_R\rangle_{0}&\geq&\frac {n}{\nu(R)}
\left(1-\mfr {2R}/{\ell}\right)^2\left[1-(1-Q)^{(n-1)}\right]\\
\langle 
W_R\rangle_{0}&\leq&
\frac 
{n}{\nu(R)}\left[1-(1-Q)^{(n-1)}\right]
\end{eqnarray}
with 
\begin{equation}
Q=A(R)/\ell^{2}
\end{equation}
  being   the relative volume occupied by the
support of the potential $U_{R}$.
Since $U_{R}^{2}=\nu(R)^{-1}U_{R}$ we also have
\begin{equation}
\langle 
W_R^{2}\rangle_{0}\leq \frac n{\nu(R)}\langle 
W_R\rangle_{0}.
\end{equation}

As in \cite{LY1998} we estimate $[1-(1-Q)^{(n-1)}]$ by
\begin{equation}
(n-1)Q\geq \left[1-(1-Q)^{(n-1)}\right]\geq \frac{(n-1)Q}{1+(n-1)Q}
\end{equation}
This gives
\begin{eqnarray}\label{firstorder2}
\langle 
W_R\rangle_{0}&\geq&\frac {n(n-1)}{\nu(R)}
\cdot\frac{Q}{1+(n-1)Q},\\
\langle 
W_R\rangle_{0}&\leq&
\frac {n(n-1)}{\nu(R)}
\cdot Q\ .
\end{eqnarray}

{F}rom  Temple's inequality \cite{TE} we 
obtain like in \eqref{temple} the estimate
\begin{equation}\label{temple2d}E_{0}(n,\ell)
\geq (1-\varepsilon)\langle 
W_R\rangle_{0}\left(1-\frac{\mu \big(\langle
W_R^2\rangle_0-\langle W_R\rangle_0^2\big)}{\langle
W_R\rangle_0\big(E_{1}^{(0)}-\mu \langle W_R\rangle_0\big)}\right)
\end{equation}
where
\begin{equation}
E_{1}^{(0)}=\frac{\varepsilon\mu}{\ell^{2}}
\end{equation}
is the energy of the lowest excited state of $\varepsilon T_n$. 
This estimate is valid for $E_{1}^{(0)}/\mu > \langle 
W_R\rangle_0$, i.e., it is important that $\ell$ is not too big. 

Putting \eqref{firstorder2}--\eqref{temple2d} together we obtain 
the estimate
\begin{equation}\label{alltogether}E_{0}(n,\ell)\geq 
\frac{n(n-1)}{\ell^{2}}\,\frac {A(R)}{\nu(R)}\,
K(n)
\end{equation}
with
\begin{equation}\label{k}
K(n)=(1-\varepsilon)\cdot 
\frac{(1-\mfr{2R}/{\ell})^{2}}{1+(n-1)Q}\cdot
\left(1-\frac n{(\varepsilon\,\nu(R)/\ell^{2})-n(n-1)\,Q}\right)
\end{equation}
Note that $Q$ depends on $\ell$ and $R$, and $K$ depends on 
$\ell$, $R$ and $\varepsilon$ besides $n$.  We have here dropped the 
term  $\langle W_R\rangle_0^2$ in the numerator in \eqref{temple2d},  
which is  appropriate  for the purpose of a lower bound. 

We note that $K$ is monotonically decreasing in $n$, so for a given 
$n$ we may replace $K(n)$ by $K(p)$ provided $p\geq n$.  As 
explained in the previous section, 
\eqref{superadd}--\eqref{estimate1}, 
convexity of $n\mapsto n(n-1)$ together with 
superadditivity of $E_{0}(n,\ell)$ in $n$ 
leads, for $p=4\rho\ell^{2}$,  to an estimate for the energy of $N$ 
particles in the 
large box when the  side length $L$ is  an integer multiple of 
$\ell$: 
\begin{equation}\label{alltogether2}E_{0}(N,L)/N\geq \frac 
{\rho A(R)}{\nu(R)}\left (1-\frac 1{\rho\ell^{2}}\right) 
K(4\rho\ell^{2})
\end{equation}
with $\rho=N/L^2$.

Let us now look at the conditions on the parameters $\varepsilon$, 
$R$ 
and $\ell$ that have to be met in order to 
obtain a lower bound with the same leading term as the upper bound 
\eqref{upperbound1}.

{F}rom \eqref{nu} we have 
\begin{equation}\label{alltogether3}
\frac{A(R)}{\nu(R)}=\frac{4\pi}
{\left(\ln(R^{2}/a^{2})-1\right)}\left(1-{\rm 
O}((R_{0}^{2}/R^{2})\ln(R/R_{0})\right)
\end{equation}
We thus see that as long as $a<R<\rho^{-1/2}$ the logarithmic factor
in the denominator in \eqref{alltogether3} has the right form for a 
lower 
bound. Moreover, for 
Temple's inequality the denominator in the third factor in \eqref{k} 
must be positive. With $n=4\rho\ell^2$ and 
        $\nu(R)\geq {\rm(const.)} R^2\ln(R^2/a^2)\ {\rm for}\ R\gg 
R_{0}$,
this condition amounts to
\begin{equation}\label{templecond}
{\rm (const.)}\varepsilon \ln(R^2/a^2) /\ell^{2}>\rho^{2}\ell^{4}.
\end{equation}
The relative error terms in \eqref{alltogether2} that have to be $\ll 
1$ 
are
\begin{equation}\label{errors}
        \varepsilon,\quad \frac{1}{\rho\ell^{2}},\quad 
\frac{R}{\ell},\quad\rho R^2,\quad
\frac{\rho\ell^4}{\varepsilon R^2\ln(R^2/a^2)}.
\end{equation}
We now choose
\begin{equation}
\varepsilon\sim|\ln(\rho a^2)|^{-1/5},
\quad \ell\sim \rho^{-1/2}|\ln(\rho a^2)|^{1/10},
\quad R\sim \rho^{-1/2}|\ln(\rho a^2)|^{-1/10}
\end{equation}

Condition \eqref{templecond} is satisfied since the left side is 
$>{\rm
(const.)}|\ln(\rho a^2)|^{3/5}$ and the right side is $\sim |\ln(\rho
a^2)|^{2/5}$. The first three error terms in \eqref{errors} are all of
the same order, $|\ln(\rho a^2)|^{-1/5}$, the last is $\sim
|\ln(\rho a^2)|^{-1/5}(\ln|\ln(\rho a^2)|)^{-1}$. With these 
choices, \eqref{alltogether2} thus leads to the following:

\begin{thm}[{\bf Lower bound}]
For all $N$ and $L$ large enough such that $L>{\rm (const.)}
\rho^{-1/2}|\ln(\rho a^2)|^{1/10}$ and 
$N>{\rm (const.)}|\ln(\rho a^2)|^{1/5}$ with $\rho=N/L^2$, the 
ground state energy with Neumann boundary condition satisfies
\begin{equation} \label{lower}
E_{0}(N,L)/N\geq \frac{4\pi\mu\rho}{|\ln(\rho a^2)|}
\left(1-{\rm O}(|\ln(\rho a^2)|^{-1/5})\right). 
\end{equation}  
\end{thm}

In combination with the upper bound \eqref{upperbound1} this also 
proves
\begin{thm}[{\bf Energy at low density in the thermodynamic 
limit}]
\begin{equation}
\lim_{\rho a^2\to 0}\frac{e_0(\rho)}{4\pi\mu\rho|\ln(\rho 
a^2)|^{-1}}=1
\end{equation}
where $e_0(\rho)=\lim_{N\to\infty} E_0(N,\rho^{-1/2}N^{1/2})/N$.
This holds irrespective of boundary conditions.
\label{limitthm}
\end{thm}

As in the three-dimensional case, Theorem \ref{limitthm} is also valid
for an infinite range potential $v$ provided that $v\geq 0$ and for
some $R$ we have $\int_{R}^{\infty} v(r)r\ dr <\infty$, which
guarantees a finite scattering length.

\section{Generalized Poincar\'e Inequalities}\label{poincare}

This section contains some lemmas that are of independent mathematical
interest, but whose significance for the physics of the Bose gas may
not be obvious at this point. They will, however, turn out to be
important tools for the discussion of Bose-Einstein condensation (BEC)
and superfluidity in the next section.

The classic Poincar\'e inequality \cite{LL01} bounds the $L^q$-norm 
of a
function, $f$, orthogonal to a given function $g$ in a domain $\K$, in
terms of some $L^p$-norm of its gradient in $\K$.  For the proof of
BEC we shall need a generalization of this inequality where the
estimate is in terms of the gradient of $f$ on a subset
$\Omega\subset\K$ and a remainder that tends to zero with the volume
of the complement $\Omega^c=\K\setminus \Omega$.  For superfluidity it
will be necessary to generalize this further by adding a vector
potential to the gradient.  This is the most complex of the lemmas
because the other two can be derived directly from the classical
Poincar\'e inequality using H\"older's inequality.  The first lemma is
the simplest variant and it is sufficient for the discussion of BEC in
the case of a homogeneous gas.  In this case the function $g$ can be
taken to be the constant function.  The same holds for the second
lemma, which will be used for the discussion of superfluidity in a
homogeneous gas with periodic boundary conditions, but the 
modification
of the gradient requires a more elaborate proof.  The last lemma, 
that will be used for the discussion of BEC
in the inhomogeneous case, is again a simple consequence of the 
classic
Poincar\'e and H\"older inequalities. For a more comprehensive 
discussion of 
generalized Poincar\'e inequalities with further generalizations we 
refer to \cite{lsy02}.

\begin{lem}[\textbf{Generalized Poincar{\'e} inequality: 
Homogeneous case}]
\label{lem2b}
Let $\K\subset \R^3$ be a cube of side length $L$, and define the
average of a function $f\in L^1(\K)$ by $$ \langle
f\rangle_\K=\frac 1{L^3} \int_\K f(\x)\, d\x \ .$$ There
exists a constant $C$ such that for all measurable sets
$\Omega\subset\K$ and all $f\in H^1(\K)$ the inequality
\begin{equation} \label{poinchom}
 \int_{\K} |f(\x)-\langle f\rangle_\K |^2 d\x \leq C  
\left(L^2\int_\Omega |\nabla f(\x)|^2 d\x
+|\Omega^c|^{2/3}\int_\K |\nabla f(\x)|^2 d\x \right)
\end{equation}
holds. Here $\Omega^c=\K\setminus\Omega$, and 
$|\cdot|$ denotes the measure of a set.
.
\end{lem}

\begin{proof} By scaling, it suffices to consider the case $L=1$. 
Using
the usual Poincar\'e-Sobolev inequality on $\K$ (see
\cite{LL01}, Thm. 8.12), we infer that there exists a $C>0$ such
that
\begin{eqnarray}\nonumber
\|f-\langle f\rangle_\K\|_{L^2(\K)}^2&\leq& \half C
\|\nabla f\|_{L^{6/5}(\K)}^2\\ &\leq&  C\left(\|\nabla
f\|_{L^{6/5}(\Omega)}^2+\|\nabla f\|_{L^{6/5}(\Omega^c)}^2\right)\
.
\end{eqnarray}
Applying H\"older's inequality $$ \|\nabla f\|_{L^{6/5}(\Omega)}
\leq \|\nabla f\|_{L^{2}(\Omega)}|\Omega|^{1/3} $$ (and the
analogue with $\Omega$ replaced by $\Omega^c$), we see that
(\ref{poinchom}) holds.
\end{proof}

In the next lemma $\K$ is again a cube of side length $L$, but we
now replace the gradient $\nabla$ by
\beq\label{nablaphi}
\nabla_{\varphi}:=\nabla+i(0,0,\varphi/L),
\eeq
where $\varphi$ is a real parameter, and require periodic boundary 
conditions on $\K$. 

\begin{lem}[\textbf{Generalized Poincar\'e inequality with a 
vector potential}]\label{L2}
For any $|\varphi|<\pi$ there  are constants 
$c>0$ and $C<\infty$  such that for all subsets 
$\Omega\subset\K$ and all functions $f\in H^1(\K)$ with periodic 
boundary 
conditions on $\K$  the 
following estimate holds:
\begin{multline}\label{poinc}
\Vert\nabla_\varphi
f\Vert_{L^2(\Omega)}^2\geq \frac{\varphi^2}{L^2}\|f\|_{L^2(\K)}^2 
+\frac c{L^2} \Vert 
f-\langle f\rangle_{\K}
\Vert_{L^2(\K)}^2\\ -C\left(\|\nabla_\varphi f\|_{L^2(\K)}^2 + \frac 
1{L^2} 
\|f\|_{L^2(\K)}^2\right) \left(\frac{|\Omega|^c}{L^3}\right)^{1/2} \ 
.
 \end{multline} Here $\vert\Omega^c\vert$ is the volume of
 $\Omega^c=\K\setminus\Omega$, the complement of $\Omega$ in $\K$.
\end{lem}

\begin{proof} 
We shall derive (\ref{poinc}) from a special form of this inequality
that holds for all functions that are orthogonal to the constant
function. Namely, for any positive $\alpha<2/3$ and some constants
$c>0$ and $\widetilde C<\infty$ (depending only on $\alpha$ and
$|\varphi|<\pi$) we claim that
\begin{equation}\label{poinc2}
\|\nabla_\varphi h\|_{L^2(\Omega)}^2  \geq 
\frac{\varphi^2+c}{L^2}
\Vert h\Vert_{L^2(\K)}^2-\widetilde 
C\left(\frac{|\Omega^c|}{L^3}\right)^\alpha\Vert\nabla_\varphi
h\Vert_{L^2(\K)}^2 \ ,
\end{equation}
provided 
$\langle 1,h\rangle_{\K} =0$. (Remark: Eq.~(\ref{poinc2}) 
holds also for $\alpha=2/3$, but the proof is slightly more 
complicated in that case. See \cite{lsy02}.) If (\ref{poinc2}) is 
known 
the 
derivation of (\ref{poinc}) is easy: For any $f$, the function
$h=f-L^{-3}\langle 1,f\rangle_{\K}$ is orthogonal to 
$1$. Moreover, 
\begin{eqnarray}\nonumber
\Vert\nabla_\varphi h\Vert^2_{L^2(\Omega)}&=& 
\Vert\nabla_\varphi h\Vert^2_{L^2(\K)}-\Vert\nabla_\varphi 
h\Vert^2_{L^2(\Omega^c)}\nonumber\\   
&=&    
\Vert\nabla_\varphi f\Vert^2_{L^2(\Omega)}-
\frac{\varphi^2}{L^2}\vert\langle L^{-3/2},f\rangle_{\K}\vert^2 
\left(1+\frac{|\Omega^c|}{L^3}\right)
\nonumber \\ &&\quad +2\frac{\varphi}{L}{\rm 
Re}\,\langle L^{-3/2},f\rangle_{\K}
\langle\nabla_\varphi f,L^{-3/2}
\rangle_{\Omega^c}\nonumber\\ \nonumber  
&\leq& \Vert\nabla_\varphi f\Vert^2_{L^2(\Omega)}-
\frac{\varphi^2}{L^2} \vert\langle L^{-3/2} ,f\rangle_{\K}\vert^2 
\\ \nonumber && \quad
+\frac{|\varphi|}{L}
\left(L \Vert\nabla_\varphi f\Vert_{L^2(\K)}^2+\frac 1L \Vert 
f\Vert_{L^2(\K)}^2\right) \left(\frac{|\Omega^c|}{L^3}\right)^{1/2} 
\\ 
\label{quadrat1} 
\end{eqnarray}
and
\begin{eqnarray}\nonumber 
\frac{\varphi^2+c}{L^2}\Vert 
h\Vert_{L^2(\K)}^2&=&\frac{\varphi^2}{L^2}\left(  \Vert 
f\Vert_{L^2(\K)}^2- \vert\langle L^{-3/2} 
,f\rangle_{\K}\vert^2\right)\\  && + 
\frac c{L^2} \Vert f-L^{-3}\langle 1 ,f\rangle_{\K} \label{quadrat2}
\Vert_{L^2(\K)}^2 \ .
\end{eqnarray}
Setting $\alpha=\half$, using $\|\nabla_\varphi h\|_{L^2(\K)}\leq 
\|\nabla_\varphi f\|_{L^2(\K)}$ in the last term in (\ref{poinc2}) 
and combining (\ref{poinc2}), (\ref{quadrat1}) and (\ref{quadrat2}) 
gives 
(\ref{poinc}) with $C=|\varphi|+\widetilde C$. 

We now turn to the proof of (\ref{poinc2}). For simplicity we set 
$L=1$. The general case  follows by scaling.  Assume that 
(\ref{poinc2}) 
is false. Then there exist sequences of constants $C_{n}\to\infty$,
functions $h_{n}$ with $\Vert h_{n}\Vert_{L^2(\K)}=1$ and 
$\langle 1,h_{n}\rangle_{\K} =0$, and domains $\Omega_{n}\subset\K$ 
such 
that
\begin{equation}\label{false}
\lim_{n\to\infty}\left\{
\Vert\nabla_\varphi
h_{n}\Vert_{L^2(\Omega_{n})}^2+C_{n}\vert\Omega_{n}^c\vert^\alpha\Vert\nabla_\varphi
h_{n}\Vert_{L^2(\K)}^2\right\}
\leq \varphi^2 \ .
\end{equation}
We shall show that this leads to a contradiction.

Since the sequence $h_{n}$ is bounded in $L^2(\K)$ it has a
subsequence, denoted again by $h_{n}$, that converges weakly to some
$h\in L^2(\K)$ (i.e., $\langle g,h_{n}\rangle_{\K}\to \langle
g,h\rangle_{\K}$ for all $g\in L^2(\K)$).  Moreover, by H\"older's
inequality the $L^p(\Omega_{n}^c)$ norm $\Vert \nabla_\varphi
h_{n}\Vert_{L^p(\Omega_{n}^c)}=(\int_{\Omega^c_{n}} \vert
\nabla_{\varphi}h(\x)\vert^pd\x)^{1/p}$ is bounded by
$\vert\Omega_{n}^c\vert^{\alpha/2}\Vert\nabla_\varphi
h_{n}\Vert_{L^2(\K)}$ for $p=2/(\alpha+1)$.  From (\ref{false}) we
conclude that $\Vert \nabla_\varphi h_{n}\Vert_{L^p(\Omega_{n}^c)}$ is
bounded and also that $\Vert \nabla_\varphi
h_{n}\Vert_{L^p(\Omega_{n})}\leq\Vert \nabla_\varphi
h_{n}\Vert_{L^2(\Omega_{n})}$ is bounded. Altogether, $\nabla_\varphi
h_{n}$ is bounded in $L^p(\K)$, and by passing to a further
subsequence if necessary, we can therefore assume that $\nabla_\varphi
h_{n}$ converges weakly in $L^p(\K)$. The same applies to $\nabla
h_{n}$. Since $p=2/(\alpha+1)$ with $\alpha<2/3$ the hypotheses of the
Rellich-Kondrashov Theorem \cite[Thm~8.9]{LL01} are fulfilled and
consequently $h_{n}$ converges {\it strongly} in $L^2(\K)$ to $h$
(i.e., $\Vert h-h_{n}\Vert_{L^2(\K)}\to 0$).  We shall now show that
\begin{equation}\label{lowersemi}
\liminf_{n\to\infty}\Vert\nabla_\varphi
h_{n}\Vert_{L^2(\Omega_{n})}^2\geq \Vert\nabla_\varphi
h\Vert_{L^2(\K)}^2 \ .
\end{equation}
This will complete the proof because the $h_{n}$ are normalized and
orthogonal to $1$ and the same holds for $h$ by strong
convergence. Hence the right side of (\ref{lowersemi}) is necessarily
$>\varphi^2$, since for $|\varphi|<\pi$ the lowest eigenvalue of
$-\nabla_\varphi^2$, with constant eigenfunction, is
non-degenerate. This contradicts (\ref{false}).

Eq.~(\ref{lowersemi}) is essentially a consequence of the weak lower
semicontinuity of the $L^2$ norm, but the dependence on $\Omega_{n}$
leads to a slight complication.  First, Eq.~(\ref{false}) and 
$C_{n}\to
\infty$ clearly imply that $\vert\Omega_{n}^c\vert\to 0$, because 
$\Vert\nabla_\varphi
h_{n}\Vert_{L^2(\K)}^2>\varphi^2$.  By choosing
a subsequence we may assume that
$\sum_{n}\vert\Omega_{n}^c\vert<\infty$.  For some fixed $N$ let
$\widetilde\Omega_{N}=\K\setminus\cup_{n\geq N}\Omega_{n}^c$. Then 
$\tilde\Omega_{N}\subset\Omega_{n}$ for $n\geq N$. 
Since $\Vert\nabla_\varphi
h_{n}\Vert_{L^2(\Omega_{n})}^2$ is bounded, $\nabla_\varphi
h_{n}$ is also bounded in $L^2(\widetilde\Omega_{N})$ and a 
subsequence 
of it converges weakly in $L^2(\widetilde\Omega_{N})$ to 
$\nabla_\varphi
h$. Hence 
\begin{equation}\label{lowersemi2}
\liminf_{n\to\infty}\Vert\nabla_\varphi
h_{n}\Vert_{L^2(\Omega_{n})}^2  \geq 
\liminf_{n\to\infty}\Vert\nabla_\varphi
h_{n}\Vert_{L^2(\widetilde \Omega_{N})}^2\geq\Vert\nabla_\varphi
h\Vert_{L^2(\widetilde\Omega_{N})}^2  \ . 
\end{equation}
Since 
$\widetilde\Omega_{N}\subset \widetilde\Omega_{N+1}$ and 
$\cup_{N}\widetilde\Omega_{N}=\K$ (up to a set of measure zero), we 
can 
now let $N\to\infty$ on the right side of (\ref{lowersemi2}). By 
monotone convergence this converges to $\Vert\nabla_\varphi
h\Vert_{L^2(\K)}^2$. This proves (\ref{lowersemi}) which, as remarked 
above, contradicts
(\ref{false}). 
\end{proof}

The last lemma is a simple generalization of Lemma \ref{lem2b}
with $\K\subset\R^m$ a bounded and connected
set that is sufficiently nice so that the Poincar\'e-Sobolev
inequality (see \cite[Thm.~8.12]{LL01}) holds on $\K$. In
particular, this is the case if $\K$ satisfies the cone property
\cite{LL01} (e.g.\, if $\K$ is a rectangular box or a cube). 
Moreover, 
the constant function on $\K$ is here replaced by a more general 
bounded function.

\begin{lem}[\textbf{Generalized Poincar{\'e} inequality: 
Inhomog. case}]\label{lem2}
For $d\geq 2$ let $\K\subset\R^d$ be as explained above, and let
$h$ be a bounded function with $\int_\K h=1$. There exists a
constant $C$ (depending only on $\K$ and $h$) such that for all
measurable sets $\Omega\subset\K$ and all $f\in H^1(\K)$  with
$\int_\K f h\, d\x=0$, the inequality
\begin{equation} \label{poinc3}
 \int_{\K} |f(\x)|^2 d\x \leq C \left(\int_\Omega |\nabla f(\x)|^2 d\x
+\left(\frac{|\Omega^c|}{|\K|}\right)^{2/d}\int_\K |\nabla
f(\x)|^2 d\x \right)
\end{equation}
holds. Here $|\cdot|$ denotes the measure of a set, and
$\Omega^c=\K\setminus\Omega$.
\end{lem}

\begin{proof} By the usual Poincar\'e-Sobolev inequality on $\K$ (see
\cite[Thm.~8.12]{LL01}),
\begin{eqnarray}\nonumber
\|f\|_{L^2(\K)}^2&\leq& \tilde C \|\nabla
f\|_{L^{2d/(d+2)}(\K)}^2\\ &\leq& 2\tilde C\left(\|\nabla
f\|_{L^{2d/(d+2)}(\Omega)}^2+\|\nabla
f\|_{L^{2d/(d+2)}(\Omega^c)}^2\right),
\end{eqnarray}
if $d\geq 2$
and $\int_\K f h=0$. Applying H\"older's inequality $$ \|\nabla
f\|_{L^{2d/(d+2)}(\Omega)} \leq \|\nabla
f\|_{L^{2}(\Omega)}|\Omega|^{1/d} $$ (and the analogue with
$\Omega$ replaced by $\Omega^c$), we see that (\ref{poinc}) holds
with $C=2|\K|^{2/d}\tilde C$.
\end{proof}

\section{Bose-Einstein Condensation and Superfluidity for Homogeneous 
Gases}\label{bec}

\subsection{Bose-Einstein Condensation}

Bose-Einstein condensation (BEC) is the phenomenon of a macroscopic 
occupation of a single one-particle quantum state, discovered by  
Einstein for thermal equilibrium states of an ideal Bose gas at 
sufficiently low temperatures \cite{Einstein}. We are here concerned 
with 
interacting Bose gases, where the question of the existence of BEC 
is highly nontrivial even for the ground state. Due to the 
interaction 
the many body ground state is not a product of one-particle states 
but the concept of a macroscopic occupation of a single state 
acquires a  
precise meaning through the {\it one-particle density} matrix. Given 
the normalized ground state wave function
this  is the operator on $L^2(\R^d)$    ($d=2$ or $3$) given
by the kernel
\begin{equation}\label{defgamma}
 \gamma(\x,\x')=N\int \Psi_0(\x,\X)
\Psi_0(\x',\X) d\X \ ,
\end{equation}
 where we introduced the short hand
notation
\begin{equation}\label{defX}
\X=(\x_2,\dots,\x_N)\qquad{\rm and}\quad  d\X=\prod\limits_{j=
2}^N d\x_j.
\end{equation}
Then $\int \gamma(\x, \x) d\x =\Tr[\gamma] = N$. BEC in the ground 
state 
means, by definition, that this operator has an eigenvalue of order 
$N$ in 
the thermodynamic limit. Since
$\gamma$ is  a positive kernel and, hopefully, translation
invariant in the thermodynamic limit, the eigenfunction belonging
to  the largest eigenvalue must be the constant function
$L^{-d/2}$. Therefore, another way to say that there is BEC in the 
ground 
state is that
\begin{equation}\label{defbec}
 \frac 1{L^d} \int\int \gamma(\x,\, \y) d\x
d\y = \textrm{O}(N)\ 
\end{equation}
as $N\to \infty$, $L\to \infty$ with $N/L^d$ fixed; more precisely 
Eq.\ \eqref{defbec} requires that there is a $c>0$ such that the left 
side is $>cN$ for all large $N$.
This is also 
referred to as {\it off-diagonal long range order}.
Unfortunately, this is something that is frequently invoked but has 
so far never been
proved for many body Hamiltonians with genuine interactions 
--- except for one special case:  hard core bosons on a
lattice at half-filling (i.e.,
$N=$ half the number of lattice sites). The proof is in \cite{KLS} 
and \cite{DLS}.

The problem remains open after more than 75 years since the first 
investigations on 
the Bose gas \cite{Bose,Einstein}. 
%It is also not at all clear
%that BEC is essential for superfluidity, as frequently claimed. 
Our
construction in Section \ref{sect3d} shows that (in 3D) BEC exists on
a length scale of order $\rho^{-1/3} Y^{ -1/17}$ which, unfortunately,
is not a `thermodynamic' length like $\textrm{volume}^{1/3}$. The same
remark applies to the 2D case of Section 3, where BEC is proved over a
length scale $\rho^{-1/10}|\ln(\rho a^2)|^{1/10}$.

In a certain limit, however, one can prove (\ref{defbec}), as has been
shown in \cite{LS02}.  In this limit the interaction
potential $v$ is varied with $N$ so that the ratio $a/L$ of the
scattering length to the box length is of order $1/N$, i.e., the
parameter $Na/L$ is kept fixed.  
Changing $a$
with $N$ can be done by scaling, i.e., we write
\begin{equation}\label{v1}
v(|\x|)=\frac 1{a^2} v_1(|\x|/a)
\end{equation}
for some $v_1$ having scattering length $1$, and vary $a$ while
keeping $v_1$ fixed. It is easily checked that the $v$ so defined has
scattering length $a$. It is important to note that, in
 the limit considered,  $a$ tends to zero (as
$N^{-2/3}$ since $L=(N/\rho)^{1/3}\sim N^{1/3}$ for $\rho$ fixed), 
and $v$ becomes a {\it hard} potential of {\it short}
range. This is the {\it opposite} of the usual mean field limit where
the strength of the potential goes to zero while its range tends to
infinity.

We shall refer to this as the {\it Gross-Pitaevskii (GP) limit} since
$Na/L$ will turn out to be the natural interaction parameter for
inhomogeneous Bose gases confined in traps, that are described by the
Gross-Pitaevskii equation discussed in Sections~\ref{sectgp}
and~\ref{becsect}. Its significance for a homogeneous gas can also be
seen by noting that $Na/L$ is the ratio of $\rho a$ to $1/L^2$, i.e.,
in the GP limit the interaction energy per particle is of the same
order of magnitude as the energy gap in the box, so that the
interaction is still clearly visible, even though $a\to 0$. Note that
$\rho a^3\sim N^{-2}$ in the GP limit, so letting $N\to\infty$ with
$\rho$ fixed and $Na/L$ fixed can be regarded as a {\it simultaneous
  thermodynamic and low density limit}.  For simplicity, we shall here
treat only the 3D case.

\begin{thm}[\textbf{BEC in a dilute limit}]\label{hombecthm}
Assume that, as $N\to\infty$, $\rho=N/L^3$ and $\g=Na/L$ stay
fixed, and impose either periodic or Neumann boundary conditions
for $H$. Then
\begin{equation}\label{xxyy}
\lim_{N\to\infty} \frac 1N \frac 1{L^3} \int\int \gamma(\x,\, \y)
d\x d\y = 1\ .
\end{equation}
\end{thm}

The reason  we do not
deal with Dirichlet boundary conditions at this point should be clear 
from the discussion preceding the theorem: There would be an
additional contribution $\sim 1/L^2$ to the energy, i.e.\
of the same order as the 
interaction energy, and the
system would not be homogeneous any more. Dirichlet boundary
conditions can, however, be treated with the methods of Section
\ref{becsect}.

By scaling, the limit in Theorem \ref{hombecthm} is equivalent to
considering a Bose gas in a {\it fixed box} of side length $L=1$, and
keeping $Na$ fixed as $N\to\infty$, i.e., $a\sim 1/N$.  The ground
state energy of the system is then, asymptotically, $N\times 4\pi Na$,
and Theorem \ref{hombecthm} implies that the one-particle reduced
density matrix $\gamma$ of the ground state converges, after division
by $N$, to the projection onto the constant function.  An analogous
result holds true for inhomogeneous systems as will be discussed
in Section \ref{becsect}.

The proof of Theorem \ref{hombecthm} has two main ingredients. One 
is  
{\it localization of the energy} that is stated as Lemma \ref{L1} 
below. 
This lemma is a refinement of the energy estimates of Section 2.2
and says essentially that the kinetic energy of the ground state is 
concentrated in a subset of configuration space where at least one 
pair of particles is close together and whose volume 
tends to zero as $a\to 0$. The other is the generalized 
Poincar\'e inequality, Lemma \ref{lem2b} from which one deduces that 
the one particle 
density matrix is approximately constant if the kinetic energy is 
localized in a small set.

The localization lemma will be proved in a slightly more general
version that is necessary for Theorem \ref{hombecthm}, namely with the
gradient $\nabla$ replaced by
$\nabla_{\varphi}=\nabla+i(0,0,\varphi/L)$, cf.\ Eq.\ 
\eqref{nablaphi}.  We
denote by $H_{N}'$ the corresponding many-body Hamiltonian \eqref{ham}
with $\nabla_{\varphi}$ in place of $\nabla$.  This generalization
will be used in the subsequent discussion of superfluidity, but a
reader who wishes to focus on Theorem \ref{hombecthm} only can simply
ignore the $\varphi$ and the reference to the diamagnetic inequality 
in the proof.  We denote the gradient with respect to $\x_{1}$
by $\nabla_1$, and the corresponding modified operator by
$\nabla_{1,\varphi}$.

\begin{lem}[\textbf{Localization of energy}]\label{L1}
Let $\K$ be a box of side length $L$.
    For all symmetric, normalized wave functions 
$\Psi(\x_{1},\dots,\x_{N})$ with periodic boundary conditions on 
$\K$, and for $N\geq Y^{-1/17}$, 
\begin{eqnarray} \nonumber 
\frac1N\langle\Psi, H_{N}'\Psi\rangle &\geq& \big(1-\const 
Y^{1/17}\big) \\ && \times \Big(4\pi\mu\rho 
a+
\mu \int _{\K^{N-1}}
d\X \int_{\Omega_{\X}}d\x_{1}\big|
\nabla_{1,\varphi}\Psi(\x_{1},\X)\big\vert^2\Big) \ , \nonumber \\ 
\label{lowerbd}
\end{eqnarray}   
where $\X=(\x_{2},\dots,\x_{N})$, $d\X=\prod_{j=2}^N d\x_j$, and 
\begin{equation}
\Omega_{\X}=\left\{\x_{1}:\ \min_{j\geq 2}\vert\x_{1}- 
\x_{j}\vert\geq R \right\} 
    \end{equation} 
with $R=a Y^{-5/17}$. 
\end{lem} 
 
\begin{proof} 
Since $\Psi$ is symmetric, the left side of 
(\ref{lowerbd}) can be written as
\begin{equation}    
\int_{\K^{N-1}} d\X \int_\K d\x_{1}
\Big[\mu\big|\nabla_{1,\varphi}\Psi(\x_{1},\X)\big\vert^2
 +\half\sum_{j\geq 2}
v(\vert\x_{1}-\x_{j}\vert)|\Psi(\x_{1},\X)\vert^2\Big]\ .
\end{equation}
For any $\varepsilon>0$ and $R>0$ this is 
\begin{equation}
\geq \varepsilon T+(1-\varepsilon)(T^{\rm in}+I)+(1-\varepsilon)
T_{\varphi}^{\rm out} \ ,
\end{equation}
with
\begin{equation}\label{15}
T=\mu\int_{\K^{N-1}} d\X\int_\K 
d\x_{1}\big|\nabla_1\vert\Psi(\x_{1},\X)
\vert\big|^2 \ ,
\end{equation}
\begin{equation}\label{16}      
\quad T^{\rm in}=\mu
\int_{\K^{N-1}} d\X\int_{\Omega^c_{\X}} d\x_{1}\big|\nabla_1
\vert\Psi(\x_{1},\X)\vert\big|^2\ ,
\end{equation}  
\begin{equation}
T_{\varphi}^{\rm out}=\mu\int_{\K^{N-1}} d\X\int_{\Omega_{\X}} d\x_{1}
\big\vert\nabla_{1,\varphi} 
\Psi(\x_{1},\X)\big\vert^2\ ,
\end{equation}
and 
\begin{equation}
I=\half\int_{\K^{N-1}}
d\X\int_\K d\x_{1}\sum_{j\geq 2}
v(\vert\x_{1}-\x_{j}\vert)|\Psi(\x_{1},\X)\vert^2 \ .   
\end{equation}
Here
\begin{equation}\Omega^c_{\X}=\left\{\x_{1}:\ \vert\x_{1}- 
\x_{j}\vert<R\ \text{for some }j\geq 2\right\}
\end{equation}   
is the complement of $\Omega_{\X}$, and the diamagnetic inequality 
\cite{LL01}
$\vert\nabla_{\varphi}f(\x)\vert^2\geq 
\left|\nabla|f(\x)|\right|^2$ 
has been used.  
The proof is completed by using the estimates used for the proof of 
Theorem 2.4,
in particular (\ref{estimate1}) and 
(\ref{Kformula2})--(\ref{exponents}),
which tell us that for
$\varepsilon=Y^{1/17}$ and $R=aY^{-5/17}$ 
\begin{equation}\label{abs}
\varepsilon T+(1-\varepsilon)(T^{\rm in}+I)\geq\big(1-\const 
Y^{1/17}\big)
4\pi\mu\rho a 
\end{equation}
as long as $N\geq Y^{-1/17}$. 
\end{proof}

\begin{proof}[Proof of Theorem \ref{hombecthm}] 
We combine Lemma \ref{L1} 
(with $\varphi=0$ and hence $H'_{N}=H_{N}$) with 
Lemma \ref{lem2b} that gives a lower bound to the second term on the 
right side of \eqref{lowerbd}. We thus infer that, for any symmetric 
$\Psi$ 
with 
$\langle \Psi,\Psi\rangle=1$ and for $N$ large enough,  
\begin{eqnarray} \nonumber 
&&\frac1N\langle\Psi,H_{N}\Psi\rangle  \big(1-\const 
Y^{1/17}\big)^{-1}\\ \nonumber && \geq  4\pi\mu\rho a  - C Y^{1/17} 
\Big(\frac 1{L^2} - \frac 1N 
\big\langle\Psi,\mbox{$ \sum_j$} \nabla_j^2 
\Psi\big\rangle\Big) 
 \\ \nonumber && \quad + \frac c{L^2}
\int_{\K^{N-1}}
d\X \int_{\K} d\x_{1}\Big|
\Psi(\x_1,\X) -L^{-3}\big[\mbox{$ \int_\K$} d\x 
\Psi(\x,\X)\big] 
\Big|^2 \ ,\\ \label{lowerbd2}
\end{eqnarray}   
where we used that $|\Omega^c|\leq \mbox{$\frac{4\pi}3$} N R^3= 
\const L^3 Y^{2/17}$. Since the kinetic energy, divided by 
$N$, is certainly bounded independent of $N$, as the upper bound 
\eqref{upperbound}
shows, and since the upper and the lower bound to $E_0$ agree in the
limit considered, the positive last term in (\ref{lowerbd2}) has to 
vanish in the limit. I.e., we get that for the ground state wave 
function
$\Psi_0$ of $ H_N$
\begin{equation}
\lim_{N\to\infty} \int_{\K^{N-1}}
d\X \int_{\K} d\x_{1}\Big|
\Psi_0(\x_1,\X)\\ -L^{-3}\big[\mbox{$ \int_\K$} d\x 
\Psi_0(\x,\X)\big] 
\Big|^2 = 0 \ .
\end{equation}
This proves (\ref{xxyy}), since 
\begin{eqnarray}\nonumber
 &&\int_{\K^{N-1}}
d\X \int_{\K} d\x_{1}\Big|
\Psi_0(\x_1,\X)-L^{-3}\big[\mbox{$ \int_\K$} d\x \Psi_0(\x,\X)\big] 
\Big|^2 \\ && = 1- \frac 1{NL^3} \int_{\K\times\K} \gamma(\x,\x') d\x 
d\x' 
\ . \label{ahaha}
\end{eqnarray}
\end{proof}

\subsection{Superfluidity}

The phenomenological two-fluid model of superfluidity (see, e.g., 
\cite{TT}) 
is based on the
idea that the particle density $\rho$ is composed of two parts, the 
density $\rho_{\rm s}$ of the inviscid superfluid and the normal 
fluid density
$\rho_{\rm n}$.  If an external velocity field is imposed on the fluid
(for instance by moving the walls of the container) only the viscous 
normal
component responds to the velocity field, while the superfluid
component stays at rest.  In accord
with these ideas the superfluid density in the ground state is
often defined as follows \cite{HM}: Let $E_0$ denote the ground
state energy of the system in the rest frame and $E_0'$ the ground
state energy, measured in the
moving frame, when a velocity field ${\bf v}$ is imposed.  
Then for small ${\bf v}$
\begin{equation}\label{rhos}
    \frac{E_0'}N=\frac {E_0}N+({\rho_{\rm s}}/\rho)\half{ m} {\bf
    v}^2+O(|{\bf v}|^4) 
\end{equation} 
where $N$ is the particle number and $m$ the particle mass.  At
positive temperatures the ground state energy should be replaced by
the free energy.  (Remark: It is important here that (\ref{rhos})
holds uniformly for all large $N$; i.e., that the error term $O(|{\bf
v}|^4)$ can be bounded independently of $N$.  For fixed $N$ and a
finite box, Eq.\ (\ref{rhos}) with $\rho_{\rm s}/\rho=1$ always holds
for a Bose gas with an arbitrary interaction if ${\bf v}$ is small
enough, owing to the discreteness of the energy spectrum.\footnote{
The ground state with ${\bf v}=0$ remains an eigenstate of the
Hamiltonian with arbitrary ${\bf v}$ (but not necesssarily a ground
state) since its total momentum is zero.  Its energy is $\half m N{\bf
v}^2$ above the ground state energy for ${\bf v}=0$.  Since in a
finite box the spectrum of the Hamiltonian for arbitrary ${\bf v}$ is
discrete and the energy gap above the ground state is bounded away
from zero for ${\bf v}$ small, the ground state for ${\bf v}=0$ is at
the same time the ground state of the Hamiltonian with ${\bf v}$ if
$\half m N{\bf v}^2$ is smaller than the gap.} There are other
definitions of the superfluid density that may lead to different
results \cite{PrSv}, but this is the one we shall use here and shall
not dwell on this issue since it is not clear that there is a \lq\lq
one-size-fits-all\rq\rq\ definition of superfluidity.  For instance,
in the definition we use here the ideal Bose gas is a perfect
superfluid in its ground state, whereas the definition of Landau in
terms of a linear dispersion relation of elementary excitations would
indicate otherwise.  Our main result is that with the definition
adopted here there is 100\% superfluidity in the ground state of a 3D
Bose gas in the GP limit explained in the previous subsection.

One of the unresolved issues in the theory of superfluidity is its
relation to Bose-Einstein condensation (BEC).  It has been argued that
in general neither condition is necessary for the other (c.f., e.g.,
\cite{huang,giorgini,KT}), but in the case considered here, i.e., the
GP limit of a 3D gas, we show that 100\% BEC into the constant wave
function (in the rest frame) prevails even if an external velocity
field is imposed.  A simple example illustrating the fact that BEC is
not necessary for superfluidity is the 1D hard-core Bose gas.  This
system is well known to have a spectrum like that of an ideal Fermi
gas \cite{gir} (see also Section 8), and it is easy to see that
it is superfluid in its ground state in the sense of (\ref{rhos}).  On
the other hand, it has no BEC \cite{Lenard,PiSt}.  The definition of
the superfluid velocity as the gradient of the phase of the condensate
wave function \cite{HM,baym} is clearly not applicable in such cases.

We consider a Bose gas with the Hamiltonian (\ref{ham}) in a box $\K$ 
of 
side length $L$, 
assuming periodic boundary 
conditions in all 
three coordinate directions.
Imposing an external velocity field ${\bf v}=(0,0,\pm|{\bf v}|)$ means
that the momentum operator ${\bf p}={-{\rm i}\hbar \nabla}$ is
replaced by by ${\bf p}-m{\bf v}$, retaining the periodic boundary
conditions. 
The Hamiltonian in the moving frame is thus 
\begin{equation}\label{hamprime}
H_N'  = -\mu\sum_{j=1}^N \nabla_{j,\varphi}^2+
\sum_{1\leq i<j\leq N}v(\vert\x_{i}-\x_{j}\vert) \ ,
\end{equation}
where  $\nabla_{j,\varphi}=\nabla_{j}+i(0,0,\varphi/L)$ and
the dimensionless phase $\varphi$ is connected to 
the velocity ${\bf v}$ by
\begin{equation}
\label{varphi}\varphi=\frac{\pm|{\bf v}|Lm}\hbar\ . 
\end{equation}

Let $E_0(N,a,\varphi)$ denote the ground state energy of 
(\ref{hamprime})
with periodic boundary conditions. Obviously it is no
restriction to consider only the case $-\pi\leq \varphi\leq \pi$,
since $E_0$ is periodic in $\varphi$ with period $2\pi$ (see Remark 1 
below).  For $\Psi_0$
the ground state of $H_N'$, let $\gamma_N$ be its one-particle reduced
density matrix. 
We are interested in the {\it Gross-Pitaevskii} (GP) limit
$N\to\infty$ with $Na/L$ fixed. We also fix the box size $L$. This
means that $a$ should vary like $1/N$ which, as explained in the 
previous subsection, can be achieved by writing
$v(r)=a^{-2}v_1(r/a)$, where $v_1$ is a fixed potential
with scattering length 1, while $a$ changes with $N$.

\begin{thm}[{\bf Superfluidity and BEC of homogeneous gas}]\label{T1a}
For $|\varphi| \leq \pi$
\begin{equation}\label{i}
\lim_{N\to\infty} \frac{E_0(N,a,\varphi)}N = 4\pi\mu  a\rho + 
\mu\frac{\varphi^2}{L^2} 
\end{equation} 
in the limit $N\to \infty$ with $Na/L$ and $L$ fixed. Here 
$\rho=N/L^3$, 
so $a\rho$ is fixed too.
In the same 
limit, 
for $|\varphi| < \pi$, 
\begin{equation}\label{ii}
\lim_{N\to\infty} \frac 1N\, \gamma_N(\x,\x')=
\frac 1{L^3}
\end{equation}
in trace class norm, i.e., $\lim_{N\to\infty} \Tr 
\big[\,\big|\gamma_N/N - |L^{-3/2}\rangle\langle L^{-3/2}|\, 
\big|\,\big]=0$. 
\end{thm}

Note that, by the definition (\ref{rhos}) of $\rho_{s}$ and Eq.\
(\ref{varphi}), Eq.\ (\ref{i}) means that $\rho_{s}=\rho$, i.e., there
is 100\% superfluidity. For $\varphi=0$, Eq.\ (\ref{i}) follows from
Eq.\ \ref{basic}, while (\ref{ii}) for $\varphi=0$ is the BEC
of Theorem \ref{hombecthm}.\footnote{The convention in
Theorem \ref{hombecthm}, where $\rho$ and $Na/L$ stay fixed,
is different from the one employed here, where $L$ and $Na/L$ are 
fixed, but these two conventions are clearly equivalent by scaling.}

\bigskip
\noindent {\it Remarks.}
1. By a unitary gauge transformation,
\begin{equation}\label{gauge}
\big(U\Psi\big)(\x_1,\dots,\x_N)= e^{{\rm i}\varphi(\sum_i z_i)/L}\, 
\Psi(\x_1,\dots,\x_N) \ , 
\end{equation}
the passage from (\ref{ham}) to (\ref{hamprime})  is equivalent to 
replacing periodic boundary
conditions in a box by the 
{\it twisted
boundary condition}
\begin{equation}\label{twist}
\Psi(\x_1 + (0,0,L), \x_2, \dots, \x_N)= e^{{\rm i}\varphi} 
\Psi(\x_1, 
\x_2, \dots, \x_N)
\end{equation}
in the direction of
the velocity field, while retaining the original Hamiltonian 
(\ref{ham}).

2. The criterion $|\varphi|\leq\pi$ means that $|{\bf v}|\leq 
\pi\hbar/(mL)$. The corresponding energy $\half m ( \pi\hbar/(mL) 
)^2$ is the gap in the excitation spectrum of the one-particle 
Hamiltonian in the finite-size system. 

3. The reason that we have to restrict ourselves to $|\varphi|<\pi$ 
in the
second part of Theorem~\ref{T1a} is that for $|\varphi|=\pi$ there are
two ground states of the operator $(\nabla+{\rm i}\vp/L)^2$ with 
periodic
boundary conditions.  All we can say in this case is
that there is a subsequence of $\gamma_N$ that converges to a density
matrix of rank $\leq 2$, whose range is spanned by these two 
functions

\begin{proof}[Proof of Theorem~\ref{T1a}]

As in the proof of Theorem \ref{hombecthm} we combine the 
localization Lemma 
\ref{L1}, this time with $\varphi\neq 0$, and a generalized 
Poincar\'e inequality, 
this time Lemma \ref{L2}.
We thus infer that, for any symmetric $\Psi$ 
with 
$\langle \Psi,\Psi\rangle=1$ and for $N$ large enough,  
\begin{eqnarray} \nonumber 
&&\frac1N\langle\Psi,H_{N}'\Psi\rangle  \big(1-\const 
Y^{1/17}\big)^{-1}\\ \nonumber && \geq  4\pi\mu\rho a + \mu 
\frac{\varphi^2}{L^2} - C Y^{1/17} \Big(\frac 1{L^2} - \frac 1N 
\big\langle\Psi,\mbox{$ \sum_j$} \nabla_{j,\varphi}^2
\Psi\big\rangle\Big) 
 \\ \nonumber && \quad + \frac c{L^2}
\int_{\K^{N-1}}
d\X \int_{\K} d\x_{1}\Big|
\Psi(\x_1,\X) -L^{-3}\big[\mbox{$ \int_\K$} d\x 
\Psi(\x,\X)\big] 
\Big|^2 \ ,\label{lowerbd2a}
\end{eqnarray}   
where we used that $|\Omega^c|\leq \mbox{$\frac{4\pi}3$} N R^3= 
\const L^3 Y^{2/17}$. From this 
we can infer two things. First, since the kinetic energy, divided by 
$N$, is certainly bounded independently of $N$, as the upper bound 
shows, we get that
\begin{equation}
\liminf_{N\to\infty} \frac{E_0(N,a,\varphi)}N \geq 4\pi \mu \rho a + 
\mu \frac{\varphi^2}{L^2}
\end{equation}
for any $|\varphi|<\pi$. By continuity this holds also for
$|\varphi|=\pi$, proving (\ref{i}). (To be precise,
$E_0/N-\mu\varphi^2L^{-2}$ is concave in $\varphi$, and therefore 
stays
concave, and in particular continuous, in the limit $N\to\infty$.) 
Secondly, since the upper and the lower bounds to $E_0$ agree in the
limit considered, the positive last term in (\ref{lowerbd2}) has to 
vanish in the limit. I.e., we get that for the ground state wave 
function $\Psi_0$ of $ H_N'$
\begin{equation}
\lim_{N\to\infty} \int_{\K^{N-1}}
d\X \int_{\K} d\x_{1}\Big|
\Psi_0(\x_1,\X)  -L^{-3}\big[\mbox{$ \int_\K$} d\x 
\Psi_0(\x,\X)\big] 
\Big|^2 = 0 \ .
\end{equation}
Using again (\ref{ahaha}), this proves (\ref{ii}) in a weak sense. 
As explained in \cite{LS02,LSSY}, this suffices for  the convergence 
$N^{-1}\gamma_N \to |L^{-3/2}\rangle \langle L^{-3/2}|$  in trace 
class norm. 
\end{proof}

Theorem~\ref{T1a} can be generalized in various ways to a physically
more realistic setting, for example replacing the periodic box by a 
cylinder centered at the origin. We shall comment on such extensions 
at the end of Section~\ref{becsect}.

\section{Gross-Pitaevskii Equation for Trapped Bosons} \label{sectgp}

In the recent experiments on Bose condensation (see, e.g.,
\cite{TRAP}), the particles are confined at very low temperatures 
in a `trap' where the particle density is {\em inhomogeneous}, 
contrary to the case of a large `box', where the 
density is essentially uniform.
We model the trap by a slowly varying confining
potential $ V$, with $V(\x)\to \infty $ as $|\x|\to \infty$.  
The
Hamiltonian becomes
\begin{equation}\label{trapham}
H =  \sum_{i=1}^{N}\left\{ -\mu \Delta_i +V(\x_i)\right\} +
 \sum_{1 \leq i < j \leq N} v(|\x_i - \x_j|) \ .
\end{equation}
Shifting the energy scale if necessary 
we can assume that $V$ is nonnegative.
The ground state energy, $\hbar\omega$, of
$- \mu \Delta + V(\x)$ is
a natural energy unit and the corresponding 
length unit, $\sqrt{\hbar/(m\omega)}=\sqrt{2\mu/(\hbar\omega)}
\equiv L_{\rm osc}$, is a measure of the extension 
of the trap. 

In the sequel we shall be considering a limit
where $a/L_{\rm osc}$ tends to zero while $N\to\infty$.  
Experimentally
$a/L_{\rm osc}$ can be changed in two ways:  One can either vary 
$L_{\rm
osc}$ or $a$.  The first alternative is usually simpler in practice 
but
very recently a direct tuning of the scattering length itself has also
been shown to be feasible \cite{Cornish}.  Mathematically, both
alternatives
are equivalent, of course.  The first corresponds to writing
$V(\x)=L_{\rm osc}^{-2}  V_1(\x/L_{\rm osc})$ and keeping $V_1$ and
$v$ fixed.  The second corresponds to writing the interaction 
potential 
as $v(|\x|)=a^{-2}v_1(|\x|/a)$ like 
in \eqref{v1}, where $v_1$ has unit scattering length, 
and keeping $V$ and $v_1$ fixed. This is equivalent to the first, 
since for given $V_1$ and $v_1$ the ground state energy of 
(\ref{trapham}),
measured 
in units of $\hbar\omega$, depends only on $N$ and $a/L_{\rm osc}$. 
In the dilute limit when $a$ is much smaller than the mean particle 
distance, the energy becomes independent of $v_1$.

We choose $L_{\rm osc}$ as a length unit. The energy unit is
$\hbar\omega=2\mu L_{\rm osc}^{-2}=2\mu$.  Moreover, we find it
convenient to regard $V$ and $v_1$ as fixed. This justifies the notion
$E_0(N,a)$ for the quantum mechanical ground state energy.

The idea is now to use the information about the thermodynamic 
limiting
energy of the dilute Bose gas in a box to find the ground state
energy of  (\ref{trapham}) in an appropriate limit. This has been done
in \cite{LSY1999,LSY2d} and in this section we 
give an account of this work.
As we saw in Sections \ref{sect3d} and
\ref{sect2d} there is a difference in the $\rho$ dependence between
two and three dimensions, so we can expect a related difference now.
We discuss 3D first.

\subsection{Three Dimensions}

Associated with the quantum mechanical ground state energy problem
is the Gross-Pitaevskii (GP) energy functional 
\cite{G1961,G1963,P1961}
\begin{equation}\label{gpfunc3d}
\E^{\rm
GP}[\phi]=\int_{\R^3}\left(\mu|\nabla\phi|^2+V|\phi|^2+4\pi \mu
a|\phi|^4\right)d\x
\end{equation}
with the subsidiary condition \begin{equation}\label{norm}
\int_{\R^3}|\phi|^2=N.\end{equation} 
As before, $a>0$ is the scattering length of $v$. 
The corresponding energy is
\begin{equation}\label{gpen3d}
E^{\rm GP}(N,a)=\inf_{\int|\phi|^2=N}\E^{\rm GP}[\phi]= \E^{\rm
GP}[\phi^{\rm GP}],\end{equation} with a unique, positive $\phi^{\rm
GP}$. The existence of the minimizer $\phi^{\rm GP}$ is proved by
standard techniques and it can be shown to be continuously
differentiable, see \cite{LSY1999}, Sect.~2 and Appendix A. The
minimizer depends on $N$ and $a$, of course, and when this is
important we denote it by $\phi^{\rm GP}_{N,a}$.

The variational equation satisfied by the minimizer is the
{\it GP equation}
\begin{equation}\label{gpeq}
-\mu\Delta\phi^{\rm GP}(\x)+ V(\x)\phi^{\rm GP}(\x)+8\pi\mu a 
\phi^{\rm
GP}(\x)^3 = \mu^{\rm GP} \phi^{\rm GP}(\x),
\end{equation}
where $\mu^{\rm GP}$ is the chemical potential, given by
\begin{equation}\label{mugp}
\mu^{\rm GP}=dE^{\rm GP}(N,a)/dN=E^{\rm GP}(N,a)/N+ 
(4\pi \mu a/N)\int |\phi^{\rm GP}(\x)|^4 d\x.       
        \end{equation}

The GP theory has the following scaling property:
\begin{equation}\label{scalen}
E^{\rm GP}(N,a)=N E^{\rm GP}(1,Na),
\end{equation}
and
\begin{equation}\label{scalphi}
\phi^{\rm GP}_{N,a}(\x)= N^{1/2} \phi^{\rm GP}_{1,Na}(\x).
\end{equation}
Hence we see that the relevant parameter in GP theory is the
combination $Na$.
%%%%%%%%%%%%%%%%%%%%%%%%%%%

We now turn to the relation of $E^{\rm GP}$ and 
$\phi^{\rm GP}$ to the quantum mechanical ground state.  
If $v=0$, then the ground state of \eqref{trapham} is 
$$\Psi_{0}(\x_{1},\dots,\x_{N})=\hbox{$\prod_{i=1}^{N}$}\phi_{0}(\x_{i})
$$
with $\phi_{0}$ the normalized ground state of $-\mu \Delta + V(\x)$.
In this case 
clearly $\phi^{\rm GP}=\sqrt{N}\ \phi_{0}$, and then
$E^{\rm GP}=N\hbar\omega = E_0$.  In the other extreme, if $V(\x)=0$ 
for
$\x$ inside a large box of volume $L^3$ and $V(\x)= \infty$ otherwise,
then $\phi^{\rm GP} \approx \sqrt{N/L^3}$ and we get $E^{\rm
GP}(N,a) = 4\pi \mu a N^2/L^3$, which is the previously considered 
energy 
$E_0$ for the homogeneous gas in the low
density regime. (In this case, the gradient term in $\E^{\rm GP}$
plays no role.)

In general, we expect that for {\it dilute} gases in a suitable limit
\begin{equation}\label{approx}E_0
        \approx E^{\rm GP}\quad{\rm and}\quad \rho^{\rm QM}(\x)\approx
\left|\phi^{\rm GP}(\x)\right|^2\equiv \rho^{\rm 
GP}(\x),\end{equation}
where the quantum mechanical particle density in the ground state is
defined by \begin{equation} \rho^{\rm
QM}(\x)=N\int|\Psi_{0}(\x,\x_{2},\dots,\x_{N})|^2d\x_{2}\cdots
d\x_{N}.  \end{equation} {\it Dilute} means here that
\begin{equation}\bar\rho a^3\ll 1,\end{equation} where
        \begin{equation}\label{rhobar}
\bar\rho=\frac 1N\int|\rho^{\rm GP}(\x)|^2 d\x
\end{equation}
is the {\it mean density}.

The limit in which \eqref{approx} can be expected to be true 
should be chosen so that {\it all three} terms in
$\E^{\rm GP}$ make a contribution. The scaling relations 
\eqref{scalen} and \eqref{scalphi}
indicate that fixing
$Na$ as $N\to\infty$ is the right thing to do (and this is quite 
relevant since
experimentally $N$ can be quite large, $10^6$ and more,  and 
$Na$ can range from about 1 to $10^4$ \cite{DGPS}).
 Fixing $Na$ 
(which we refer to as the GP
case) also means that
we really are dealing with a dilute limit, because the mean density 
$\bar \rho$ is then of the order $N$ (since 
$\bar\rho_{N,a}=N\bar\rho_{1,Na}$) and hence 
\begin{equation}
a^3\bar\rho\sim N^{-2}.
\end{equation}

The precise statement of \eqref{approx} is:

\begin{thm}[\textbf{GP limit of the QM ground state energy and 
density}]\label{thmgp3}  
If $N\to\infty$ with $Na$ fixed, then
\begin{equation}\label{econv}
\lim_{N\to\infty}\frac{{E_0(N,a)}}{ {E^{\rm GP}(N,a)}}=1,
\end{equation}
and
\begin{equation}\label{dconv}
\lim_{N\to\infty}\frac{1}{ N}\rho^{\rm QM}_{N,a}(\x)= \left
|{\phi^{\rm GP}_{1,Na}}(\x)\right|^2
\end{equation}
in the weak $L^1$-sense.
\end{thm}

%%%%%%%%%%%%%%

Convergence can not only be proved for the ground state energy and 
density, but also for the individual energy components:

\begin{thm}[\textbf{Asymptotics of the energy 
components}]\label{compthm}
Let $\psi_0$ denote the solution to the zero-energy 
scattering equation for
$v$ (under the boundary condition
$\lim_{|\x|\to\infty}\psi_0(\x)=1$) and
$s=\int|\nabla\psi_0|^2/(4\pi a)$. Then $0<s\leq 1$ and, in the same 
limit as in Theorem~\ref{thmgp3} above, 
\begin{subequations}\label{parttwo}
\begin{eqnarray}\nonumber
&&\!\!\!\!\!\!\!\!\!\! \lim_{N\to\infty} \int  |\nabla_{\x_1} 
\Psi_0(\x_1,\X)|^2 d\x_1\,
d\X
\\ \label{3a}
 &&\qquad= \int|\nabla\phi^{\rm GP}_{1,Na}(\x)|^2d\x + 4\pi Na s 
\int|\phi^{\rm GP}_{1,Na}(\x)|^4
d\x,\\  &&\!\!\!\!\!\!\!\!\!\!\lim_{N\to\infty} \int  
V(\x_1)|\Psi_0(\x_1,\X)|^2 d\x_1\,
d\X = \int V(\x) |\phi^{\rm GP}_{1,Na}(\x)|^2 d\x, \\ \nonumber
&&\!\!\!\!\!\!\!\!\!\!\lim_{N\to\infty} \half\sum_{j=2}^N \int
v(|\x_1-\x_j|)|\Psi_0(\x_1,\X)|^2 d\x_1\, d\X
\\ \label{part2} &&\qquad=(1-s) 4\pi Na \int|\phi^{\rm 
GP}_{1,Na}(\x)|^4 
d\x.
\end{eqnarray}
\end{subequations}
\end{thm}

Here we introduced again the short hand notation (\ref{defX}).
Theorem~\ref{compthm} is a simple consequence of Theorem~\ref{thmgp3}
by variation with respect to the different components of the energy,
as was also noted in \cite{CS01a}. More precisely, Eq.\ \eqref{econv}
can be written as
\begin{equation}\label{econv22}
\lim_{N\to \infty}\frac1N E_{0}(N,a)=E^{\rm GP}(1,Na).    
\end{equation}    
The ground state
energy is a concave function of the mass parameter $\mu$, so it is
legitimate to differentiate both sides of \eqref{econv22} with respect
to $\mu$. In doing so, it has to be noted that $Na$ depends on $\mu$
through the scattering length. Using \eqref{partint} one sees that
\begin{equation}\label{dmua}
\frac{d(\mu a)}{d\mu}=\frac1 {4\pi}\int |\nabla\psi_0|^2d\x
\end{equation}
by the Feynman-Hellmann principle, since $\psi_0$ minimizes the left
side of \eqref{partint}.

We remark that in the case of a
two-dimensional Bose gas, where the relevant parameter to be kept 
fixed in
the GP limit is $N/|\ln (a^2 \bar\rho_N)|$ (c.f.\ 
Sections~\ref{sect2d} 
and~\ref{sub2d}.), the parameter
$s$ in Theorem \ref{compthm} can be shown to be always equal to
$1$. I.e., in 2D the interaction energy is purely kinetic in the GP
limit (see \cite{CS01b}).

%%%%%%%%%%%%%%
\bigskip

To describe situations where $Na$ is very large, it is appropriate 
to consider a limit where, as $N\to\infty$,  $a\gg
N^{-1}$, i.e. $Na\to\infty$, but still $\bar\rho a^3\to 0$.
 In
this case, the gradient term in the GP functional becomes
negligible compared to the other terms and the so-called {\it
Thomas-Fermi (TF) functional}
\begin{equation}\label{gtf}
\E^{\rm TF}[\rho]=\int_{\R^3}\left(V\rho+4\pi \mu a\rho^2\right)d\x
\end{equation}
arises. (Note that this functional has nothing to do with the
fermionic theory invented by Thomas and Fermi in 1927, except 
for a certain formal analogy.) It is
defined for nonnegative functions $\rho$ on $\R^3$. Its ground
state energy $E^{\rm TF}$ and density $\rho^{\rm TF}$ are defined
analogously to the GP case. (The TF functional is especially
relevant for the two-dimensional Bose gas. There  $a$ has to
decrease exponentially with $N$ in the GP limit, so the TF limit
is more adequate; see Subsection \ref{sub2d} below).

Our second main result of this section is that minimization of
(\ref{gtf}) reproduces correctly the ground state energy and
density of the many-body Hamiltonian in the limit when
$N\to\infty$, $a^3\bar \rho\to 0$, but $Na\to \infty$ (which we
refer to as the TF case), provided the external potential is
reasonably well behaved. We will assume that $V$ is asymptotically
equal to some function $W$ that is homogeneous of some order $s>0$, 
i.e., 
$W(\lambda\x)=\lambda^s W(\x)$ for all $\lambda>0$, 
and locally H\"older continuous (see \cite{LSY2d} for a precise
definition). This condition can be relaxed, but it seems adequate
for most practical applications and simplifies things
considerably.

\begin{thm}[\textbf{TF limit of the QM ground state energy
and   density}]\label{thm2} Assume that $V$ satisfies 
the conditions
stated above. If $\g\equiv Na\to\infty$ as $N\to\infty$, but still
$a^3\bar\rho\to 0$, then
\begin{equation}\label{econftf}
\lim_{N\to\infty}\frac{E_0(N,a)} {E^{\rm TF}(N,a)}=1,
\end{equation}
and
\begin{equation}\label{dconvtf}
\lim_{N\to\infty}\frac{\g^{3/(s+3)}}{N}\rho^{\rm
QM}_{N,a}(\g^{1/(s+3)}\x)= \tilde\rho^{\rm TF}_{1,1}(\x)
\end{equation}
in the weak $L^1$-sense, where $\tilde\rho^{\rm TF}_{1,1}$ is the
minimizer of the TF functional under the condition $\int\rho=1$,
$a=1$, and with $V$ replaced by $W$.
\end{thm}

%\noindent{\it Remark.} The theorems are independent of actual form
%of the interaction potential $v_1$ in (\ref{v1}), they depend only
%on the scattering length $a$. This means that in the limit we
%consider only the scattering length effects the ground state
%properties, and not the details of the potential. Note also that
%the particular limit we consider is {\it not} a mean field limit,
%since the interaction potential is very hard in this limit; in
%fact the term $4\pi a|\phi|^4$ is mostly kinetic energy (see also
%Theorem \ref{compthm} below).
%\medskip

In the following, we will present the essentials of the proofs 
Theorems \ref{thmgp3} and \ref{thm2}.
We will derive appropriate upper and lower bounds on the ground
state energy $E_0$. 

The proof of the lower bound in Theorem \ref{thmgp3} 
presented here is a modified version of (and partly simpler than)
the original proof in \cite{LSY1999}.

The convergence of the densities follows from
the convergence of the energies in the usual way by variation with
respect to the external potential. For simplicity, we set $\mu\equiv 
1$ in
the following.

\begin{proof}[Proof of Theorems \ref{thmgp3} and \ref{thm2}] {\it 
Part 1: 
Upper bound to the QM energy.} To derive an upper bound on $E_0$ we
use a generalization of a trial wave function of Dyson \cite{dyson},
who used this function to give an upper bound on the ground state
energy of the homogeneous hard core Bose gas (c.f.\ Section
\ref{upsec}). It is of the form
\begin{equation}\label{ansatz}
\Psi(\x_{1},\dots,\x_{N})
    =\prod_{i=1}^N\phi^{\rm
GP}(\x_{i})F(\x_{1},\dots,\x_{N}),
\end{equation}
where $F$ is constructed in the following way:
\begin{equation}F(\x_1,\dots,\x_N)=\prod_{i=1}^N
f(t_i(\x_1,\dots,\x_i)),\end{equation} where $t_i =
\min\{|\x_i-\x_j|, 1\leq j\leq i-1\}$ is the distance of $\x_{i}$
to its {\it nearest neighbor} among the points
$\x_1,\dots,\x_{i-1}$, and $f$ is a  function of $r\geq 0$. As in 
\eqref{deff} we
choose it to be
\begin{equation}
f(r)=\left\{\begin{array}{cl} f_{0}(r)/f_0(b) \quad
&\mbox{for}\quad r<b\\ 1 &\mbox{for}\quad r\geq b,
\end{array}\right.
\end{equation}
where $f_0$ is the solution of the zero energy scattering equation
(\ref{3dscatteq}) and $b$ is some cut-off parameter of order
$b\sim \bar\rho^{-1/3}$. The function (\ref{ansatz}) is not
totally symmetric, but for an upper bound it is nevertheless an
acceptable test wave function since the bosonic ground state
energy is equal to the {\it absolute} ground state energy.

The result of a somewhat lengthy computation (see \cite{LSY1999} for 
details) 
is the upper bound
\begin{equation}\label{ubd}
E_0(N,a)\leq E^{\rm GP}(N,a) \left( 1+O(a\bar\rho^{1/3})\right).
\end{equation}

\bigskip\noindent {\it Part 2: Lower bound to the QM energy, GP case.}
To obtain a lower bound for the QM ground state energy the
strategy is to divide space into boxes and use the estimate on the
homogeneous gas, given in Theorem \ref{lbthm2}, in each box with
{\it Neumann} boundary conditions. One then minimizes over all
possible divisions of the particles among the different boxes.
This gives a lower bound to the energy because discontinuous wave
functions for the quadratic form defined by the Hamiltonian are
now allowed. We can neglect interactions among particles in
different boxes because $v\geq 0$. Finally, one lets the box size
tend to zero. However, it is not possible to simply approximate
$V$ by a constant potential in each box. To see this consider the
case of noninteracting particles, i.e., $v=0$ and hence $a=0$.
Here $E_0=N\hbar\omega$, but a `naive' box method gives only $\min_\x
V(\x)$ as lower bound, since it clearly pays to put all the
particles with a constant wave function in the box with the lowest
value of $V$.

For this reason we start by separating out the GP wave function in
each variable and write a general wave function $\Psi$ as
\begin{equation}\label{5.23}
\Psi(\x_{1},\dots,\x_{N})=\prod_{i=1}^N\phi^{\rm
GP}(\x_{i})F(\x_{1},\dots,\x_{N}).
\end{equation}
Here $\phi^{\rm GP}=\phi^{\rm GP}_{N,a}$ is normalized so that 
$\int|\phi^{\rm GP}|^2=N$.
Eq.\ \eqref{5.23} defines $F$ for a given $\Psi$ because $\phi^{\rm 
GP}$ is
everywhere strictly positive, being the ground state of the
operator $- \Delta + V+8\pi a|\phi^{\rm GP}|^2$. We now compute
the expectation value of $H$ in the state $\Psi$. Using partial
integration and the variational equation (\ref{gpeq}) for
$\phi^{\rm GP}$, we see that 
\begin{equation}\label{ener2}
\frac{\langle\Psi|H\Psi\rangle}{\langle\Psi|\Psi\rangle}-E^{\rm 
GP}(N,a)=4\pi a
\int |\rho^{\rm GP}|^2 +Q(F),
\end{equation}
with
\begin{multline}
Q(F)=\\ \sum_{i=1}^{N} \frac{\int\prod_{k=1}^{N}\rho^{\rm GP}(\x_k)
\left(|\nabla_i F|^2+\left[\half\sum_{j\neq i} v(|\x_i-\x_j|)-8\pi a
\rho^{\rm GP}(\x_i)\right]|F|^2\right)} {\int\prod_{k=1}^{N}\rho^{\rm
GP}(\x_k)|F|^2}. \label{ener3}
\end{multline}
We recall that $\rho^{\rm GP}(\x)=|\phi^{\rm
GP}_{N,a}(\x)|^2$. For computing the ground state energy
of $H$ we have to minimize the normalized quadratic form $Q$.
 Compared to the expression for the energy involving
$\Psi$ itself we have thus obtained the replacements
\begin{equation}\label{repl}
V(\x)\to -8\pi a\rho^{\rm GP}(\x) \quad\mbox{and}\quad
\prod_{i=1}^Nd\x_i \to \prod_{i=1}^N\rho^{\rm GP}(\x_{i})d\x_{i}\ .
\end{equation}
We now use the box method on {\it this} problem. More precisely,
labeling the boxes by an index $\alpha$, we have
\begin{equation}\label{5.26}
\inf_F Q(F)\geq \inf_{\{n_\al\}} \sum_\al \inf_{F_\al}Q_\al
(F_\al),
\end{equation}
where $Q_\al$ is defined by the same formula as $Q$  but with the
integrations limited to the box $\alpha$,  $F_{\alpha}$ is a wave
function with particle number $n_\alpha$, and the infimum is taken
over all distributions of the particles with $\sum n_\al=N$.

We now fix some $M>0$, that will eventually tend to $\infty$, and
restrict ourselves to boxes inside a cube $\Lambda_M$ of side length
$M$. Since $v\geq 0$ the contribution to
\eqref{5.26} of boxes outside this cube is easily estimated from 
below by 
$-8\pi Na \sup_{\x\notin \Lambda_M}\rho^{\rm GP}(\x)$, which, divided
by $N$, is arbitrarily small for $M$ large, since $Na$ is fixed and
$\pgp/N^{1/2}=\pgp_{1,Na}$ decreases faster than exponentially at
infinity (\cite{LSY1999}, Lemma A.5).

%%%%%%%%%%%%%%
For the boxes inside the cube $\Lambda_M$ we want to use Lemma
\ref{dysonl} and therefore we must approximate $\rho^{\rm GP}$ by
constants in each box. Let $\rmax$ and $\rmin$, respectively, denote
the maximal and minimal values of $\rho^{\rm GP}$ in box $\al$. Define
\begin{equation}
\Psi_\alpha(\x_1,\dots, \x_{n_\al})=F_\alpha(\x_1,\dots, \x_{n_\al}) 
\prod_{k=1}^{n_\al}\phi^{\rm GP}(\x_k),
\end{equation}
and
\begin{equation}
\Psi^{(i)}_\alpha(\x_1,\dots, \x_{n_\al})=F_\alpha(\x_1,\dots, 
\x_{n_\al}) 
\prod_{\substack{k=1 \\ k\neq 
i}}^{n_\al}\phi^{\rm GP}(\x_k).
\end{equation}
We have, for all $1\leq i\leq n_\al$,
\begin{equation}\label{5.29}
\begin{split}
&\int\prod_{k=1}^{n_\alpha}\rho^{\rm GP}(\x_k)\left(|\nabla_i 
F_\alpha|^2\right.+\half\sum_{j\neq i}\left.
v(|\x_i-\x_j|)|F_\alpha|^2\right) 
\\&\geq
\rmin\int \left(|\nabla_i\Psi^{(i)}_\alpha|^2\right. 
+\half\sum_{j\neq i}\left.
v(|\x_i-\x_j|)|\Psi^{(i)}_\alpha|^2\right).
\end{split}
\end{equation}
We now use 
Lemma \ref{dysonl} to get, for all $0\leq \eps\leq 1$,
\begin{equation}\label{5.30}
(\ref{5.29})\geq \rmin\int \left(\varepsilon 
|\nabla_i\Psi^{(i)}_\alpha|^2 
+a(1-\eps)U(t_i)|\Psi^{(i)}_\alpha|^2\right)
\end{equation}
where $t_i$ is the distance to the nearest neighbor of $\x_i$, c.f., 
\eqref{2.29}, and $U$ the potential \eqref{softened}.

Since $\Psi_\alpha=\pgp(\x_i)\Psi^{(i)}_\al$ we can estimate
\begin{equation}\label{tpsial}
|\nabla_i\Psi_\alpha|^2 \leq 2\rmax |\nabla_i\Psi^{(i)}_\alpha|^2 +
2|\Psi^{(i)}_\alpha|^2 N C_M
\end{equation}
with
\begin{equation}
C_M=\frac1N\sup_{\x\in\Lambda_M}|\nabla\pgp(\x)|^2=\sup_{\x\in\Lambda_M}
|\nabla\pgp_{1,Na}(\x)|^2.  \end{equation} Since $Na$ is fixed, $C_M$
is independent of $N$. Inserting \eqref{tpsial} into
\eqref{5.30}, summing over $i$ and using $\rho^{\rm GP}(\x_i)\leq 
\rmax$ in 
the 
last term of \eqref{ener3} (in the box $\al$), we get
\begin{equation}\label{qalfal}
Q_\al(F_\alpha)\geq \frac{\rmin}{\rmax}E^{U}_\eps(n_\al,L)-8\pi 
a\rmax n_\al -
\eps C_M n_\al,
\end{equation}
where $L$ is the side length of the box and $E^{U}_\eps(n_\al,L)$ is
the ground state energy of
\begin{equation}\label{eueps}
\sum_{i=1}^{n_\al}(-\half\eps \Delta_i+(1-\eps)aU(t_i)) 
\end{equation}
in the box (c.f.\ \eqref{halfway}). We want to minimize \eqref{qalfal}
with respect to $n_\al$ and drop the subsidiary condition
$\sum_\al{n_\al}=N$ in \eqref{5.26}. This can only lower the minimum.
For the time being we also ignore the last term in
\eqref{qalfal}. (The total contribution of this term for all boxes is
bounded by $\eps C_M N$ and will be shown to be negligible compared to
the other terms.)

Since the lower bound for the energy of Theorem
\ref{lbthm2} was obtained precisely from a lower bound to the operator
\eqref{eueps}, we can use the statement and proof of Theorem 
\ref{lbthm2}. 
{F}rom this we see that
\begin{equation}\label{basicx}
E^{U}_\eps(n_\al,L)\geq (1-\varepsilon)\frac{4\pi
an_\al^2}{L^3}(1-CY_\al^{1/17})
\end{equation} 
with $Y_\al=a^3n_\al/L^3$, provided  $Y_\al$ is small enough, 
$\eps\geq Y_\al^{1/17}$ and
$n_\al\geq {\rm (const.)} Y_\al^{-1/17}$. The condition on $\eps$ is
certainly fulfilled if we choose $\eps=Y^{1/17}$ with
$Y=a^3N/L^3$. We now want to show that the $n_\alpha$ minimizing 
the right side of \eqref{qalfal} is large enough for \eqref{basicx} 
to apply.

If the minimum of the right side of \eqref{qalfal}
(without the last term) is taken for some $\bar n_\al$, we have
\begin{equation}\label{minnal}
\frac{\rmin}{\rmax}
\left(E^{U}_\eps(\bar n_\al+1,L)-E^{U}_\eps(\bar n_\al,L)\right)\geq 
8\pi a\rmax.
\end{equation}
On the other hand, we  claim that
\begin{lem} For any $n$ 
\begin{equation}\label{chempot}
E^{U}_\eps( n+1,L)-E^{U}_\eps(n,L)\leq 8\pi a\frac{ 
n}{L^3}.
\end{equation}
\end{lem}
\begin{proof}
Denote the operator \eqref{eueps} by $\tilde H_n$,  with 
$n_\alpha=n$, and 
let $\tilde 
\Psi_n$ 
be its ground state. Let $t_i'$ be the distance to the nearest 
neighbor
of $\x_i$ among the $n+1$ points $\x_1,\dots,\x_{n+1}$ (without 
$\x_i$) and $t_i$ the
corresponding distance excluding $\x_{n+1}$. Obviously, for $1\leq 
i\leq n$,
\begin{equation}
U(t_i')\leq U(t_i)+U(|\x_i-\x_{n+1}|)
\end{equation}
and
\begin{equation}
U(t_{n+1}')\leq \sum_{i=1}^nU(|\x_i-\x_{n+1}|).
\end{equation}
Therefore
\begin{equation}
\tilde H_{n+1}\leq \tilde 
H_{n}-\half\eps\Delta_{n+1}+2a\sum_{i=1}^nU(|\x_i-\x_{n+1}|).
\end{equation}
Using $\tilde\Psi_n/L^{3/2}$  as trial function for $\tilde H_{n+1}$ 
we arrive at
\eqref{chempot}.
\end{proof}
Eq.\ \eqref{chempot} together with \eqref{minnal} shows that
 $\bar n_\al$ is at least $\sim \rmax L^3$.
We shall choose 
$L\sim N^{-1/10}$, so
the conditions needed for (\ref{basicx}) are fulfilled for $N$ large 
enough, 
since $\rmax\sim N$ and hence
$\bar n_\al\sim N^{7/10}$ and
$Y_\al\sim N^{-2}$.

In order to obtain a lower bound on $Q_\al$ we therefore have to 
minimize
\begin{equation}\label{qalpha}
 4\pi 
a\left(\frac{\rmin}{\rmax}\frac{n_\al^2}{L^3}\left(1-CY^{1/17}\right)
-2n_\al\rmax\right).
\end{equation}
We can drop the 
requirement that $n_\al$ has to 
be an integer. The minimum of (\ref{qalpha}) is obtained for
\begin{equation}
n_\al= \frac{\rmax^2}{\rmin}\frac{L^3}{(1-CY^{1/17})}.
\end{equation}
By Eq.\ (\ref{ener2}) this gives the following lower bound, 
including now the last term in \eqref{qalfal} as well as the 
contributions from the 
boxes outside $\Lambda_M$, 
\begin{equation}\label{almostthere}
\begin{split}
&E_0(N,a)-E^{\rm GP}(N,a)\geq \\
&4\pi a\int|\rho^{\rm GP}|^2-4\pi a\sum_{\al\subset\Lambda_M} 
\rmin^2
L^3\left(\frac{\rmax^3}{\rmin^3}\frac{1}{(1-CY^{1/17})}\right)\\ 
&-Y^{1/17}NC_M-4\pi 
aN\sup_{\x\notin\Lambda_M}\rho^{\rm GP}(\x).
\end{split}
\end{equation}
Now $\rho^{\rm GP}$ is differentiable and strictly 
positive. Since all the boxes are in the fixed cube $\Lambda_M$ there 
are 
constants 
$C'<\infty$, $C''>0$,
such that
\begin{equation}
\rmax-\rmin\leq NC'L,\quad \rmin\geq NC''.
\end{equation}
Since $L\sim N^{-1/10}$ and $Y\sim N^{-17/10}$ we therefore have, for 
large $N$,
\begin{equation}
\frac{\rmax^3}{\rmin^3}\frac{1}{(1-CY^{1/17})}\leq 
1+{\rm (const.)}N^{-1/10}
\end{equation}
Also,
\begin{equation}
4\pi a\sum_{\al\subset\Lambda_M} \rmin^2 L^3\leq 4\pi a\int 
|\rho^{\rm GP}|^2\leq E^{\rm 
GP}(N,a).
\end{equation}
Hence, noting that $ E^{\rm GP}(N,a)=N E^{\rm GP}(1,Na)\sim N$ since 
$Na$ is fixed,
\begin{equation}\label{there}
\frac{E_0(N,a)}{E^{\rm GP}(N,a)}\geq 1-{\rm 
(const.)}(1+C_M)N^{-1/10}-{\rm (const.)}
\sup_{\x\notin \Lambda_M}|\pgp_{1,Na}|^2,
\end{equation}
where the constants depend on $Na$. We can now take $N\to\infty$ and 
then $M\to\infty$.

\bigskip
\noindent {\it Part 3: Lower bound to the QM energy, TF case.}
In the above proof of the lower bound in the GP case we did not 
attempt to
keep track of the dependence of the constants on $Na$. In the TF case
$Na\to\infty$, so one would need to take a closer look at this
dependence if one wanted to carry the proof directly over to this
case. But we don't have to do so, because there is a simpler direct
proof. Using the explicit form of the TF minimizer, namely
\begin{equation}\label{tfminim}
\rho^{\rm TF}_{N,a}(\x)=\frac 1{8\pi a}[\mu^{\rm TF}-V(\x)]_+,
\end{equation}
where  $[t]_+\equiv\max\{t,0\}$ and $\mu^{\rm TF}$ is chosen so
that the normalization condition $\int \rho^{\rm TF}_{N,a}=N$
holds, we can use 
\begin{equation}\label{vbound}
V(\x)\geq \mu^{\rm TF}-8\pi a \rho^{\rm
TF}(\x) 
\end{equation}
to get a replacement as in (\ref{repl}), but without
changing the measure. Moreover, $\rho^{\rm TF}$ has compact
support, so, applying again the box method described above, the
boxes far out do not contribute to the energy. However, $\mu^{\rm
TF}$ (which depends only on the combination $Na$) tends to
infinity as $Na\to\infty$. We need to control the
asymptotic behavior of $\mtf$, and this leads to
the restrictions on $V$ described in the paragraph preceding
Theorem \ref{thm2}. For simplicity, we shall here only consider the 
case when $V$ 
itself is homogeneous, i.e., $V(\lambda\x)=\lambda^sV(\x)$ for all 
$\lambda>0$ with 
some $s>0$.  

In the same way as in \eqref{mugp} we have, with $g=Na$, 
\begin{equation}\label{mutf}
\mu^{\rm TF}(g)=dE^{\rm TF}(N,a)/dN=E^{\rm TF}(1,g)+ 
4\pi g\int |\rho^{\rm TF}_{1,g}(\x)|^2 d\x.     
        \end{equation}
The TF energy, chemical potential and minimizer 
satisfy the scaling relations
\begin{equation}
E^{\rm TF}(1,g)=g^{s/(s+3)}E^{\rm TF}(1,1),
\end{equation}
\begin{equation}
\mu^{\rm TF}(g)=g^{s/(s+3)} \mu^{\rm TF}(1)  ,
\end{equation}
and
\begin{equation}
g^{3/(s+3)}\rho^{\rm TF}_{1,g}(g^{1/(s+3)}\x)= \rho^{\rm 
TF}_{1,g}(\x)  .
\end{equation}
We also introduce the scaled interaction potential, $\widehat v$, by
\begin{equation}
\widehat v(\x)  =g^{2/(s+3)}v(g^{1/(s+3)}\x)
\end{equation}
with scattering length 
\begin{equation}
\widehat a=g^{-1/(s+3)}a.
\end{equation}
 Using \eqref{vbound}, \eqref{mutf} and 
the scaling relations we obtain
\begin{equation}
E_0(N,a)\geq E^{\rm TF}(N,a)+4\pi N g^{s/(s+3)}\int |\rho^{\rm 
TF}_{1,1}|^2 +
g^{-2/(s+3)}Q
\end{equation}
with
\begin{equation}
Q=\inf_{\int|\Psi|^2=1}\sum_{i}\int\left(|\nabla_i\Psi|^2\right.+\half 
\sum_{j\neq i}
\left.\widehat v(\x_i-\x_j)|\Psi|^2-8\pi
N\widehat a \rtf_{1,1}(\x_i)|\Psi|^2\right).
\end{equation}
We can now proceed exactly as in Part 2 to arrive at the analogy 
of 
Eq.\ \eqref{almostthere}, which in the present case becomes
\begin{equation}\label{almosttherex}
\begin{split}
&E_0(N,a)-E^{\rm TF}(N,a)\geq \\
&4\pi N g^{s/(s+3)}\int |\rho^{\rm TF}_{1,1}|^2-4\pi N\widehat 
a\sum_{\al} 
\rmax^2
L^3(1-C\widehat Y^{1/17})^{-1}.
\end{split}
\end{equation}
Here $\rmax$ is the maximum of $\rho^{\rm TF}_{1,1}$ in the box
$\alpha$, and $\widehat Y=\widehat a^3 N/L^3$. This holds as long as 
$L$ does not
decrease too fast with $N$. In particular, if $L$ is simply fixed, 
this
holds for all large enough $N$. Note that
\begin{equation}
\bar\rho=N\bar\rho_{1,g}\sim N g^{-3/(s+3)} \bar\rho_{1,1},
\end{equation}
so that $\widehat a^3 N\sim a^3 \bar
\rho$ goes to zero as $N\to\infty$ by assumption. Hence, if we first 
let
$N\to\infty$ (which implies $\widehat Y\to 0$) and then take $L$ to 
zero, we
of arrive at the desired 
result
\begin{equation}\label{lowertf}
\liminf_{N\to\infty}\frac{E_0(N,a)}{E^{\rm TF}(N,a)}\geq 1
\end{equation}
in the limit $N\to\infty$, $a^3\bar\rho\to 0$. Here
we used the fact that (because $V$, and hence $\rtf$, is continuous by
assumption) the Riemann sum $\sum_\al\rmax^2 L^3$ converges to
$\int|\rtf_{1,1}|^2$ as $L\to 0$.  Together with the upper bound 
(\ref{ubd}) 
and
the fact that $E^{\rm GP}(N,a)/E^{\rm TF}(N,a)=E^{\rm GP}(1,Na)/E^{\rm
TF}(1,Na)\to 1$ as $Na\to\infty$, which holds under our regularity
assumption on $V$ (c.f.\ Lemma 2.3 in \cite{LSY2d}), this proves
(\ref{econv}) and (\ref{econftf}).

\bigskip
\noindent {\it Part 4: Convergence of the densities.} The
convergence of the energies implies the convergence of the
densities in the usual way by variation of the external potential.
We show here the TF case, the GP case goes analogously. Set again
$\g=Na$. Making the replacement
\begin{equation}
V(\x)\longrightarrow V(\x)+\delta\g^{s/(s+3)}Z(\g^{-1/(s+3)}\x)
\end{equation}
for some positive $Z\in C_0^\infty$ and redoing the upper and
lower bounds we see that (\ref{econftf}) holds with $W$ replaced
by $W+\delta Z$. Differentiating with respect to $\delta$ at
$\delta=0$ yields
\begin{equation}
\lim_{N\to\infty}\frac{\g^{3/(s+3)}}N\rho^{\rm
QM}_{N,a}(\g^{1/(s+3)}\x) =\tilde\rho^{\rm TF}_{1,1}(\x)
\end{equation}
in the sense of distributions. Since the functions all have
$L^1$-norm 1, we can conclude that there is even weak
$L^1$-convergence.
\end{proof}

\subsection{Two Dimensions}\label{sub2d}

%%%%%%%%%%%%%%%%
In contrast to the three-dimensional case the energy per particle for
a dilute gas in two dimensions is {\it nonlinear} in $\rho$. In view
of Schick's formula \eqref{2den} for the energy of the homogeneous gas
it would appear natural to take the interaction into account in two
dimensional GP theory by a term
\begin{equation}
4\pi\int_{\R^2} |\ln(|\phi(\x)|^2 a^2)|^{-1}|\phi(\x)|^4{
d}\x,\end{equation} 
and such a term has, indeed, been suggested in
\cite{Shev} and \cite{KoSt2000}.  However, since the nonlinearity
appears only in a logarithm, this term is unnecessarily complicated
as far as leading order computations are concerned.  For dilute gases
it turns out to be sufficient, to leading order, to use an interaction
term of the same form as in the three-dimensional case, i.e, define 
the 
GP functional as (for simplicity we put $\mu=1$ in this section)
\begin{equation}\label{2dgpfunc}
\E^{\rm
GP}[\phi]=\int_{\R^2}\left(|\nabla\phi|^2+V|\phi|^2+4\pi 
\alpha|\phi|^4\right)d\x,
\end{equation}
where instead of $a$ the coupling constant is now
\begin{equation}\label{alpha}\alpha=|\ln(\bar\rho_N 
a^2)|^{-1}\end{equation}
with $\bar\rho_N$ the {\em mean density}
for the GP functional 
at coupling constant 
$1$ and particle number $N$. This is defined analogously to 
\eqref{rhobar} 
as
\begin{equation}
\bar\rho_N=\frac1N\int|\phi^{\rm GP}_{N,1}|^4d\x
\end{equation}
where $\phi^{\rm GP}_{N,1}$ is the minimizer of \eqref{2dgpfunc} with 
$\alpha=1$ and subsidiary condition $\int|\phi|^2=N$.
Note that $\alpha$ in \eqref{alpha} depends on 
$N$ through the mean density.

%%%%%%%%%%%%%%%%%%%
Let us denote the GP energy 
for a given $N$ and coupling constant 
$\alpha$ by $E^{\rm GP}(N,\alpha)$ and the corresponding minimizer by
$\phi^{\rm GP}_{N,\alpha}$.
As in three dimensions the scaling relations
\begin{equation}E^{\rm GP}(N,\alpha)=NE^{\rm 
GP}(1,N\alpha)\end{equation}
and
    \begin{equation}N^{-1/2}\phi^{\rm GP}_{N,\alpha}=\phi^{\rm 
GP}_{1,N\alpha},
\end{equation}
hold, and the relevant parameter is
\begin{equation}g\equiv N\alpha.\end{equation}

In three dimensions, where $\alpha=a$, 
it is natural to consider the limit $N\to\infty$ with $g=Na$= const.
The analogue of Theorem \ref{thmgp3} in two dimensions is
\begin{thm}[{\bf Two-dimensional GP limit 
    theorem}] 
\label{2dlimit}
If, for $N\to\infty$,\linebreak $a^2\brtf_N\to 0$ with
$g=N/|\ln(a^2\brtf_N)|$ fixed, then
\begin{equation}\label{econv2}
\lim_{N\to\infty}\frac{E_{0}(N,a)}{\Egp(N,1/|\ln(a^2\brtf_N)|)}=
1
\end{equation}
and
\begin{equation}\label{dconv2}
\lim_{N\to\infty}\frac{1}{ N}\rho^{\rm QM}_{N,a}(\x)= \left
|{\phi^{\rm GP}_{1,g}}(\x)\right|^2
\end{equation}
in the weak $L^1$-sense.
\end{thm}

This result, however, is of rather limited use in practice.  The 
reason is
that in two dimensions the scattering length has to
decrease exponentially with $N$ if $g$ is fixed.  
The parameter $g$ is
typically {\it very large} in two dimensions 
so it is more appropriate to consider the
limit $N\to\infty$ and $g\to\infty$ (but still $\bar\rho_N a^2\to
0$).

For  potentials $V$ that are {\it homogeneous} functions of $\x$, 
i.e., 
\begin{equation}\label{homog}V(\lambda 
\x)=\lambda^sV(\x)\end{equation}
for some $s>0$, this limit can be described by the a
`Thomas-Fermi' energy functional like \eqref{gtf} with coupling 
constant unity:
\begin{equation}\label{tffunct}
\E^{\rm TF}[\rho]=\int_{\R^2}\left(V(\x)\rho(\x)+4\pi 
\rho(\x)^2\right)
{ d}\x.
\end{equation}
This is just the GP functional without the gradient term and 
$\alpha=1$.
Here $\rho$ is a nonnegative function on $\R^2$ and the normalization 
condition is 
\begin{equation}\label{norm2}\int\rho(\x)d\x=1.\end{equation}

The minimizer of \eqref{tffunct} can be given explicitly.  It is
\begin{equation}\label{tfminim2}\rho^{\rm 
TF}_{1,1}(\x)=(8\pi)^{-1}[\mu^{\rm TF}-V(\x)]_+\end{equation}
where the chemical potential 
$\mu^{\rm TF}$ is determined by the normalization condition 
\eqref{norm2} 
and $[t]_{+}=t$ 
for $t\geq 0$ and zero otherwise.
We denote the corresponding energy by $E^{\rm TF}(1,1)$.
By scaling one obtains
\begin{equation}\lim_{g\to\infty} 
    E^{\rm GP}(1,g)/g^{s/(s+2)}=E^{\rm TF}(1,1),\end{equation}
 \begin{equation}\label{gptotf}\lim_{g\to\infty}g^{2/(s+2)}
\rho^{\rm 
GP}_{1,g}(g^{1/(s+2)}\x)=\rho^{\rm TF}_{1,1}(\x),\end{equation}
with the latter limit in the  strong $L^2$ sense.

Our main result about two-dimensional 
Bose gases in  external potentials satisfying \eqref{homog}  
is that analogous limits also hold for the many-particle quantum 
mechanical 
ground state at
low densities:
\begin{thm}[{\bf Two-dimensional TF limit theorem}]\label{thm22}
In two dimensions, if
$a^2\bar\rho_N\to 0$, but $g=N/|\ln(\bar\rho_N a^2)|\to \infty$ as 
$N\to\infty$ 
then
\begin{equation}\lim_{N\to \infty}\frac{E_0(N,a)}{g^{s/s+2}}= 
E^{\rm TF}(1,1)\end{equation}
and, in the weak $L^1$ sense,
\begin{equation}\label{conv}\lim_{N\to\infty}\frac{g^{2/(s+2)}}N
\rho^{\rm 
QM}_{N,a}(g^{1/(s+2)}\x)=\rho^{\rm TF}_{1,1}(\x).\end{equation}
\end{thm}

\noindent {\it Remarks:} 1. As in Theorem \ref{thm2}, it is 
sufficient that $V$ is
asymptotically equal to some homogeneous potential, $W$. In this case,
$E^{\rm TF}(1,1)$ and $\rho^{\rm TF}_{1,1}$ in Theorem \ref{thm22}
should be replaced by the corresponding quantities for $W$.

2. From Eq.\ \eqref{gptotf} it follows that
\begin{equation}\bar\rho_N\sim N^{s/(s+2)}\end{equation} for large 
$N$.
Hence the low density criterion 
$a^2\bar\rho\ll 1$, means that
$a/L_{\rm osc}\ll  N^{-s/2(s+2)}$.

%%%%%%%%%%%%%%%%%
We shall now comment briefly on the proofs of Theorems 
\ref{2dlimit} and \ref{thm22}, 
mainly pointing out the differences from the 3D case considered 
previously.

The upper bounds for the energy are obtained exactly in a same way as
in three dimensions. For the lower bound in Theorem \ref{2dlimit} the
point to notice is that the expression
 \eqref{qalpha}, that has to be minimized over $n_\al$, is in 2D
 replaced by
\begin{equation}\label{qalpha2}
 4\pi 
\left(\frac{\rmin}{\rmax}\frac{n_\al^2}{L^2}\frac1{|\ln(a^2n_\alpha/L^2)|}
\left(1-\frac C{|\ln(a^2N/L^2)|^{1/5}}\right)
-\frac{2n_\al\rmax}{|\ln(a^2\bar\rho_N)|}\right),
\end{equation}
since Eq.\ \eqref{basicx} has to be 
replaced by the analogous inequality for 2D (c.f.\ \eqref{lower}).
To minimize \eqref{qalpha2} we use the following lemma:

\begin{lem}\label{xb}
For $0<x,b<1$ and $k\geq 1$ we have
\begin{equation}
\frac{x^2}{|\ln x|}-2\frac b{|\ln b|}xk\geq -
\frac{b^2}{|\ln b|}\left(1+\frac 1{(2|\ln b|)^2}\right)k^2.
\end{equation}
\end{lem}

\begin{proof} Replacing $x$ by $xk$ and using the monotonicity of 
$\ln$ we 
see that it suffices to consider $k=1$.
Since $\ln x\geq-\frac 1{de}x^{-d}$ for
all $d>0$ we have
\begin{equation}
\frac{x^2}{b^2}\frac{|\ln b|}{|\ln x|}
-2\frac xb\geq\frac{|\ln b|}{b^2}ed x^{2+d}-\frac{2x}{b}
\geq c(d)(b^ded\,|\ln b|)^{-1/(1+d)}
\end{equation}
with
\begin{equation}
c(d)=2^{(2+d)/(1+d)}\left(\frac 1{(2+d)^{(2+d)/(1+d)}}-\frac 1
{(2+d)^{1/(1+d)}}\right)\geq -1-\frac 14d^2.
\end{equation}
Choosing $d=1/|\ln b|$ gives the desired result.
\end{proof}

Applying this lemma with $x=a^2n_\al/L^2$, $b=a^2\rmax$ and 
\begin{equation}k=\frac{\rmax}{\rmin}\,
\left(1-\frac 
C{|\ln(a^2N/L^2)|^{1/5}}\right)^{-1}\frac{|\ln(a^2\rmax)|}
{|\ln(a^2\bar\rho_N)|}
\end{equation}
we get the bound
\begin{equation}
\eqref{qalpha2}\geq -4\pi\frac{\rmax^2L^2}{|\ln(a^2\bar\rho_N)|}
\left(1+\frac1{4|\ln(a^2\rmax)|^2}\right) k.
\end{equation}
In the limit considered, $k$ and the factor in parenthesis both tend 
to 1 and 
the Riemann sum over the boxes $\alpha$ converges to the integral as 
$L\to 0$.

The TF case, Thm.\ \ref{thm22}, is treated in the same way as in 
three 
dimensions, with modifications analogous to those just discussed when 
passing 
from 3D to 2D in GP theory.

\section{Bose-Einstein Condensation and Superfluidity for Dilute 
Trapped Gases}\label{becsect}

It was shown in the previous section that, for each  fixed $Na$, the 
minimization
of the GP functional correctly reproduces the large $N$
asymptotics of the ground state energy and density of $H$ -- but
no assertion about BEC in this limit was made. We will now extend
this result by showing that in the Gross-Pitaevskii limit  there
is indeed 100\% Bose condensation in the ground state. This is a
generalization of the homogeneous case considered in Theorem
\ref{hombecthm} and although it is not the same as BEC in the 
thermodynamic limit it is quite relevant for the actual experiments 
with Bose gases in traps. 
In the following, we concentrate on the 3D case,
but analogous considerations apply also  to the 2D case. We also 
discuss briefly some extensions of Theorem \ref{T1a} pertaining to 
superfluidity in trapped gases.

As in the last section we choose to keep the length scale $L_{\rm
osc}$ of the confining potential fixed and thus write $Na$ instead of
$Na/L_{\rm osc}$.  Consequently the powers of $N$ appearing in the
proofs are different from those in the proof Theorem \ref{hombecthm}, 
where we
kept $Na/L$ {\it and} $N/L^3$ fixed.

For use later, we define the projector
\begin{equation}
P^{\rm GP}= |\phi^{\rm GP}\rangle\langle \phi^{\rm GP}|\ .
\end{equation}
Here (and everywhere else in this section) we denote $\phi^{\rm
GP}\equiv\phi^{\rm GP}_{1,Na}$ for simplicity, where $\phi^{\rm
GP}_{1,Na}$ is the minimizer of the GP functional (\ref{gpfunc3d})
with parameter $Na$ and normalization condition $\int|\phi|^2=1$
(compare with (\ref{scalphi})). Moreover, we set $\mu\equiv 1$.

In the following, $\Psi_0$ denotes the (nonnegative and normalized)
ground state of the Hamiltonian (\ref{trapham}). BEC refers to the
reduced one-particle density matrix $ \gamma(\x,\x')$ of $\Psi_0$,
defined in (\ref{defgamma}). The precise definition of BEC is is that 
for some 
$c>0$ this integral operator has for all large $N$ an
an eigenfunction with eigenvalue $\geq cN$.

Complete (or 100\%) BEC is defined to be the property that
$\mbox{$\frac{1}{N}$}\gamma(\x,\x')$ not only has an eigenvalue of
order one, as in the general case of an incomplete BEC, but in the
limit it has only one nonzero eigenvalue (namely 1). Thus,
$\mbox{$\frac{1}{N}$}\gamma(\x,\x')$ becomes a simple product
$\varphi(\x)^*\varphi(\x')$ as $N\to \infty$, in which case $\varphi$ 
is called
the {\it condensate wave function}.  In the GP limit, i.e.,
$N\to\infty$ with $N a$ fixed, we can show that this is the case, and
the condensate wave function is, in fact, the GP minimizer 
$\phi^{\rm GP}$.

\begin{thm}[\textbf{Bose-Einstein condensation in a 
trap}]\label{becthm}
For each fixed $Na$ $$ \lim_{N\to\infty} \frac 1 N \gamma(\x, \x')
= \phi^{\rm GP}(\x)\phi^{\rm GP}(\x')\ . $$ in trace norm, 
i.e., $\Tr \left|\frac 1
N \gamma - P^{\rm GP} \right| \to 0$.
\end{thm}

We remark that Theorem \ref{becthm} implies that there is also 100\%
condensation for all $n$-particle reduced density matrices 
\begin{eqnarray}\nonumber
&&\gamma^{(n)}(\x_1,\dots,\x_n;\x_1',\dots,\x_n')\\&&=n!\binom{N}{n}\int
\Psi_0(\x_1,\dots,\x_N)\Psi_0(\x_1',\dots,\x_n',\x_{n+1},
\dots\x_N)d\x_{n+1}\cdots d\x_N\nonumber \\
\end{eqnarray}
of
$\Psi_0$, i.e., they converge, after division by the normalization 
factor,  
to the one-dimensional projector onto
the $n$-fold tensor product of $\phi^{\rm GP}$. In other words, 
for 
$n$ fixed particles the probability of finding them all in the same 
state 
$\phi^{\rm GP}$ tends to 1 in the 
limit considered. To see this,
let $a^*, a$ denote the boson creation and annihilation operators
for the state $\phi^{\rm GP}$, and observe that
\begin{equation}
1\geq \lim_{N\to\infty} N^{-n}\langle \Psi_0 | (a^*)^n 
a^n|\Psi_0\rangle =
\lim_{N\to\infty} N^{-n} \langle \Psi_0 | (a^*a)^n|\Psi_0\rangle  \  ,
\end{equation}
since the terms coming from the commutators $[a, a^*]=1$ are of
lower order as $N\to \infty$ and vanish in the limit. From
convexity it follows that
\begin{equation}
N^{-n}  \langle \Psi_0 | (a^*a)^n|\Psi_0\rangle \geq N^{-n} \langle
\Psi_0 | a^*a|\Psi_0\rangle ^n \,
\end{equation}
which converges to $1$ as $N\to\infty$, proving our claim.

Another corollary, important for the interpretation of
experiments, concerns the momentum distribution of the ground
state.

\begin{corollary}[\textbf{Convergence of momentum distribution}] Let
$$\widehat\rho (\k)=\int \int\gamma(\x, \x') \exp [i \k\cdot (\x
-\x')]
 d\x d\x'$$
denote the one-particle momentum  density of $\Psi_0$. Then, for
fixed $Na$, $$ \lim_{N\to\infty} \frac 1N
\widehat\rho(\k)=|\widehat\phi^{\rm GP}(\k)|^2 $$ strongly in
$L^1(\R^3)$. Here, $\widehat\phi^{\rm GP}$ denotes the Fourier
transform of $\phi^{\rm GP}$.
\end{corollary}

\begin{proof} If ${\mathcal F}$ denotes the (unitary) operator 
`Fourier
transform' and if $h$ is an arbitrary $L^\infty$-function,
then
\begin{eqnarray}\nonumber
\left|\frac 1N\int \widehat\rho h-\int |\widehat\phi^{\rm
GP}|^2 h\right|&=&\left|\Tr[{\mathcal F}^{-1}
(\gamma/N-P^{\rm GP}){\mathcal F}h]\right|\\ \nonumber
&\leq& \|h\|_\infty \Tr |\gamma/N-P^{\rm GP}|,
\end{eqnarray}
from which we conclude that $$\|\widehat\rho/N-|\widehat\phi^{\rm
GP}|^2 \|_1\leq \Tr|\gamma/N-P^{\rm GP}|\ .$$
\end{proof}

As already stated, Theorem \ref{becthm} is a generalization of Theorem
\ref{hombecthm}, the latter corresponding to the case that $V$ is a
box potential.  It should be noted, however, that we use different
scaling conventions in these two theorems: In Theorem \ref{hombecthm}
the box size grows as $N^{1/3}$ to keep the density fixed, while in
Theorem \ref{becthm} we choose to keep the confining external
potential fixed.  Both conventions are equivalent, of course, c.f.\
the remarks in the second paragraph of Section \ref{sectgp}, but
when comparing the exponents of $N$ that appear in the proofs of the
two theorems the different conventions should be born in mind.

As in Theorem \ref{hombecthm} there are two essential components of
our proof of Theorem \ref{becthm}.  The first is a proof that the part
of the kinetic energy that is associated with the interaction $v$
(namely, the second term in (\ref{3a})) is mostly located in small
balls surrounding each particle.  More precisely, these balls can be
taken to have radius roughly $N^{-5/9}$, which is much smaller than
the mean-particle spacing $N^{-1/3}$.  (The exponents differ from
those of Lemma \ref{L1} because of different scaling conventions.)  
This
allows us to conclude that the function of $\x$ defined for each fixed
value of $\X$ by
\begin{equation}\label{defff}
f_\X(\x)=\frac 1{\phi^{\rm GP}(\x)} \Psi_0(\x,\X)\geq 0
\end{equation}
has the property that $\nabla_\x f_\X(\x)$ is almost zero outside
the small balls centered at points of $\X$.

The complement of the small balls has a large volume but it can be
a weird set; it need not even be connected. Therefore, the
smallness of $\nabla_\x f_\X(\x)$ in this set does not guarantee
that $f_\X(\x)$ is nearly constant (in $\x$), or even that it is
continuous. We need $f_\X(\x)$ to be nearly constant in order to
conclude BEC. What saves the day is the knowledge that the total
kinetic energy of $f_\X(\x)$ (including the balls) is not huge.
The result that allows us to combine these two pieces of
information in order to deduce the almost constancy of $f_\X(\x)$
is the generalized Poincar\'e inequality in Lemma \ref{lem2}.
The important point in this lemma is  that there is no
restriction on $\Omega$ concerning regularity or connectivity.

Using the results of Theorem \ref{compthm}, partial integration
and the GP equation (i.e., the variational equation for $\phi^{\rm 
GP}$,
see Eq. (\ref{gpeq})) we see that
\begin{equation}\label{bound}
\lim_{N\to\infty} \int  |\phi^{\rm GP}(\x)|^2 |\nabla_\x 
f_\X(\x)|^2 d\x\,
d\X
 = 4\pi Na s\int |\phi^{\rm GP}(\x)|^4 d\x\ .
\end{equation}
The following Lemma shows that to leading order all the energy in
(\ref{bound}) is concentrated in small balls.

\begin{lem}[\textbf{Localization of the energy in a trap}]\label{lem1}
For fixed $\X$ let
\begin{equation}\label{defomega} \Omega_\X=\left\{\x\in \R^3
\left| \, \min_{k\geq 2}|\x-\x_k|\geq
N^{-1/3-\delta}\right\}\right.
\end{equation} for some $0<\delta< 2/9$. Then $$ \lim_{N\to\infty}
\int d\X \int_{\Omega_\X} d\x |\phi^{\rm GP}(\x)|^2 |\nabla_\x 
f_\X(\x)|^2
= 0\ . $$
\end{lem}

\noindent {\it Remark.} In the proof 
of Theorem \ref{hombecthm} we chose $\delta$ to be 4/51, but the
following proof shows that one can extend the range of $\delta$ beyond
this value.

\begin{proof}
We shall show that
\begin{eqnarray} \nonumber &&\int
d\X \int_{\Omega_\X^c} d\x\, |\phi^{\rm GP}(\x)|^2 |\nabla_\x 
f_\X(\x)|^2\\
\nonumber &&+\int d\X \int d\x  |\phi^{\rm GP}(\x)|^2  |f_\X(\x)|^2 
\left[ 
\half \sum_{k\geq 2}
v(|\x-\x_k|) - 8\pi Na |\phi^{\rm GP}(\x)|^2\right]  \\ 
\label{lowbound}&& \geq 
-4\pi Na
\int|\phi^{\rm GP}(\x)|^4 d\x - o(1)
\end{eqnarray}
as $N\to \infty$. We claim that this implies the assertion of the
Lemma. To see this, note that the left side of (\ref{lowbound}) can be
written as
\begin{equation}
\frac 1N E_0 - \mu^{\rm GP} - \int d\X \int_{\Omega_\X} d\x 
|\phi^{\rm GP}(\x)|^2 |\nabla_\x 
f_\X(\x)|^2 \ ,
\end{equation}
where we used partial integration and the GP equation (\ref{gpeq}), 
and
also the symmetry of $\Psi_0$. The convergence of the energy in
Theorem~\ref{thmgp3} and the relation~(\ref{mugp}) now imply the 
desired result.

The proof of
(\ref{lowbound}) is actually just a detailed examination of the
lower bounds to the energy derived in \cite{LSY1999} and
\cite{LY1998} and described in Sections~\ref{sect3d} and~\ref{sectgp}.
We use the same methods as there,
just describing the differences from the case considered here.

Writing
\begin{equation}
f_\X(\x)=\prod_{k\geq 2}\phi^{\rm GP}(\x_k)F(\x,\X)
\end{equation}
and using that $F$ is symmetric in the particle coordinates, we
see that (\ref{lowbound}) is equivalent to
\begin{equation}\label{qf}
\frac 1N Q_\delta(F)\geq -4\pi Na \int|\phi^{\rm GP}|^4 - o(1),
\end{equation}
where $Q_\delta$ is the quadratic form
\begin{eqnarray}\nonumber Q_\delta(F)&=&\sum_{i=1}^{N} 
\int_{\Omega_i^c} |\nabla_i
F|^2\prod_{k=1}^{N}|\phi^{\rm GP}(\x_k)|^2d\x_k\\ \nonumber 
&&+\sum_{1\leq
i<j\leq N} \int
v(|\x_i-\x_j|)|F|^2\prod_{k=1}^{N}|\phi^{\rm GP}(\x_k)|^2d\x_k\\
\label{qf2} &&-8\pi Na\sum_{i=1}^{N} \int
|\phi^{\rm GP}(\x_i)|^2|F|^2\prod_{k=1}^{N}|\phi^{\rm 
GP}(\x_k)|^2d\x_k.
\end{eqnarray}
Here $\Omega_i^c$ denotes the set
$$\Omega_i^c=\{(\x_1,\X)\in\R^{3N}| \, \min_{k\neq
i}|\x_i-\x_k|\leq N^{-1/3-\delta}\}.$$

While  (\ref{qf}) is not true for all conceivable $F$'s satisfying
the normalization condition $$\int
|F(\x,\X)|^2\prod_{k=1}^{N}|\phi^{\rm GP}(\x_k)|^2d\x_k=1,$$ it 
{\it is}
true for an $F$, such as ours, that has bounded kinetic energy
(\ref{bound}). Looking at Section \ref{sectgp}, we see that Eqs.
\eqref{ener2}--\eqref{ener3}, \eqref{almostthere}--\eqref{there} 
are similar to (\ref{qf}),
(\ref{qf2}) and almost establish (\ref{qf}), but there are
differences which we now explain.

In our case, the kinetic energy of particle $i$ is restricted
to the subset of $\R^{3N}$ in which $\min_{k\neq i}|\x_i-\x_k|\leq
N^{-1/3-\delta}$. However, looking at the proof of the lower bound
to the ground state energy of a homogeneous Bose gas discussed in 
Section 2,
which enters the proof of Theorem \ref{thmgp3}, we
see that if we choose $\delta\leq 4/51$ only this part of the
kinetic energy is needed for the lower bound, except for
some part with a relative magnitude of the order
$\eps=O(N^{-2\alpha})$ with $\alpha=1/17$. (Here we use the a
priori knowledge that the kinetic energy is bounded by
\eqref{bound}.)
%See also the
%analogous discussion in Section \ref{sectbe}, p. \pageref{qffff}.) 
We can even do better and choose some
$4/51<\delta<2/9$, if $\alpha$ is chosen small enough. (To be
precise, we choose $\beta=1/3+\alpha$ and $\gamma=1/3-4\alpha$ in
the notation of  (\ref{ans}), and $\alpha$ small enough). The
choice of $\alpha$ only affects the magnitude of the error term,
however, which is still $o(1)$ as $N\to\infty$. 
\end{proof}

\begin{proof}[Proof of Theorem \ref{becthm}]
For some $R>0$ let $\K=\{\x\in\R^3, |\x|\leq R\}$, and define $$
\langle f_\X\rangle_\K=\frac 1{\int_\K |\phi^{\rm GP}(\x)|^2 d\x} 
\int_\K
|\phi^{\rm GP}(\x)|^2 f_\X(\x)\, d\x \  . $$ We shall use Lemma 
\ref{lem2},
with $d=3$, $h(\x)=|\phi^{\rm GP}(\x)|^2/\int_\K|\phi^{\rm 
GP}|^2$,
$\Omega=\Omega_\X\cap\K$ and $f(\x)= f_\X(\x)-\langle f_\X
\rangle_\K$ (see (\ref{defomega}) and (\ref{defff})). Since 
$\phi^{\rm GP}$
is bounded on $\K$ above and below by some positive constants,
this Lemma also holds (with a different constant $C'$) with $d\x$
replaced by $|\phi^{\rm GP}(\x)|^2d\x$ in (\ref{poinc}). Therefore,
\begin{eqnarray}\nonumber
&& \int d\X \int_\K d\x |\phi^{\rm GP}(\x)|^2 
\left[f_\X(\x)-\langle
f_\X\rangle_\K\right]^2
\\ \nonumber && \leq C'\int d\X\left[\int_{\Omega_\X\cap \K}
|\phi^{\rm GP}(\x)|^2|\nabla_{\x} f_\X(\x)|^2 d\x\right. \\ 
&&\left.
\qquad\quad\qquad + \frac {N^{-2\delta}}{R^2} \int_\K
|\phi^{\rm GP}(\x)|^2|\nabla_{\x} f_\X(\x)|^2 d\x \right], 
\label{21}
\end{eqnarray}
where we used that $|\Omega_\X^c\cap\K|\leq (4\pi/3)
N^{-3\delta}$. The first integral on the right side of (\ref{21})
tends to zero as $N\to\infty$ by Lemma \ref{lem1}, and the second
is bounded by (\ref{bound}). We conclude, since $$\int_\K
|\phi^{\rm GP}(\x)|^2 f_\X(\x) d\x\leq \int_{\R^3} |\phi^{\rm 
GP}(\x)|^2
f_\X(\x)d\x$$ because of the positivity of $f_\X$, that
\begin{eqnarray}\nonumber \liminf_{N\to\infty} \frac 1N \langle
\phi^{\rm GP}|\gamma|\phi^{\rm GP}\rangle &\geq& \int_\K 
|\phi^{\rm GP}(\x)|^2 d\x \,
\lim_{N\to\infty}\int d\X \int_\K d\x |\Psi_0(\x,\X)|^2
\\ \nonumber &=&\left[\int_\K |\phi^{\rm GP}(\x)|^2 d\x\right]^2,
\end{eqnarray}
where the last equality follows from (\ref{dconv}). Since the
radius of $\K$ was arbitrary, we conclude that
$$\lim_{N\to\infty}\frac 1 N \langle\phi^{\rm 
GP}|\gamma|\phi^{\rm GP}\rangle= 1,$$
implying convergence of $\gamma/N$ to $P^{\rm GP}$ in
Hilbert-Schmidt norm. Since the traces are equal, convergence even
holds in trace norm  (cf. \cite{S79}, Thm. 2.20), and Theorem
\ref{becthm} is proven.
\end{proof}

We remark that the method presented here also works in the case of a
two-dimensional Bose gas. The relevant parameter to be kept fixed in
the GP limit is $N/|\ln (a^2 \bar\rho_N)|$, all other considerations
carry over without essential change, using the results in
\cite{LSY2d,LY2d}, c.f.\ Sections~\ref{sect2d} and~\ref{sub2d}. It
should be noted that the existence of BEC in the ground state in 2D is
not in conflict with its absence at positive temperatures \cite{Ho,M}.
In the hard core lattice gas at half filling precisely this phenomenon
occurs \cite{KLS}.

%We also
%point out that our method necessarily fails for the one-dimensional
%Bose gas, where there is presumably no BEC \cite{PiSt}. An analogue 
of
%Lemma \ref{lem1} cannot hold in the 1D case since even a hard core
%potential with arbitrarily small range produces an interaction energy
%that is not localized on scales smaller than the mean particle 
%spacing.

\bigskip 
Finally, we remark on generalizations of Theorem~\ref{T1a} on
superfluidity from a torus to some physically more realistic settings.
As an example, let ${\mathcal C}$ be a finite cylinder based on an
annulus centered at the origin. Given a bounded, real function
$a(r,z)$ let $A$ be the vector field (in polar coordinates)
$A(r,\theta,z)=\varphi a(r,z) \widehat e_\theta$, where $\widehat
e_\theta $ is the unit vector in the $\theta$ direction. We also allow
for a bounded external potential $V(r,z)$ that does not depend
on~$\theta$.

Using the methods of Appendix~A in \cite{LSY1999}, it is not 
difficult to
see that there exists a $\varphi_0>0$, depending only on ${\mathcal 
C}$ and
$a(r,z)$, such that for all $|\varphi|<\varphi_0$ there is a unique
minimizer $\phi^{\rm GP}$ of the Gross-Pitaevskii functional
\begin{equation}\label{defgp}
\E^{\rm 
GP}[\phi]=\int_{\mathcal C}\Big(\big|\big(\nabla+{\rm i}A(\x)\big)
\phi(\x)\big|^2  + V(\x) 
|\phi(\x)|^2 + 4\pi\mu N a 
|\phi(\x)|^4\Big)d\x 
\end{equation}
under the normalization condition $\int|\phi|^2=1$. This minimizer
does not depend on $\theta$, and can be chosen to be positive, for the
following reason: The relevant term in the kinetic energy is $T=
-r^{-2}[\partial/\partial \theta + {\rm i}\varphi\, r\, a(r,z)]^2$. If
$|\varphi\, r\, a(r,z)| < 1/2$, it is easy to see that $T\geq 
\varphi^2
a(r,z)^2$, in which case, without raising the energy, we can replace
$\phi$ by the square root of the $\theta$-average of $|\phi|^2$.  This
can only lower the kinetic energy \cite{LL01} and, by convexity of
$x\to x^2$, this also lowers the $\phi^4$ term.

We denote the ground state energy of $\E^{\rm GP}$ by
$E^{\rm GP}$, depending on $Na$ and $\varphi$. 
The following Theorem \ref{T2} 
concerns the ground state energy $E_0$ of
\begin{equation}
H_N^{A}=\sum_{j=1}^N\Big[- \big(\nabla_j+{\rm i}A(\x_j)\big)^2 +
V(\x_j)\Big]
 +\sum_{1\leq i<j\leq N}v(\vert\x_{i}-\x_{j}\vert) \ ,
\end{equation}
with Neumann boundary conditions on ${\mathcal C}$, and the 
one-particle reduced
density matrix $\gamma_N$ of the ground state, respectively. Different
boundary conditions can be treated in the same manner, if they are
also used in (\ref{defgp}).

\medskip
\noindent {\it Remark.}  As a special case, consider a uniformly 
rotating
system. In this case $A(\x)=\varphi r \widehat e_\theta$, where $2
\varphi$ is the angular velocity. $H_N^A$ is the Hamiltonian in
the rotating frame, but with external potential $V(\x)+ A(\x)^2$
(see e.g. \cite[p.~131]{baym}).

\begin{thm}[{\bf Superfluidity in a cylinder}]\label{T2} 
For $|\varphi|<\varphi_0$  
\begin{equation}\label{gpone}
\lim_{N\to\infty} \frac{E_0(N,a,\varphi)}N = E^{\rm GP}(Na,\varphi) 
\end{equation} 
in the limit $N\to \infty$ with $Na$ fixed. In the same limit, 
\begin{equation}
\lim_{N\to\infty} \frac 1N\, \gamma_N(\x,\x')=
\phi^{\rm GP}(\x)\phi^{\rm GP}(\x') 
\end{equation}
in trace class norm, i.e., $\lim_{N\to\infty} \Tr 
\big[\,\big|\gamma_N/N - |\phi^{\rm GP}\rangle\langle \phi^{\rm 
GP}|\, \big|\,\big]=0$. 
\end{thm}

In the case of a uniformly rotating system, where $2\varphi$ is the
angular velocity, the condition $|\varphi|< \varphi_0$ in
particular means that the angular velocity is smaller than the
critical velocity for creating vortices \cite{rot1,fetter}. 

\medskip
\noindent{\it Remark.} 
In the special case of 
the curl-free vector potential $A(r,\theta)=\varphi r^{-1} \widehat
e_\theta$, i.e., $a(r,z)=r^{-1}$, one can say more about the role of 
$\varphi_0$.  In this case, there is a unique GP
minimizer for all $\varphi\not\in \Z+\half$, whereas there are two
minimizers for $\varphi\in \Z+\half$. Part two of Theorem \ref{T2}
holds in this special case for all $\varphi\not\in \Z+\half$, and
(\ref{gpone}) is true even for all $\varphi$.

\section{One-Dimensional Behavior of Dilute Bose Gases in Traps}

Recently it has become possible to do experiments in highly elongated
traps on ultra-cold Bose gases that are effectively one-dimensional
\cite{bongs,goerlitz,greiner,schreck,esslinger}. These experiments 
show
peculiar features predicted by a model of a one-dimensional Bose gas
with repulsive $\delta$-function pair interaction, analyzed long ago
by Lieb and Liniger \cite{LL}. These include quasi-fermionic behavior
\cite{gir}, the absence of Bose-Einstein condensation (BEC) in a 
dilute
limit \cite{Lenard,PiSt,girardeau}, and an excitation spectrum 
different
from that predicted by Bogoliubov's theory \cite{LL,jackson,komineas}.
The theoretical work on the dimensional cross-over for the ground 
state in
elongated traps has so far been based either on variational 
calculations,
starting from a 3D delta-potential \cite{olshanii,das2,girardeau2}, or
on numerical Quantum Monte Carlo studies \cite{blume,astra} with more
realistic, genuine 3D potentials, but particle numbers limited to the
order of 100.  This work is important and has led to valuable 
insights,
in particular about different parameter regions \cite{petrov,dunjko},
but a more thorough theoretical understanding is clearly desirable 
since
this is not a simple problem.  In fact, it is evident that for a 
potential
with a hard core the true 3D wave functions do not approximately 
factorize
in the longitudinal and transverse variables (otherwise the energy 
would
be infinite) and the effective 1D potential can not be obtained by 
simply
integrating out the transverse variables of the 3D potential (that 
would
immediately create an impenetrable barrier in 1D).  It is important to
be able to demonstrate rigorously, and therefore unambiguously, that 
the
1D behavior really follows from the fundamental Schr\"odinger 
equation.
It is also important to delineate, as we do here, precisely what can
be seen in the different parameter regions.  The full proofs of our
assertions are long and are given in \cite{LSY}. Here we state our 
main
results and outline the basic ideas for the proofs.

We start by describing the setting more precisely. 
It is convenient to write the
Hamiltonian in the following way (in units where $\hbar=2m =1$): 
\begin{equation} \label{3dham}
H_{N,L,r,a}=\sum_{j=1}^N \left( -\Delta_j +
V^{\perp}_{r}(\x^\perp_j)  + V_{L} (z_j) \right) + \sum_{1\leq i<j\leq
N} v_{a}(|\x_i-\x_j|) 
\end{equation}
with $\x=(x,y,z)=(\x^\perp,z)$ and with
\begin{align}
V^{\perp}_{r}(\x^\perp)&=\frac 1{r^2}
 V^{\perp}(\x^\perp/r)\ , \notag \\
V_L(z)=\frac 1{L^2} V (z/L)&\ , \quad v_{a}(|\x|)=\frac 
1{a^2}v(|\x|/a)\ .
\end{align}
Here, $r, L, a$ are variable scaling parameters while $V^{\perp}$, $V$
and $v$ are fixed.  

We shall be concerned with the ground state of this Hamiltonian for 
large particle number $N$, which is appropriate for the consideration 
of actual 
experiments. The other parameters of the problem are the
scattering length, $a$, of the two-body interaction potential, $v$,
and two lengths, $r$ and $L$, describing the transverse and the
longitudinal extension of the trap potential, respectively.

The interaction potential $v$ is supposed to be
nonnegative, of finite range and have scattering length 1; the scaled
potential $v_{a}$ then has scattering length $a$.  The external trap
potentials $V$ and $V^\perp$ confine the motion in the longitudinal
($z$) and the transversal ($\x^{\perp}$) directions, respectively, and
are assumed to be continuous and tend to $\infty$ as $|z|$ and $|
\x^{\perp}|$ tend to $\infty$.  To simplify the discussion we find it
also convenient to assume that $V$ is homogeneous of some order 
$s>0$, namely
$V(z)=|z|^s$, but weaker assumptions, e.g. asymptotic homogeneity
(cf. Section~\ref{sectgp})}, would in fact suffice.  The case of a 
simple box with
hard walls is realized by taking $s=\infty$, while the usual harmonic 
approximation is $s=2$. It is understood that the
lengths associated with the ground states of $-d^2/dz^2+V(z)$ and
$-\Delta^\perp+V^\perp(\x^\perp)$ are both of the order $1$ so that
$L$ and $r$ measure, respectively, the longitudinal and the transverse
extensions of the trap.  We denote the ground state energy of
(\ref{3dham}) by $E^{\rm QM}(N,L,r,a)$ and the ground state particle
density by $\rho^{\rm QM}_{N,L,r,a}(\x)$.
On the average, this 3D density will always be low in the parameter 
range considered here (in the sense that
distance between particles is large compared to the 3D
scattering length). The effective 1D density can be either high or
low, however. 

In parallel with the 3D Hamiltonian we consider the 
Hamiltonian for $n$  Bosons in 1D
with delta interaction
and coupling constant $g\geq 0$ , i.e., 
\beq\label{13}
H_{n,g}^{\rm 1D}=\sum_{j=1}^n-\partial^2/\partial z_{j}^2  
+ g \sum_{1\leq i<j\leq n} 
\delta(z_i-z_j)\ .
\eeq
We consider this 
Hamiltonian 
for the $z_{j}$ in
an interval of length $\ell$ in the 
thermodynamic limit, $\ell\to\infty$, $n\to\infty$ with $\rho=n/\ell$ 
fixed. 
The ground state energy per particle in this limit is independent of 
boundary conditions and can, according to \cite{LL}, be written as 
\beq \label{1dendens}
e_{0}^{\rm 1D}(\rho)=\rho^2e(g/\rho) \ ,
\eeq
with a function $e(t)$ determined by a certain 
integral equation. Its asymptotic form is $e(t)\approx 
\half t$ 
for $t\ll 1$ and $e(t)\to \pi^2/3$ for $t\to \infty$. Thus
\beq\label{e0limhigh}
e_{0}^{\rm 1D}(\rho)\approx \half g\rho\ \ \hbox{\rm for}\ \  
g/\rho\ll 
1
\eeq
and
\beq\label{e0limlow}
e_{0}^{\rm 1D}(\rho)\approx (\pi^2/3)\rho^2\ \ 
\hbox{\rm for}\ \
g/\rho\gg
1\ .
\eeq
This latter energy is the same as for non-interacting fermions in 1D, 
which can be understood from the fact that (\ref{13}) with $g=\infty$ 
is equivalent to a Hamiltonian describing free fermions. 

Taking $\rho e_{0}^{\rm 1D}(\rho)$ as a local energy density for an 
inhomogeneous 1D system we can form the energy functional 
\beq\label{genfunc}
\E[\rho]=\int_{-\infty}^{\infty} \!\!\!\! \left( 
|\nabla\sqrt\rho(z)|^2 +
V_{L}(z)\rho(z) + \rho(z)^3 e(g/\rho(z)) \right) dz \ .
\eeq
Its ground state energy is obtained by minimizing over all normalized 
densities, i.e., 
\beq\label{genfuncen}
E^{\rm 1D}(N,L,g)=\inf \left\{ \E[\rho] \, : \, \rho(z)\geq 0 , \, 
\int_{-\infty}^{\infty}
\rho(z)dz = N \right\} .
\eeq
Using convexity of the map $\rho\mapsto \rho^3 e(g/\rho)$, it is 
standard
to show that there exists a unique minimizer of (\ref{genfunc}) (see, 
e.g., \cite{LSY1999}). It will be denoted by $\rho_{N,L,g}$. We also 
define the {\it mean 1D density} of this minimizer to be
\beq
\bar\rho= \frac 1N\int_{-\infty}^{\infty}
\left(\rho_{N,L,g}(z)\right)^2 dz \ .
\eeq
In a rigid box, i.e., for $s=\infty$, $\bar \rho$ is simply $N/L$
(except for boundary corrections), but in more general traps it
depends also on $g$ besides $N$ and $L$.  The order of magnitude of
$\bar\rho$ in the various parameter regions will be described below.

Our main result relates the 3D ground state energy of (\ref{3dham}) 
to the 1D density functional energy $E^{\rm
1D}(N,L,g)$ in the large $N$ limit with $g\sim a/r^2$ provided $r/L$
and $a/r$ are sufficiently small. To state this precisely, let
$e^\perp$ and $b(\x^\perp)$, respectively, denote the ground state
energy and the normalized ground state wave function of
$-\Delta^\perp+V^\perp(\x^\perp)$. The corresponding quantities
for $-\Delta^\perp+V^\perp_{r}(\x^\perp)$ are $e^\perp/r^2$ and
$b_{r}(\x^\perp)=(1/r)b(\x^\perp/r)$. In the case that the trap is a
cylinder with hard walls $b$ is a Bessel function; for a quadratic
$V^\perp$ it is a Gaussian.

Define $g$ by
\beq\label{defg}
g=\frac {8\pi a}{r^2} \int |b(\x^\perp)|^4 d\x^\perp={8\pi a}
\int |b_{r}(\x^\perp)|^4d\x^\perp.
\eeq
Our main result of this section is:

\begin{thm}[{\bf From 3D to 1D}]\label{T1}
Let $N\to\infty$ and simultaneously $r/L\to 0$ and $a/r\to 0$ in such 
a way that 
$r^2\bar\rho\cdot\min\{\bar\rho,g\}\to 0$. Then
\beq\label{lim}
\lim    \frac {E^{\rm QM}(N,L,r,a)-Ne^\perp /r^2 }{E^{\rm 1D}(N,L,g)} 
= 1.
\eeq
\end{thm}

An analogous result hold for the ground state density. Define the 1D 
QM density by averaging
over the transverse variables, i.e., 
\beq
\hat\rho^{\rm QM}_{N,L,r,a}(z)\equiv \int \rho^{\rm QM}_{N,L,r,a}
(\x^\perp,z)d\x^\perp \ .
\eeq
Let $\bar L:= N/\bar\rho$ denote the extension of the system in 
$z$-direction, and define the rescaled density $\widetilde\rho$ by
\beq
\rho_{N,L,g}^{\rm 1D}(z)=\frac NL \widetilde \rho (z/\bar L) \ .
\eeq
Note that, although $\widetilde \rho$ depends on $N$, $L$ and $g$, 
$\|\widetilde \rho\|_1=\|\widetilde \rho\|_2=1$, which shows in 
particular that $\bar L$ is the relevant scale in $z$-direction. The 
result for the ground state density is:

\begin{thm}[{\bf 1D limit for density}]\label{T1dens}
In the same limit as considered in Theorem~\ref{T1},
\beq
\lim \left( \frac {\bar L}N \hat\rho^{\rm QM}_{N,L,r,a}(z\bar L)  - 
\widetilde \rho(z)\right) = 0
\eeq
in weak $L^1$ sense. 
\end{thm}

Note that because of (\ref{e0limhigh}) and (\ref{e0limlow}) the 
condition 
$r^2\bar\rho\cdot 
\min\{\bar\rho,g\}\to 0$ is the same as 
\beq \label{condition}
e_{0}^{\rm 1D}(\bar \rho)\ll 1/r^2 \ ,
\eeq
i.e., the average energy per particle associated with the longitudinal
motion should be much smaller than the energy gap between the ground
and first excited state of the confining Hamiltonian in the transverse
directions. Thus, the basic physics is highly quantum-mechanical and 
has no
classical counterpart. The system can be described by a 
1D functional (\ref{genfunc}), {\it even though the 
transverse trap dimension is much larger than the range of the atomic 
forces.} 

\subsection{Discussion of the results}

We will now give a discussion of the various parameter regions that
are included in the limit considered in Theorems~\ref{T1} 
and~\ref{T1dens}
above. We begin by describing the division of the space of parameters
into two basic regions.  This decomposition will eventually be refined
into five regions, but for the moment let us concentrate on the basic
dichotomy.

In Section~\ref{sectgp} we proved that the 3D
Gross-Pitaevskii formula for the energy is correct to leading order in
situations in which $N$ is large but $a$ is small compared to the mean
particle distance. This energy has two parts: The energy necessary to
confine the particles in the trap, plus the internal energy of
interaction, which is $N 4\pi a
\rho^{\rm 3D}$.  This formula was proved to be correct for a {\it
fixed} confining potential in the limit $N\to \infty$ with $a^3
\rho^{\rm 3D}\to 0$. However, this limit does not hold uniformly 
if $r/L$ gets small as $N$ gets large.  In other words, new physics
can come into play as $r/L\to 0$ and it turns out that this depends on
the ratio of $a/r^2$ to the 1D density, or, in other words, on
$g/\bar\rho$.  There are two basic regimes to consider in highly
elongated traps, i.e., when $r \ll L$. They are
\begin{itemize}
\item The 1D limit of the 
3D Gross-Pitaevskii regime
\item The `true' 1D regime.
\end{itemize}
The former is characterized by $g/\bar\rho\ll 1$, while in the latter
regime $g/\bar\rho$ is of the order one or even tends to infinity. (If
$g/\bar\rho\to\infty$ the particles are effectively impenetrable; this
is usually referred to as the Girardeau-Tonks region.) These two
situations correspond to high 1D density (weak interaction) and low 1D
density (strong interaction), respectively. Physically, the main
difference is that in the strong interaction regime the motion of the
particles in the longitudinal direction is highly correlated, while in
the weak interaction regime it is not. Mathematically, this
distinction also shows up in our proofs. The first region is correctly
described by both the 3D and 1D theories because the two give the same
predictions there. That's why we call the second region the `true' 1D
regime.
 
In both regions the internal energy of the gas is small compared to
the energy of confinement.  However,
this in itself does not imply a specifically 1D behavior. (If $a$ is
sufficiently small it is satisfied in a trap of any shape.)  1D
behavior, when it occurs, manifests itself by the fact that the
transverse motion of the atoms is uncorrelated while the longitudinal
motion is correlated (very roughly speaking) in the same way as pearls
on a necklace.  Thus, the true criterion for 1D behavior is that
$g/\bar\rho$ is of order unity or larger and not merely the condition
that the energy of confinement dominates the internal energy.

We shall now briefly describe the finer division of these two regimes
into five regions altogether. Three of them (Regions 1--3) belong to
the weak interaction regime and two (Regions 4--5) to the strong
interaction regime. They are characterized by the behavior of
$g/\bar\rho$ as $N\to \infty$. In each of these regions the general
functional (\ref{genfunc}) can be replaced by a different, simpler
functional, and the energy $E^{\rm 1D}(N,L,g)$ in Theorem~\ref{T1} by the
ground state energy of that functional. Analogously, the density in
Theorem~\ref{T1dens} can be replaced by the minimizer of the 
functional
corresponding to the region considered.

The five regions are
\medskip

\noindent $\bullet$
{\bf Region 1, the Ideal Gas case:} In the trivial case where the
interaction is so weak that it effectively vanishes in the large $N$
limit and everything collapses to the ground state of $-d^2/dz^2+V(z)$
with ground state energy $e^{\parallel}$, the energy $E^{\rm 1D}$ in
(\ref{lim}) can be replaced by $N e^{\parallel} /L^2 $. This is the
case if $g/\bar\rho\ll N^{-2}$, and the mean density is just
$\bar\rho\sim N/L$. Note that $g/\bar\rho\ll N^{-2}$ means that the 3D
interaction energy per particle $\sim a
\rho^{\rm 3D}\ll 1/L^2$.
\medskip

\noindent $\bullet$
{\bf Region 2, the 1D GP case:} In this region $g/\bar\rho\sim 
N^{-2}$,
with $\bar\rho\sim N/L$. This case is described by a 1D
Gross-Pitaevskii energy functional of the form
\beq\label{GPfunct}
\E^{\rm GP}[\rho]=\int_{-\infty}^\infty \left( 
|\nabla\sqrt\rho(z)|^2+ 
V_L(z)\rho(z) + \half g\rho(z)^2 \right) dz \ ,
\eeq
corresponding to the high density approximation (\ref{e0limhigh}) of 
the 
interaction energy in (\ref{genfunc}). Its ground state energy, 
$E^{\rm GP}$, fulfills the scaling relation $E^{\rm GP}(N,L,g) = 
NL^{-2} E^{\rm GP}(1,1,NgL)$. 
\medskip

\noindent $\bullet$
{\bf Region 3, the 1D TF case:}  $N^{-2}\ll g/\bar\rho \ll 1$, 
with $\bar\rho$ being of the order $\bar\rho\sim (N/L) 
(NgL)^{-1/(s+1)}$, where $s$ is the degree of
homogeneity of the longitudinal confining potential $V$. This region
is described by a Thomas-Fermi type functional 
\beq\label{TFfunct}
\E^{\rm TF}[\rho]=\int_{-\infty}^\infty \left( 
V_L(z)\rho(z) + \half g\rho(z)^2 \right) dz \ .
\eeq
It is a limiting case of Region 2 in the sense that $NgL\gg 1$, but
$a/r$ is sufficiently small so that $g/\bar\rho \ll 1$, i.e., the high
density approximation in (\ref{e0limhigh}) is still valid.  The
explanation of the factor $(NgL)^{1/(s+1)}$ is as follows: The linear
extension $\bar L$ of the minimizing density of (\ref{GPfunct}) is for
large values of $NgL$ determined by $V_{L}(\bar L)\sim g(N/\bar L)$,
which gives $\bar L\sim (NgL)^{1/(s+1)} L$.  In addition condition
(\ref{condition}) requires $g\bar\rho \ll r^{-2}$, which means that
$Na/L(NgL)^{1/(s+1)}\ll 1$. The minimum energy of (\ref{TFfunct}) has
the scaling property $E^{\rm TF}(N,L,g) = NL^{-2}(NgL)^{s/(s+1)}
E^{\rm TF}(1,1,1)$.
\medskip

\noindent $\bullet$
{\bf Region 4, the LL case:}  $g/\bar\rho\sim 1$ , with
$\bar\rho\sim  (N/L )
N^{-2/(s+2)}$, described by an energy functional 
\beq\label{llfunct}
\E^{\rm LL}[\rho]=\int_{-\infty}^\infty \left( V_{L}(z)\rho(z) 
+ \rho(z)^3 
e(g/\rho(z)) \right) dz \ .
\eeq
This region corresponds to the case $g/\bar\rho\sim 1$, so that
neither the high density (\ref{e0limhigh}) nor the low density
approximation (\ref{e0limlow}) is valid and the full LL energy
(\ref{1dendens}) has to be used. The extension $\bar L$ of the system
is now determined by $V_L(\bar L)\sim (N/\bar L)^2$ which leads to
$\bar L\sim L N^{2/(s+2)}$. Condition (\ref{condition}) means in
this region that $Nr/\bar L\sim N^{s/(s+2)}r/L\to 0$. Since
$Nr/\bar L\sim(\bar\rho/g)(a/r)$, this condition is
automatically fulfilled if $g/\bar\rho$ is bounded away from zero and
$a/r\to 0$.  The ground state energy of (\ref{llfunct}), $E^{\rm
LL}(N,L,g)$, is equal to $N\gamma^2 E^{\rm LL}(1,1,g/\gamma)$, where
we introduced the density parameter $\gamma := (N/L) N^{-2/(s+2)}$.

\medskip
\noindent $\bullet$
{\bf Region 5, the GT case:} $g/\bar\rho\gg 1$, 
with $\bar\rho\sim  (N/L) N^{-2/(s+2)}$, described by a functional 
with energy density $\sim \rho^3$, corresponding to the
Girardeau-Tonks limit of the  LL energy density. 
It corresponds to impenetrable particles, i.e, the limiting 
case $g/\bar\rho\to\infty$ and hence formula 
(\ref{e0limlow}) for the energy density. As in Region 4, the mean 
density is here
 $\bar\rho\sim  \gamma$.
The energy functional is
\beq
\E^{\rm GT}[\rho]=\int_{-\infty}^\infty \left( V_{L}(z)\rho(z) + 
({\pi^2}/3 )\rho(z)^3  \right) dz \ ,
\eeq
with minimum energy  $E^{\rm GT}(N,L)= N \gamma^2 E^{\rm GT}(1,1)$.

\medskip
As already mentioned above, Regions 1--3 can be reached as limiting
cases of a 3D Gross-Pitaevskii theory.  In this sense, the behavior in
these regions contains remnants of the 3D theory, which also shows up
in the fact that BEC prevails in Regions 1 and 2 (See \cite{LSY} 
for details.)
Heuristically, these traces
of 3D can be understood from the fact that in Regions 1--3 the 1D
formula for energy per particle, $g\rho\sim aN/(r^2 L)$, gives the 
same
result as the 3D formula, i.e., scattering length times
3D density.  This is no longer so in Regions 4 and 5 and different
methods are required. 

\subsection{Outline of Proof}

We now outline the main steps in the proof of Theorems~\ref{T1} 
and~\ref{T1dens}, referring to \cite{LSY} for full details.
To prove (\ref{lim}) one has to establish upper and lower bounds, with
controlled errors, on the QM many-body energy in terms of the energies
obtained by minimizing the energy functionals appropriate for the
various regions.  The limit theorem for the densities can be derived
from the energy estimates in a standard way by variation with respect
to the external potential $V_{L}$.

The different parameter regions have to be treated by different 
methods, a watershed lying between Regions 1--3 on the one hand 
and Regions 4--5 on the other. In Regions 1--3, 
similar methods as in the proof of the 3D Gross-Pitaevskii limit 
theorem discussed in Section~\ref{sectgp}  can be used. 
This 3D proof needs some  
modifications, however,  because there the external 
potential was fixed and the estimates are not uniform in the ratio 
$r/L$. We will not go into the details here, but mainly focus on 
Regions 4 and 5, where new methods are needed. It turns out to be
necessary to localize the particles by dividing the trap into finite
`boxes' (finite in $z$-direction), with a controllable particle 
number in
each box. The particles
are then distributed optimally among the boxes to minimize the 
energy, in a similar way as Eq. (\ref{estimate1}) was derived from
Eq. (\ref{estimate4}).

A core lemma for Regions 4--5 is an estimate of the 3D ground state 
energy in a finite box in 
terms of the 1D energy of the Hamiltonian (\ref {13}). 
I.e., we will consider the ground state energy of (\ref{3dham})
with the external potential $V_L(z)$ replaced by a finite box (in 
$z$-direction) with length $\ell$. Let $E_{\rm D}^{\rm
QM}(n,\ell,r,a)$ and $E_{\rm N}^{\rm QM}(n,\ell,r,a)$ denote its 
ground state energy with Dirichlet and Neumann
boundary conditions, respectively. 

\begin{lem} \label{finthm}
Let $E_{\rm D}^{\rm 1D}(n,\ell,g)$ and $E_{\rm N}^{\rm
1D}(n,\ell,g)$ denote the ground state energy of (\ref{13}) on
$L^2([0,\ell]^n)$, with Dirichlet and Neumann boundary conditions,
respectively, and let $g$ be given by (\ref{defg}). Then there is a
finite number $C>0$ such that
\beq\label{lbthm}
E_{\rm N}^{\rm QM}(n,\ell,r,a)-\frac{ne^\perp}{r^2} \geq E_{\rm 
N}^{\rm
1D}(n,\ell,g)
\left( 1 -C n 
\left(\frac{a}{r}\right)^{1/8}\left[1+\frac {nr}{\ell}
\left(\frac{a}{r}\right)^{1/8}   \right]\right)  \ .
\eeq
Moreover, 
\beq\label{ubthm}
E_{\rm D}^{\rm QM}(n,\ell,r,a)-\frac {ne^\perp}{r^2} \leq E_{\rm 
D}^{\rm
1D}(n,\ell,g)
\left(1+ C \left[ \left(\frac{n
a}{r}\right)^{2}\left( 1+ \frac {a\ell}{r^2}\right) 
\right]^{1/3}\right) \ ,
\eeq
provided the term in square brackets is less than $1$. 
\end{lem}

This Lemma is the key to the proof of Theorems~\ref{T1} 
and~\ref{T1dens}. The reader interested in the details is referred to 
\cite{LSY}. Here we only give a sketch of the proof of 
Lemma~\ref{finthm}. 

\begin{proof}[Proof of Lemma~\ref{finthm}]

We start with the upper bound (\ref{ubthm}).  
Let $\psi$ denote the ground state of (\ref{13}) with Dirichlet 
boundary conditions, normalized by $\langle\psi|\psi\rangle=1$, and 
let $\rho^{(2)}_\psi$ denote its two-particle density, normalized by 
$\int \rho_\psi^{(2)}(z,z')dzdz'=1$. Let $G$ and $F$ be given by 
$G(\x_1,\dots,\x_n)=\psi(z_1,\dots,z_n)\prod_{j=1}^n 
b_r(\x^\perp_j)$ and $F(\x_1,\dots,\x_n)=\prod_{i<j} f(|\x_i-\x_j|)$.
Here $f$ is a monotone increasing function, with $0\leq f\leq 1$
and $f(t)=1$ for $t\geq R$ for some $R\geq R_0$. For $t\leq R$ we
shall choose $f(t)=f_0(t)/f_0(R)$, where $f_0$ is the solution to
the zero-energy scattering equation for $v_a$ (\ref{3dscatteq}). 
Note that
$f_0(R)=1-a/R$ for $R\geq R_0$, and $f'_0(t)\leq
t^{-1}\min\{1,a/t\}$. We use as a trial wave function 
\beq
\Psi(\x_1,\dots,\x_n)=G(\x_1,\dots,\x_n)F(\x_1,\dots,\x_n) \ .
\eeq

We first have to estimate the norm of $\Psi$. Using the fact that $F$ 
is $1$
whenever no pair of particles is closer together than a distance
$R$, we obtain
\begin{equation}
\langle\Psi|\Psi\rangle \geq 1- \frac{n(n-1)}2
\frac {\pi R^2}{r^2} \|b\|_4^4 \ .
\end{equation}
To evaluate the expectation value of the Hamiltonian, we use
\beq
\langle\Psi|-\Delta_j|\Psi\rangle=-\int F^2 G\Delta_j G + \int G^2
|\nabla_j F|^2
\eeq
and the Schr\"odinger equation $H_{n,g}\psi=E_{\rm D}^{\rm 1D}\psi$. 
This gives
\begin{eqnarray}\nonumber
\langle\Psi|H|\Psi\rangle &=&\left(E_{\rm D}^{\rm 1D}+
\frac{n}{r^2}e^\perp\right)\langle\Psi|\Psi\rangle-g\langle\Psi|
\sum_{i<j}\delta(z_i-z_j)|\Psi\rangle
\\ \label{exp} && + \int G^2 \left( \sum_{j=1}^n |\nabla_j F|^2 +
\sum_{i<j} v_a(|\x_i-\x_j|)|F|^2\right) \ .
\end{eqnarray}
Now, since $0\leq f\leq 1$ and $f'\geq 0$ by assumption, $F^2\leq
f(|\x_i-\x_j|)^2$, and
\beq\label{fpp}
\sum_{j=1}^n |\nabla_j F|^2 \leq 2 \sum_{i<j} f'(|\x_i-\x_j|)^2 +
4 \sum_{k<i<j} f'(|\x_k-\x_i|)f'(|\x_k-\x_j|)  \ .
\eeq
Consider the first term on the right side of (\ref{fpp}), together
with the last term in (\ref{exp}). These terms are bounded above by
\begin{equation}
n(n-1) \int
b_r(\x^\perp)^2b_r(\y^\perp)^2\rho^{(2)}_\psi(z,z')\left(
f'(|\x-\y|)^2+\half v_a(|\x-\y|) f(|\x-\y|)^2\right) .
\label{cha}
\end{equation}
Let
\beq
h(z)=\int \left( f'(|\x|)^2+\half v_a(|\x|)
f(|\x|)^2\right)d\x^\perp \ .
\eeq
Using Young's
inequality for the integration over the $\perp$-variables, we get
\beq\label{31}
(\ref{cha})\leq \frac{n(n-1)}{r^2} \|b\|_4^4 \int_{\R^2}
\rho^{(2)}_\psi(z,z') h(z-z') dz dz'\ .
\eeq
By similar methods, one can show that the contribution from the last 
term in (\ref{fpp}) is bounded by 
\beq\label{ahaa}
 \frac 23
n(n-1)(n-2)\frac{\|b\|_\infty^2}{r^2}\frac{\|b\|_4^4}{r^2}\|k\|_\infty
\int_{\R^2} \rho^{(2)}_\psi(z,z') k(z-z') dz
dz' \ ,
\eeq
where
\beq
k(z)=\int f'(|\x|)d\x^\perp \ .
\eeq
Note that both $h$ and $k$ are  supported in $[-R,R]$. 

Now, for any $\phi\in H^1(\R)$,
\begin{equation}
\left| |\phi(z)|^2-|\phi(z')|^2\right| \leq  2 |z-z'|^{1/2} 
\left(\int_\R
|\phi|^2\right)^{1/4}\left(\int_\R \left|\frac
{d\phi}{dz}\right|^2\right)^{3/4} \ .
\end{equation}
Applying this to $\rho_\psi^{(2)}(z,z')$, considered as a function of 
$z$ only, we  get
\begin{eqnarray}\nonumber
&&\int_{\R^2} \rho^{(2)}_\psi(z,z') h(z-z') dz
dz'- \int_\R h(z) dz \int \rho_\psi^{(2)}(z,z)dz \\ 
&& \leq 2
R^{1/2} \int_\R h(z) dz 
\left\langle\psi\left|-\frac{d^2}{dz_1^2}\right|\psi\right
\rangle^{3/4} \ , \label{cha2}
\end{eqnarray}
where we used Schwarz's inequality, the normalization of
$\rho^{(2)}_\psi$ and the symmetry of $\psi$. The same argument is
used for (\ref{ahaa}) with $h$ replaced by $k$.

It remains to bound the second term in (\ref{exp}). As in the 
estimate for the norm of $\Psi$, we use again the fact
that $F$ is equal to 1 as long as the particles are not within a
distance $R$. We obtain
\begin{equation}
\langle\Psi|\sum_{i<j}\delta(z_i-z_j)|\Psi\rangle 
\geq \frac{n(n-1)}{2} \int \rho_\psi^{(2)}(z,z) dz 
\left(1-\frac{\pi
R^2}{r^2} \|b\|_4^4\right) \label{cha3} \ .
\end{equation}
We also estimate  $g\half n(n-1)\int
\rho^{(2)}_\psi(z,z)dz\leq E_{\rm D}^{\rm 1D}$ and
$\langle\psi|-d^2/dz_1^2|\psi\rangle\leq E_{\rm D}^{\rm 1D}/n$. We 
have 
$\int h(z)dz=
4\pi a (1 - a/R)^{-1}$, and the terms containing $k$ can be bounded 
by $\|k\|_\infty \leq 2\pi a (1+\ln(R/a))/(1-a/r)$ and $\int k(z) dz 
\leq  2\pi aR (1-a/(2R))/(1-a/r)$. Putting together all the bounds, 
and 
choosing
\beq\label{Rin3.1}
R^3 = \frac {a r^2} {n^2(1+ g \ell )} \ ,
\eeq
this proves the desired result.

We are left with the lower bound (\ref{lbthm}). 
We write a general wave function $\Psi$ as
\beq
\Psi(\x_1,\dots,\x_n)=f(\x_1,\dots,\x_n)\prod_{k=1}^n
b_r(\x^\perp_k) \ ,
\eeq
which can always be done, since $b_r$ is a strictly positive 
function. 
Partial integration gives
\beq\label{329}
\langle\Psi|H|\Psi\rangle=\frac{n e^\perp}{r^2}+
\sum_{i=1}^n \int \left[ |\nabla_i f|^2 +\half \sum_{j,\, j\neq i}
v_a(|\x_i-\x_j|)|f|^2 \right] \prod_{k=1}^n b_r(\x^\perp_k)^2 d
\x_k. 
\eeq
Choose some $R>R_0$, fix $i$ and $\x_j$, $j\neq i$, and consider
the Voronoi cell $\Omega_j$ around particle $j$,  
i.e., $\Omega_j=\{\x:\, |\x-\x_{j}|\leq  |\x-\x_{k}| \hbox{ for all } 
k\neq j\}$. If ${\mathcal B}_j$ denotes the ball of radius $R$ around 
$\x_j$, we can estimate with the aid of Lemma~\ref{dysonl}
\begin{eqnarray}\nonumber
&&\int_{\Omega_j\cap {\mathcal B}_j} b_r(\x^\perp_i)^2\left( 
|\nabla_i f|^2+
\half v_a(|\x_i-\x_j|)|f|^2\right) d\x_i \\ && \geq \frac{\min_{\x\in
{\mathcal B}_j}b_r(\x^\perp)^2}{\max_{\x\in {\mathcal 
B}_j}b_r(\x^\perp)^2} a
\int_{\Omega_j\cap {\mathcal B}_j} b_r(\x^\perp_i)^2 
U(|\x_i-\x_j|)|f|^2 \ .
\end{eqnarray}
Here $U$ is given in (\ref{softened}). 
For some $\delta>0$ let ${\mathcal B}_\delta$ be the subset of $\R^2$ 
where 
$b(\x^\perp)^2\geq \delta$, and let $\chi_{{\mathcal B}_\delta}$ 
denote its
characteristic function. 
Estimating $\max_{\x\in {\mathcal B}_j}b_r(\x^\perp)^2\leq
\min_{\x\in {\mathcal B}_j}b_r(\x^\perp)^2 + 2(R/r^3) \|\nabla
b^2\|_\infty$, we obtain
\beq\label{334}
\frac{\min_{\x\in {\mathcal B}_j}b_r(\x^\perp)^2}{\max_{\x\in
{\mathcal B}_j}b_r(\x^\perp)^2}\geq
\chi_{{\mathcal B}_\delta}(\x^\perp_j/r)\left(1-2\frac R{r} 
\frac{\|\nabla
b^2\|_\infty}{\delta}\right) \ .
\eeq
Denoting $k(i)$ the nearest
neighbor to particle $i$, we conclude that, for $0\leq \eps\leq 1$,
\begin{eqnarray}\nonumber
&&\sum_{i=1}^n \int \left[ |\nabla_i f|^2 +\half \sum_{j,\, j\neq
i} v_a(|\x_i-\x_j|)|f|^2 \right] \prod_{k=1}^n b_r(\x^\perp_k)^2
d \x_k \\ \nonumber && \geq \sum_{i=1}^n \int \Big[ \eps
|\nabla_i f|^2 +(1-\eps)|\nabla_i f|^2 \chi_{\min_k |z_i-z_k|\geq
R}(z_i)\\ \label{negl} &&  \qquad + a'
U(|\x_i-\x_{k(i)}|)\chi_{{\mathcal B}_\delta}(\x^\perp_{k(i)}/r)|f|^2 
\Big]
\prod_{k=1}^n b_r(\x^\perp_k)^2 d \x_k \ ,
\end{eqnarray}
where $a'=a(1-\eps)(1-2 R \|\nabla b^2\|_\infty/ r \delta)$.

Define $F(z_1,\dots,z_n)\geq 0$ by
\beq\label{defF}
|F(z_1,\dots,z_n)|^2=\int |f(\x_1,\dots,\x_n)|^2\prod_{k=1}^n
b_r(\x_k^\perp)^2 d\x_k^\perp \ .
\eeq
Neglecting the kinetic energy in $\perp$-direction in the second term
in (\ref{negl}) and using the Schwarz inequality to bound the
longitudinal kinetic energy of $f$ by the one of $F$, we get the
estimate
\begin{eqnarray}\nonumber
&&\langle\Psi|H|\Psi\rangle- \frac{n e^\perp}{r^2}\geq \\ \nonumber
&& \sum_{i=1}^n \int\Big[\eps |\partial_i F|^2 +
(1-\eps)|\partial_i F|^2 \chi_{\min_k |z_i-z_k|\geq R}(z_i)
\Big]\prod_{k=1}^n dz_k \\ \nonumber &&+\sum_{i=1}^n \int \left[
\eps|\nabla^\perp_i f|^2 + a'
U(|\x_i-\x_{k(i)}|)\chi_{{\mathcal B}_\delta}(\x^\perp_{k(i)}/r)|f|^2
\right] \prod_{k=1}^n b_r(\x^\perp_k)^2 d \x_k , \\ \label{57}
\end{eqnarray}
where $\partial_j=d/dz_j$, and $\nabla^\perp$ denotes the gradient
in $\perp$-direction. We now investigate the last term in
(\ref{57}). Consider, for fixed $z_1,\dots,z_n$, the expression
\beq\label{44}
\sum_{i=1}^n \int \left[ \eps|\nabla^\perp_i f|^2 + a'
U(|\x_i-\x_{k(i)}|)\chi_{{\mathcal B}_\delta}(\x^\perp_{k(i)}/r)|f|^2
\right] \prod_{k=1}^n b_r(\x^\perp_k)^2 d \x_k^\perp \ .
\eeq
To estimate this term from below, we use Temple's inequality, as in
Subsect.~\ref{subsect22}. Let $\widetilde e^\perp$ denote the gap 
above zero
in the spectrum of $-\Delta^\perp+V^\perp-e^\perp$, i.e., the lowest
non-zero eigenvalue. By scaling, $\widetilde e^\perp/r^2$ is the gap
in the spectrum of $-\Delta^\perp+V^\perp_r-e^\perp/r^2$. Note that
under the transformation $\phi \mapsto b_r^{-1} \phi$ this latter
operator is unitarily equivalent to $\nabla^{\perp *}\cdot
\nabla^\perp$ as an operator on 
$L^2(\R^2,b_r(\x^\perp)^2 d\x^\perp)$, as considered in
(\ref{44}). Hence also this operator has $\widetilde e^\perp/r^2$ as
its energy gap. Denoting 
\beq
\langle U^k\rangle=  \int \left(\sum_{i=1}^n
U(|\x_i-\x_{k(i)}|)\chi_{{\mathcal 
B}_\delta}(\x^\perp_{k(i)}/r)\right)^k
\prod_{k=1}^n b_r(\x^\perp_k)^2 d \x_k^\perp \ ,
\eeq
Temple's inequality implies 
\beq\label{60}
(\ref{44})\geq |F|^2 a'\langle U \rangle \left(1- a'\frac{\langle
U^2\rangle}{\langle U\rangle}\frac{1}{\eps \widetilde
e^\perp/r^2-a'\langle U\rangle}\right) \ .
\eeq
Now, using (\ref{softened}) and Schwarz's inequality, $\langle
U^2\rangle\leq 3n(R^3-R_0^3)^{-1}\langle U\rangle$, and
\begin{equation}
\langle U\rangle \leq
n(n-1)\frac{\|b\|_4^4}{r^2}\frac{3\pi R^2}{R^3-R_0^3} \ .
\end{equation}
Therefore 
\beq
(\ref{60})\geq |F|^2 a''\langle U \rangle \ ,
\eeq
where we put all the error terms into the modified coupling constant 
$a''$. It remains to derive a lower bound on $\langle U\rangle$. Let
\beq
d(z-z')=\int_{\R^4} b_r(\x^\perp)^2 b_r(\y^\perp)^2 U(|\x-\y|)
\chi_{{\mathcal B}_\delta}(\y^\perp/r) d\x^\perp d\y^\perp \ .
\eeq
Note that $d(z)=0$ if $|z|\geq R$. An estimate similar to 
(\ref{firstorder}) gives
\begin{equation}
\langle U\rangle \geq
\sum_{i\neq j}d(z_i-z_j) \left(1-(n-2)
\frac{\pi R^2}{r^2} \|b\|_\infty^2\right) \ .
\end{equation}
Note that, for an appropriate choice of $R$, $d$ is close to a 
$\delta$-function with the desired coefficient. To make the 
connection with the $\delta$-function, we can use a bit of the 
kinetic energy saved in (\ref{57}) to obtain
\begin{eqnarray}\nonumber 
&&\int \left[ \frac\eps{n-1} |\partial_i F|^2 + a'''  d(z_i-z_j)
|F|^2\right] dz_i \\ && \geq \half g' \max_{|z_i-z_j|\leq R}
|F|^2 \chi_{[R,\ell-R]}(z_j) 
\left(1-\left(\frac{2(n-1)}{\eps}g'R\right)^{1/2}\right) \ .
\end{eqnarray}

Putting all the previous estimates together, we arrive at 
\begin{eqnarray}\nonumber
\langle\Psi|H|\Psi\rangle- \frac{ne^\perp}{r^2} &\geq& \sum_{i=1}^n 
\int\Big[(1-\eps)|\partial_i F|^2 \chi_{\min_k
|z_i-z_k|\geq R}(z_i) \Big] \prod_{k=1}^n dz_k \\ \nonumber 
\label{putt} & & +\sum_{i\neq j} \half g'' \int
\max_{|z_i-z_j|\leq R}|F|^2 \chi_{[R,\ell-R]}(z_j) \prod_{k,\, k\neq 
i}dz_k \\ 
\end{eqnarray}
for an appropriate coupling constant $g''$ that contains all the 
error terms. 
Now assume that $(n+1)R<\ell$. Given an $F$ with $\int |F|^2 
dz_1\cdots 
dz_n=1$, define, for
$0\leq z_1\leq z_2\leq \dots \leq z_n\leq \ell-(n+1)R$,
\beq
\psi(z_1,\dots,z_n)=F(z_1+R,z_2+2 R,z_3+3R,\dots,z_n+n R) \ ,
\eeq
and extend the function to all of $[0,\ell-(n+1)R]^n$ by symmetry. A
simple calculation shows that
\begin{eqnarray}\nonumber
(\ref{putt})\geq \langle\psi |H' |\psi\rangle &\geq& (1-\eps)
E_{\rm N}^{\rm 1D}(n,\ell-(n+1)R,g'') \langle\psi|\psi\rangle  \\ 
\label{puutt}
&\geq&
(1-\eps) E_{\rm N}^{\rm 1D}(n,\ell,g'') \langle\psi|\psi\rangle\ ,
\end{eqnarray}
where $H'$ is the Hamiltonian (\ref{13}) with a factor $(1-\eps)$ in 
front of the kinetic energy term. 

It remains to estimate $\langle\psi|\psi\rangle$. Using that $F$ is 
related to the true ground state $\Psi$ by (\ref{defF}), we can 
estimate it in terms of the total QM energy, namely 
\begin{eqnarray}\nonumber
\langle\psi|\psi\rangle &\geq& 1- \frac {2R}{g''} \left(E^{\rm
  QM}_{\rm N}(N,\ell,r,a)-\frac {ne^\perp}{r^2}\right) \\ && -2 n 
\frac 
R\ell - 4 n R \left(\frac 1n E^{\rm QM}_{\rm N}(n,\ell,r,a)-\frac 
{e^\perp}{r^2} \right)^{1/2} \ . \label{ddss}
\end{eqnarray} 
Collecting all the error terms and choosing  
\beq\label{chopara}
R=r\left(\frac{a}{r}\right)^{1/4}\ , \quad
\eps=\left(\frac{a}{r}\right)^{1/8}\ , \quad
\delta=\left(\frac{a}{r}\right)^{1/8} \ ,
\eeq
(\ref{puutt}) and (\ref{ddss}) lead to the desired lower bound.
\end{proof}

As already noted above, Lemma~\ref{finthm} is the key to the proof of 
Theorems~\ref{T1} and~\ref{T1dens}. The estimates are used in each 
box, and the particles are distributed optimally among the boxes. For 
the global lower bound, superadditivity of the energy and 
convexity of the energy density $\rho^3 e(g/\rho)$ are used, 
generalizing corresponding arguments in Section~\ref{sect3d}. We 
refer to \cite{LSY} for details.

\section{The Charged Bose Gas, the One- and Two-Component Cases}

The setting now changes abruptly. Instead of particles interacting
with a short-range potential $v(|\x_i-\x_j|)$ they interact via the
Coulomb potential $$v(|\x_i-\x_j|) = |\x_i-\x_j|^{-1} $$
(in 3 dimensions). The unit of electric charge is 1 in our units.

We will here consider both the one- and two-component gases.  In the
one-component gas (also referred to as the one-component plasma or
bosonic jellium) we consider positively charged particles confined to
a box with a uniformly charged background. In the two-component gas we
have particles of both positive and negative charges moving in all of
space.

\subsection{The One-Component Gas}
In the one-component gas there are $N$ positively charged particles in
a large box $\Lambda$ of volume $L^3$ as before, with $\rho =N/L^{3}$.

To offset the huge Coulomb repulsion (which would drive the particles
to the walls of the box) we add a uniform negative background of
precisely the same charge, namely density $\rho$. Our Hamiltonian is
thus
\begin{equation}\label{foldyham}
H_N^{(1)}=  \sum_{i=1}^{N} - \mu \Delta_i -V(\x_i) +
 \sum_{1 \leq i < j \leq N} v(|\x_i - \x_j|)  +C
\end{equation}
with $$
V(\x)=\rho \int_{\Lambda} |\x-\y |^{-1}d\y \qquad \qquad {\rm
and}\qquad \qquad C= \frac{1}{2} \rho \int_{\Lambda} V(\x)d\x\ .
$$
We shall use Dirichlet boundary conditions. As before the
Hamiltonian acts on  symmetric wave functions in $L^2(\Lambda^{N},
d\x_1\cdots d\x_N)$.

Each
particle interacts only with others and not with itself. Thus, 
despite 
the fact that the Coulomb potential is positive definite, the
ground state energy $E_0$ can
be (and is) negative (just take $\Psi=$const.). This time, {\it
large} $\rho$ is the `weakly interacting' regime.

We know from the work in \cite{LN} that the thermodynamic limit
$e_0(\rho)$ defined as in (\ref{eq:thmlimit}) exists. It also follows
from this work that we would, in fact, get the same thermodynamic
energy if we did not restrict the number of particles $N$, but
considered the grand-canonical case where we minimize the energy over
all possible particle numbers, but keeping the background charge
$\rho$ fixed.

Another way in which this problem is different from the previous one
is that {\it perturbation theory is correct to leading order}. If
one computes $(\Psi, H \Psi)$ with  $\Psi=$const, one gets the right
first order answer, namely $0$. It is the next order in $1/\rho$ that
is interesting, and this is {\it entirely} due to correlations.
In 1961 Foldy \cite{FO} calculated this correlation energy according
to the prescription of Bogolubov's 1947 theory. That theory was not
exact for the dilute Bose gas, as we have seen, even to first order. 
We are now looking at {\it second} order, which should be even
worse. Nevertheless, there was good physical intuition that this
calculation should be asymptotically {\it exact}. Indeed it is, as
proved in \cite{LS} and \cite{So}.

The Bogolubov theory states that the main contribution to the
energy comes from pairing of particles into momenta $\k, -\k$ and
is the bosonic analogue of the BCS theory of superconductivity
which came a decade later. I.e., $\Psi_0$ is a sum of products of
terms of the form $\exp\{i\k \cdot (\x_i-\x_j)\}$. 

The following theorem is the main result for the one-component gas.

\begin{thm}[\textbf{Foldy's law for the one-component gas}] 
\label{thm:Foldy}
\begin{equation}\label{foldyen}
\lim_{\rho\to\infty}\rho^{-1/4}e_0(\rho)=
-\frac{2}{5}\frac{\Gamma(3/4)}{\Gamma(5/4)}\left(\frac{2}{\mu\pi}\right)^{1/4}.
\end{equation}
\end{thm}

This is the {\it first example} (in more than 1 dimension) in which
Bogolubov's pairing theory has been rigorously validated. It has to be
emphasized, however, that Foldy and Bogolubov rely on the existence of
Bose-Einstein condensation. We neither make such a hypothesis nor does
our result for the energy imply the existence of such condensation. As
we said earlier, it is sufficient to prove condensation in small boxes
of fixed size.

Incidentally, the one-dimensional example for which Bogolubov's
theory is asymptotically exact to the first two orders  (high
density) is the repulsive delta-function Bose gas \cite{LL}, for 
which there 
is no Bose-Einstein condensation.

To appreciate the  $-\rho^{1/4}$ nature of (\ref{foldyen}), it is 
useful to
compare it with what one would get if the bosons had infinite mass,
i.e., the first term in (\ref{foldyham}) is dropped. Then the energy 
would
be proportional to $-\rho^{1/3}$ as shown in \cite{LN}. Thus, the 
effect
of quantum mechanics is to lower $\frac{1}{3}$ to $\frac{1}{4}$.

A problem somewhat related to bosonic jellium is \textit{fermionic}
jellium.  Graf and Solovej \cite{GS} have proved that the first two
terms are what one would expect, namely
\begin{equation}
        e_{0}(\rho)=C_{\rm TF}\rho^{5/3}-C_{\rm 
D}\rho^{4/3}+o(\rho^{4/3}),
\end{equation}
where $C_{\rm TF}$ is the usual Thomas-Fermi constant and $C_{\rm D}$ 
is the
usual Dirac exchange constant.

It is supposedly true, for both bosonic and fermionic particles, that
there is a critical mass above which the ground state should show
crystalline ordering (Wigner crystal), but this has never been proved
and it remains an intriguing open problem, even for the infinite mass
case. A simple scaling shows that large mass is the same as small
$\rho$, and is thus outside the region where a Bogolubov approximation
can be expected to hold.

As for the dilute Bose gas, there are several relevant length scales 
in
the problem of the charged Bose gas. For the dilute gas there were
three scales. This time there are just two. Because of the long range
nature of the Coulomb problem there is no scale corresponding to the
scattering length $a$. One relevant length scale is again the
interparticle distance $\rho^{-1/3}$. The other is the correlation
length scale $\ell_{\rm cor}\sim \rho^{-1/4}$ (ignoring the dependence
on $\mu$).  The order of the correlation length scale can be
understood heuristically as follows. Localizing on a scale $\ell_{\rm
  cor}$ requires kinetic energy of the order of $\ell_{\rm cor}^{-2}$.
The Coulomb potential from the particles and background on the scale
$\ell_{\rm cor}$ is $(\rho\ell_{\rm cor}^3)/\ell_{\rm cor}$. Thus the
kinetic energy and the Coulomb energy balance when $\ell_{\rm
  cor}\sim\rho^{-1/4}$. This heuristics is however much too simplified
and hides the true complexity of the situation.

Note that in the high density limit $\ell_{\rm cor}$ is long compared
to the interparticle distance.  This is analogous to the dilute gas
where the scale $\ell_c$ is also long compared to the interparticle
distance [see (\ref{scales})]. There is however no real analogy
between the scale $\ell_{\rm cor}$ for the charged gas and the scale
$\ell_c$ for the dilute gas. In particular, whereas $e_0(\rho)$ for
the dilute gas is, up to a constant, of the same order as the kinetic
energy $\sim\mu\ell_c^{-2}$ we have for the charged gas that
$e_0(\rho)\not\sim \ell_{\rm cor}^{-2}=\rho^{1/2}$. The reason for
this difference is that on average only a small fraction of the
particles in the charged gas actually correlate.

\subsection{The Two-Component Gas}
Now we consider $N$ particles with char\-ges $\pm1$. The Hamiltonian 
is thus
$$
H^{(2)}_N=\sum_{i=1}^N-\mu\Delta_i+\sum_{1\leq i<j\leq
  N}\frac{e_ie_j}{|\x_i-\x_j|}.
$$
This time we are interested in $E^{(2)}_0(N)$ the ground state energy 
of
$H^{(2)}_N$ minimized over all possible combination of charges
$e_i=\pm1$, i.e., we do not necessarily assume that the minimum occurs
for the neutral case. Restricting to the neutral case would however
not change the result we give below.

An equivalent formulation is to say that $E^{(2)}_0(N)$ is the ground
state energy of the Hamiltonian acting on all wave functions of space
and charge, i.e., functions in
$L^2\left((\R^3\times\{-1,1\})^N\right)$.  As mentioned in the
introduction, and explained in the beginning of the proof of
Thm.~\ref{ub}, for the calculation of the ground state energy we may
as usual restrict to symmetric functions in this Hilbert space.

For the two-component gas there is no thermodynamic limit.  In
fact, Dyson \cite{D2} proved that $E^{(2)}_0(N)$ was at least as
negative as $-(\mathrm{const})N^{7/5}$ as $N\to \infty$.  Thus,
thermodynamic stability (i.e., a linear lower bound) fails for this
gas. Years
later, a lower bound of this $-N^{7/5}$ form was finally established
in \cite{CLY}, thereby proving that this law is correct.

The connection of this $-N^{7/5}$ law with the jellium $-\rho^{1/4}$
law (for which a corresponding lower bound was also given in
\cite{CLY}) was pointed out by Dyson \cite{D2} in the following way.
Assuming the correctness of the $-\rho^{1/4}$ law, one can treat the
2-component gas by treating each component as a background for the
other. What should the density be? If the gas has a radius $L$ and if
it has $N$ bosons then $\rho = N L^{-3}$. However, the extra kinetic
energy needed to compress the gas to this radius is $N L^{-2}$. The
total energy is then $N L^{-2} - N \rho^{1/4}$, and minimizing this
with respect to $L$ gives $L\sim N^{-1/5}$ and leads to the $-N^{7/5}$
law. The correlation length scale is now $\ell_{\rm
  cor}\sim\rho^{-1/4}\sim N^{-2/5}$.

In \cite{D2} Dyson conjectured an exact asymptotic expression
for $E^{(2)}_0(N)$ for large $N$. That this
asymptotics, as formulated in the next theorem, is indeed correct is
proved in \cite{{LSo02}} and \cite{So}.

\begin{thm}[\textbf{Dyson's law for the two-component gas}] 
\label{thm:Dyson}
\begin{equation}\label{eq:Dyson}
  \lim_{N\to\infty}\frac{E_0^{(2)}(N)}{N^{7/5}}=\inf\biggl\{\mu\int_{\R^3}|\nabla\Phi|^2-
  I_0\int_{\R^3} \Phi^{5/2}\ 
  \biggr|\ 0\leq \Phi,\ \int_{\R^3}\Phi^2=1\biggr\} ,
\end{equation}
where $I_0$ is the constant from Foldy's law:
$$
I_0=\frac{2}{5}\frac{\Gamma(3/4)}{\Gamma(5/4)}\left(\frac{2}{\mu\pi}\right)^{1/4}.
$$
\end{thm}

This asymptotics can be understood as a mean field theory for the gas
density, very much like the Gross-Pitaevskii functional for dilute
trapped gases, where the local energy described by Foldy's law should
be balanced by the kinetic energy of the gas density. Thus if we let
the gas density be given by $\phi^2$ then the ``mean field'' energy 
should be
\begin{equation}\label{eq:Dysonunscaled}
  \mu\int_{\R^3}|\nabla\phi|^2-
  I_0\int_{\R^3} \phi^{5/2}.
\end{equation}
Here $\int\phi^2=N$. If we now define
$\Phi(\x)=N^{-8/5}\phi(N^{-1/5}\x)$ we see that $\int\Phi^2=1$ and 
that the above energy is
$$
N^{7/5}\left(\mu\int_{\R^3}|\nabla\Phi|^2-
  I_0\int_{\R^3} \Phi^{5/2}\right).
$$

It may be somewhat surprising that it is exactly the same constant 
$I_0$
that appears in both the one- and two-component cases. The reason that
there are no extra factors to account for the difference between one
and two components is, as we shall see below, a simple consequence
of Bogolubov's method. The origin of this equivalence, while clear
mathematically, does not appear to have a simple physical 
interpretation.

\subsection{The Bogolubov Approximation}

In this section we shall briefly explain the Bogolubov approximation 
and
how it is applied in the case of the charged Bose gas.  The Bogolubov
method relies on the exact diagonalization of a Hamiltonian, which is
quadratic in creation and annihilation operators. For the charged Bose
gas one only needs a very simple case of the general diagonalization
procedure.  On the other hand, the operators that appear are not
exact creation and annihilation operators. A slightly more general
formulation is needed.  
\begin{thm}[\textbf{Simple case of Bogolubov's
method}]\label{thm:FoldyBogolubov} \hfill\\
  Assume that $\bn_{\pm,\pm}$ are four (possibly unbounded)
  commuting operators satisfying the operator inequality $$
  \quad\left[\bn_{\tau,e},b^*_{\tau,e}\right]\leq1  \quad\hbox{for
    all }e,\tau=\pm.
  $$ Then for all real numbers $\cA,\cB_+,\cB_-\geq0$ we have \begin{eqnarray*}
    \lefteqn{\cA\sum_{\tau,e=\pm1}b_{\tau,e}^*\bn_{\tau,e} }&&\\&&+
    \sum_{e,e'=\pm1}\sqrt{\cB_e\cB_{e'}}ee'(b^*_{+,e}\bn_{+,e'}+b^*_{-,e}
\bn_{-,e'}+b^*_{+,e}b^*_{-,e'}
    +\bn_{+,e}\bn_{-,e'})\\ &\geq&
    -(\cA+\cB_++\cB_-)+\sqrt{(\cA+\cB_++\cB_-)^2-(\cB_++\cB_-)^2}.
  \end{eqnarray*} If $\bn_{\pm,\pm}$ are four annihilation operators
  then the lower bound is sharp.
\end{thm}

\begin{proof} Let us introduce $$
d_\pm^*=(\cB_++\cB_-)^{-1/2}(\cB_+^{1/2}b^*_{\pm 
,+}-\cB_-^{1/2}b^*_{\pm
  ,-}),
$$ and $$ c_\pm^*=(\cB_++\cB_-)^{-1/2}(\cB_-^{1/2}b^*_{\pm
,+}+\cB_+^{1/2}b^*_{\pm
  ,-}).
$$ Then these operators satisfy $$
 [\dn_+,d_+^*]\leq1,\quad [\dn_-,d_-^*]\leq1.
$$ The operator that we want to estimate from below may be rewritten
as \begin{eqnarray*}
  &&\cA(d_+^*\dn_++d_-^*\dn_-+c_+^*\cn_++c_-^*\cn_-)\\
&&+(\cB_++\cB_-)\left(d_+^*\dn_++d_-^*\dn_-+d^*_+d^*_- +\dn_+\dn_-
\right).  \end{eqnarray*} We may now complete the squares to write
this as
\begin{eqnarray*}
&&\cA(c_+^*\cn_++c_-^*\cn_-)+D(d_+^*+\lambda \dn_-)(d_+^*+\lambda
\dn_-)^*\\ &&
+D(d^*_-+\lambda\dn_+)(d^*_-+\lambda\dn_+)^*-D\lambda^2([\dn_+,d^*_+]+[\dn_-,d^*_-])
\end{eqnarray*} if $$
D(1+\lambda^2)=\cA+\cB_++\cB_-,\quad 2D\lambda=\cB_++\cB_-.  $$ We
choose the solution $
\lambda=1+\frac{\cA}{\cB_++\cB_-}-\sqrt{\left(1+\frac{\cA}
{(\cB_++\cB_-)}\right)^2-1}.  $ Hence $$ D\lambda^2=
\mfr{1}/{2}\left(\cA+\cB_++\cB_--\sqrt{(\cA+\cB_++\cB_-)^2-(\cB_++\cB_-)^2}\right).
 $$ \end{proof}

In the theorem above one could of course also have included linear
terms in $b_{\tau,e}$ in the Hamiltonian. In the technical proofs in
\cite{LS,LSo02} the Bogolubov diagonalization with linear terms is
indeed being used to control certain error terms.  Here we shall not
discuss the technical details of the proofs. We have therefore stated
the theorem in the simplest form in which we shall need it to derive
the leading contribution.

In our applications to the charged Bose gas the operators $b_{\pm,e}$
will correspond to the annihilation of particles with charge $e=\pm$
and momenta $\pm k$ for some $k\in\R^3$. Thus, only equal and opposite
momenta couple. In a translation invariant case this would be a simple
consequence of momentum conservation. The one-component gas is not
translation invariant, in our formulation.  The two-component gas is
translation invariant, but it is natural to break translation 
invariance
by going into the center of mass frame. In both cases it is only in 
some
approximate sense that equal and opposite momenta couple.

In the case of the one-component gas we only need particles of one
sign. In this case we use the above theorem with $b_{\pm,-}=0$ and
$\cB_-=0$.

We note that the lower bounds in Theorem \ref{thm:FoldyBogolubov} for 
the
one- and two-component gases are the same except for the replacement 
of
$\cB_+$ in the one-component case by $\cB_+ +\cB_-$ in the 
two-component
case. In the application to the two-component gas $\cB_+$ and $\cB_-$
will be proportional to the particle densities for respectively the
positive or negatively charged particles. For the one-component gas
$\cB_+$ is proportional to the background density.

The Bogolubov diagonalization method cannot be immediately
applied to the operators $H^{(1)}_N$ or $H^{(2)}_N$
since these operators are {\it not} quadratic in creation and
annihilation operators. In fact, they are quartic.  They have
the general form \begin{equation}\label{eq:generalform}
\sum_{\alpha,\beta}t_{\alpha\beta}a^*_\alpha\an_\beta
+\mfr{1}/{2}\sum_{\alpha,\beta,\mu,\nu}w_{\alpha\beta\mu\nu}a^*_\alpha
a^*_\beta\an_{\nu}\an_\mu, \end{equation} with $$
t_{\alpha\beta}=\langle\alpha|T|\beta\rangle,\qquad
w_{\alpha\beta\mu\nu}=\langle\alpha\beta|W|\mu\nu\rangle, $$ where $T$
is the one-body part of the Hamiltonian and $W$ is the two-body-part 
of
the Hamiltonian.

The main step in Bogolubov's approximation is now to assume 
Bose-Einstein
condensation, i.e., that almost all particles are in the same 
one-particle
state. In case of the two-component gas this means that almost half 
the
particles are positively charged and in the same one-particle state as
almost all the other half of negatively charged particles.  We denote
this condensate state by the index $\alpha=0$ in the sums above.
Based on the assumption of condensation Bogolubov now argues that one
may ignore all terms in the quartic Hamiltonian above which contain
3 or 4 non-zero indices and at the same time replace all creation and
annihilation operators of the condensate by their expectation values.
The result is a quadratic Hamiltonian (including linear terms) in the 
creation
and annihilation with non-zero index.  This Hamiltonian is of
course not particle number preserving, reflecting the simple fact
that particles may be created out of the condensate or annihilated 
into
the condensate.

In Section~\ref{sec:chargedupper} below it is explained how to
construct trial wave functions for the one- and two-component charged
gases whose expectations agree essentially with the prescription in
the Bogolubov approximation. The details are to appear in \cite{So}. 
This will imply upper bounds on the energies corresponding to the 
asymptotic forms given in Theorems~\ref{thm:Foldy} and 
\ref{thm:Dyson}.

In \cite{LS,LSo02} it is proved how to make the steps in the Bogolubov
approximation rigorous as lower bounds.  The main difficulty is to
control the degree of condensation.  As already explained it is not
necessary to prove condensation in the strong sense described above.
We shall only prove condensation in small boxes. Put differently, we
shall not conclude that most particles are in the same one-particle
state, but rather prove that most particles occupy one-particle states
that look the same on short scales, i.e., that vary slowly.
Here the short scale is the correlation length scale $\ell_{\rm cor}$.

\subsection{The Rigorous Lower Bounds} 

As already mentioned we must localize into small boxes of some fixed
size $\ell$. This time we must require $\ell_{\rm cor}\ll\ell$. For
the one-component gas this choice is made only in order to control
the degree of condensation. For the two-component gas it is required
both to control the order of condensation, and also to make a local
constant density approximation. The reason we can control the degree
of condensation in a small box is that the localized kinetic energy
has a gap above the lowest energy state. In fact, the gap is of order
$\ell^{-2}$. Thus if we require that $\ell$ is such that $N\ell^{-2}$
is much greater than the energy we may conclude that most
particles are in the lowest eigenvalue state for the localized kinetic
energy. We shall always choose the localized kinetic energy in such a
way that the lowest eigenstate, and hence the condensate, is simply a
constant function.

\subsubsection{Localizing the interaction}
In contrast to the dilute gas the long range Coulomb potential
prevents us from simply ignoring the interaction
between the small boxes.  To overcome this problem we use a sliding
technique first introduced in \cite{CLY}.
\begin{thm}[\textbf{Controlling interactions by 
sliding}]\label{thm:sliding}
  Let $\upchi$ be a smooth approximation to the characteristic
  function of the unit cube centered at the origin. For $\ell>0$ 
  and $\z\in\R^3$ let $\upchi_{\z}(\x)=\upchi((\x-\z)/\ell)$. There 
exists
  an $\omega>0$ depending on $\upchi$ (in such a way that it tends to
  infinity as $\upchi$ approximates the characteristic function) such
  that
$$
\sum_{1\leq i<j\leq
  N}\frac{e_ie_j}{|\x_i-\x_j|}\geq\left(\int\upchi^2\right)^{-1}
\int_{\R^3}\sum_{1\leq i<j\leq
  N}e_ie_jw_{\ell\z}(\x_i,\x_j)d\z-\frac{N\omega}{2\ell},
$$
for all $\x_1,\ldots\in\R^3$ and $e_1,\ldots=\pm1$,
where 
$$
w_\z(\x,\y)=\upchi_{\z}(\x)Y_{\omega/\ell}(\x-\y)\upchi_{\z}(\y)
$$
with
$Y_\mu(\x)=|\x|^{-1}\exp(-\mu |\x|)$ being the Yukawa
potential. 
\end{thm}

The significance of this result is that the two-body potential $w_\z$
is localized to the cube of size $\ell$ centered at $\ell\z$. The
lower bound above is thus an integral over localized interactions
sliding around with the integration parameter.

We have stated the sliding estimate in the form relevant to the
two-component problem. There is an equivalent version for the
one-component gas, where the sum of the particle-particle,
particle-background, and back\-ground-back\-ground interactions may be
bounded below by corresponding localized interactions.

Since $\ell\gg\ell_{\rm cor}$ the error in the sliding
estimate is much smaller than $\omega N/\ell_{\rm cor}$, which for
both the one and two-component gases is of order $\omega$ times the
order of the energy. Thus, since $\ell$ is much
bigger than $\ell_{\rm cor}$, we have room to let $\omega$ be very
large, i.e., $\upchi$ is close to the characteristic function.

\subsubsection{Localizing the kinetic energy}
Having described the technique to control the interaction between
localized regions we turn next to the {\it localization of the kinetic
  energy}.

For the two-component gas this is done in two steps. As already
mentioned it is natural to break the translation invariance of the
two-component gas. We do this by localizing the system into a box of
size $L'\gg N^{-1/5}$ (which as we saw is the expected size of the
gas) as follows. By a partition of unity we can divide space into
boxes of this size paying a localization error due to the kinetic
energy of order $NL'^{-2}\ll N^{7/5}$. We control the interaction
between these boxes using the sliding technique.

We may now argue, as follows, that the energy is smallest if all the
particles are in just one box.  For simplicity we give this argument
for the case of two boxes.  Suppose the two boxes have respective wave
functions $\psi$ and $\widetilde\psi$. The total energy of these two
non-interacting boxes is $E+\widetilde E$. Now put all the particles
in one box with the trial function $\Psi=\psi\widetilde\psi$.  The
fact that this function is not bosonic (i.e., it is not symmetric with
respect to all the variables) is irrelevant because the true bosonic
ground state energy is never greater than that of any trial state
(Perron-Frobenius Theorem). The energy of $\Psi$ is 
$$E+\widetilde
E+\iint \uprho_\psi(\x)|\x-\y|^{-1}\uprho_{\widetilde\psi}(\y)d\x d\y,
$$
where $\uprho_\psi$ and $\uprho_{\widetilde\psi}$ are the
respective {\it charge} densities of the states $\psi$ and
$\widetilde\psi$.  We claim that the last Coulomb term can be made
non-positive. How? If it is positive then we simply change the state
$\widetilde\psi$ by interchanging positive and negative charges (only
in $\widetilde\psi$ and not in $\psi$). The reader is reminded that we
have not constrained the number of positive and negative particles but
only their sum.  This change in $\widetilde \psi$ reverses the 
relative
charge of the states $\psi$ and $\widetilde\psi$ so, by symmetry the
energies $E$ and $\widetilde E$ do not change, whereas the Coulomb
interaction changes sign.

The localization into smaller cubes of size $\ell$ can however not be
done by a crude partition of unity localization. Indeed, this would
cost a localization error of order $N\ell^{-2}$, which as explained is
required to be of much greater order than the energy.

For the one-component charged gas we may instead use a Neumann
localization of the kinetic energy, as for the dilute Bose gas. If we
denote by $\Delta_\ell^{(\z)}$ the Neumann Laplacian for the cube of
size $\ell$ centered at $\z$ we may, in the spirit of the sliding
estimate, write the Neumann localization Laplacian in all of $\R^3$ as
$$
-\Delta=\int -\Delta_\ell^{(\ell\z)} d\z.
$$
In order to write the localized kinetic energy in the same form as 
the localized interaction we must introduce the smooth localization
$\upchi$ as in Theorem~\ref{thm:sliding}. This can be achieved 
by ignoring the low momentum part of the kinetic energy.  

More precisely, there exist $\varepsilon(\upchi)$ and $s(\upchi)$
such that $\varepsilon(\upchi)\to0$ and $s(\upchi)\to0$ as $\upchi$
approaches the characteristic function of the unit cube and such that
(see Lemma 6.1 in \cite{LS}) 
\begin{equation}\label{eq:1kineticloc}
  -\Delta_\ell^{(\z)}\geq
  (1-\varepsilon(\upchi))\cP_{\z}\upchi_\z(\x)F_{\ell 
s(\chi)}(-\Delta)\upchi_\z(\x)\cP_{\z}
\end{equation}
where $\cP_\z$ denotes the projection orthogonal to constants in
the cube of size $\ell$ centered at $z$ and
$$
F_s(u)=\frac{u^2}{u+s^{-2}}.
$$
For $u\ll s^{-2}$ we have that $F_s(u)\ll u$. Hence the
effect of $F$ in the operator estimate above is to ignore the low
momentum part of the Laplacian.

For the two-component gas one cannot use the Neumann localization as
for the one-component gas. Using a Neumann localization ignores the
kinetic energy corresponding to long range variations in the wave
function and one would not get the kinetic energy term
$\int\mu|\nabla\Phi|^2$ in (\ref{eq:Dyson}).  This is the essential
difference between the one- and two-component cases.  This problem is
solved in \cite{{LSo02}} where a new kinetic energy localization
technique is developed. The idea is again to separate the high and low
momentum part of the kinetic energy. The high momentum part is then
localized as before, whereas the low momentum part is used to connect
the localized regions by a term corresponding to a discrete Laplacian.
(For details and the proof the reader is referred to \cite{{LSo02}}.)
\begin{thm}[\textbf{A many body kinetic energy 
localization}]\label{thm:kinloc}
  Let $\upchi_\z$, $\cP_\z$ and $F_s$ be as above. There exist
  $\varepsilon(\upchi)$ and $s(\upchi)$ such that
  $\varepsilon(\upchi)\to0$ and $s(\upchi)\to0$ as $\upchi$ approaches
  the characteristic function of the unit cube and such that for all
  normalized symmetric wave functions $\Psi$ in
  $L^2((\R^3\times\{-1,1\})^N)$ and all $\Omega\subset\R^3$ we have
  \begin{eqnarray*}
    (1+\varepsilon(\upchi))\left(\Psi,\sum_{i=1}^N-\Delta_i\Psi\right)
    &\geq& \int_\Omega 
\Bigl[(\Psi,\cP_{\ell\z}\upchi_{\ell\z}(\x)F_{\ell
      s(\chi)}(-\Delta)\upchi_{\ell\z(\x)}\cP_{\ell\z}\Psi)\\&&
    +\mfr{1}/{2}\ell^{-2}\sum_{\y\in\Z^3\atop
      |\y|=1}(S_\Psi(\ell(\z+\y))-S_\Psi(\ell\z))^2 
\Bigr]d\z\\&&-\const\ell^{-2}{\rm Vol}(\Omega),
  \end{eqnarray*}
  where
  $$
  S_\Psi(\z)=\sqrt{\left(\Psi,(a^*_{0+}(\z)\an_{0+}(\z)+a^*_{0-}(\z)\an_{0-}(\z))\Psi\right)+1}-1
  $$
  with $\an_{0\pm}(z)$ being the annihilation of a particle of
  charge $\pm$ in the state given by the normalized characteristic
  function of the cube of size $\ell$ centered at $\z$.
\end{thm}
The first term in the kinetic energy localization in this theorem is
the same as in (\ref{eq:1kineticloc}). The second term gives rise to a
discrete Laplacian for the function $S_\Psi(\ell\z)$, which is
essentially the number of condensate particles in the cube of size
$\ell$ centered at $\ell\z$. Since we will eventually conclude that
most particles are in the condensate this term will after
approximating the discrete Laplacian by the continuum Laplacian lead 
to
the term $\int\mu|\nabla\phi|^2$ in (\ref{eq:Dysonunscaled}). We 
shall not
discuss this any further here.

When we apply this theorem to the two-component gas the set
$\ell\Omega$ will be the box of size $L'$ discussed above. Hence the
error term $\ell^{-2}{\rm Vol}(\Omega)$ will be of order
$L'^3/\ell^{-5}\ll (N^{2/5}\ell)^{-5}(N^{1/5}L')^3N^{7/5}$. Thus since
$\ell\gg N^{-2/5}$ we may still choose $L'\gg N^{-1/5}$, as required,
and have this error term be lower order than $N^{7/5}$.

\subsubsection{Controlling the degree of condensation} 
After now having localized the problem into smaller cubes we are ready
to control the degree of condensation.  We recall that the condensate
state is the constant function in each cube. Let us denote by $\hn_\z$
the number of excited (i.e., non-condensed particles) in the box of
size $\ell$ centered at $\z$. Thus for the two-component gas $\hn_\z
+a^*_{0+}(\z)\an_{0+}(\z)+a^*_{0-}(\z)\an_{0-}(\z)$ is the total
number of particles in the box and a similar expression gives the
particle number for the one-component gas.

As discussed above we can use the fact that the kinetic energy
localized to a small box has a gap above its lowest eigenvalue 
to control the number of excited particles. Actually, this will
show that the expectation $(\Psi,\hn_\z\Psi)$ is much smaller than the
total number of particles in the box for any state
$\Psi$ with negative energy expectation. 

One needs, however, also a good bound on $(\Psi,\hn_\z^2\Psi)$ to
control the Coulomb interaction of the non-condensed particles.  This
is more difficult.  In \cite{LS} this is not achieved directly through
a bound on $(\Psi,\hn_\z\Psi)$ in the ground state.  Rather it is
proved that one may change the ground state without changing its
energy very much, so that it only contains values of $\hn_\z$
localized close to $(\Psi,\hn_\z\Psi)$. The following theorem gives
this very general localization technique. Its proof can be found in
\cite{LS}.
\begin{thm}[\textbf{Localizing large matrices}]\label{local} 
  Suppose that ${\cA}$ is
  an $N+1\times N+1$ Hermitean matrix and let ${\cA}^k$, with 
$k=0,1,...,N$,
  denote the matrix consisting of the $k^{\rm th}$ supra- and
  infra-diagonal of ${\cA}$.  Let $\psi \in {\bf C}^{N+1}$ be a 
normalized vector
  and set $d_k = (\psi , {\cA}^k \psi) $ and $\lambda = (\psi , {\cA} 
\psi) =
  \sum_{k=0}^{N} d_k$.  \  ($\psi$ need not be an eigenvector of 
${\cA}$.) \ 

  Choose some positive integer $M \leq N+1$.
  Then, with $M$ fixed, there is some $n \in [0, N+1-M]$ and some 
normalized
  vector $ \phi \in   {\bf C}^{N+1}$ with the property that
  $\phi_j =0$ unless $n+1 \leq j \leq n+M$ \ (i.e., $\phi $ has 
length $M$)
  and such that
  \begin{equation}\label{localerror}
    (\phi , {\cA} \phi) \leq \lambda + \frac{C}{ M^2}\sum_{k=1}^{M-1} 
k^2 |d_k|
    +C\sum_{k=M}^{N} |d_k|\ ,
  \end{equation}
  where $C>0 $ is a  universal constant. (Note that the first sum 
starts
  with $k=1$.)
\end{thm}
To use this theorem we start with a ground state (or approximate
ground state) $\Psi$ to the many
body problem. We then consider the projections of $\Psi$ onto the
eigenspaces of $\hn_\z$. Since the possible eigenvalues run from $0$
to $N$ these projections span an at most $N+1$ dimensional space.   

We use the above theorem with $\cA$ being the many body Hamiltonian
restricted to this $N+1$ dimensional subspace. Since the Hamiltonian
can change the number of excited particles by at most two we see that
$d_k$ vanishes for $k\geq3$. We shall not here discuss the estimates
on $d_1$ and $d_2$ (see \cite{LS,LSo02}). The conclusion is that we
may, without changing the energy expectation of $\Psi$ too much,
assume that the values of $\hn_\z$ run in an interval of length much
smaller than the total number of particles.  We would like to conclude
that this interval is close to zero.  This follows from the fact that
any wave function with energy expectation close to the minimum must
have an expected number of excited particles much smaller than the
total number of particles.

\subsubsection{The quadratic Hamiltonian}

Using our control on the degree of condensation it is now possible to
estimate all unwanted terms in the Hamiltonian, i.e., terms that
contain 3 or more creation or annihilation operators corresponding to
excited (non-condensate) states. The proof which is a rather
complicated bootstrapping argument is more or less the same for the
one- and two-component gases.  The result, in fact, shows that we can
ignore other terms too. In fact if we go back to the general form
(\ref{eq:generalform}) of the Hamiltonian it turns out that we can 
control all
quartic terms except the ones with the coefficients:
$$
w_{\alpha\beta00},\ w_{00\alpha\beta},\ w_{\alpha00\beta},\hbox{ and 
} w_{0\alpha\beta0}.
$$
To be more precise, let $u_\alpha$, $\alpha=1,\ldots$ be an
orthonormal basis of real functions for the subspace of functions on
the cube of size $\ell$ centered at $\z$ orthogonal to constants, i.e,
with vanishing average in the cube. We shall now omit the subscript
$\z$ and let $a_{0\pm}$ be the annihilation of a particle of charge
$\pm1$ in the normalized constant function in the cube (i.e., in the
condensate).  Let $a_{\alpha\pm}$ with $\alpha\ne0$ be the
annihilation operator for a particle of charge $\pm1$ in the state
$u_\alpha$.  We can then show that the main contribution to the
localized energy of the two-component gas comes from the Hamiltonian
\begin{eqnarray*}
H_{\rm local} \!\!\!\!&=&  \!\!\!\!\!\! \sum_{\alpha,\beta=1\atop e=\pm1}^\infty
t_{\alpha\beta}a^*_{\alpha e}\an_{\beta e}\\&&  \!\!\!\!\!\!+ 
\mfr{1}/{2}\sum_{\alpha,\beta=1\atop
  e,e'=\pm1}ee'w_{\alpha\beta}(2a_{0e}^*a^*_{\alpha
  e'}\an_{0e'}\an_{\beta e} +a_{0e}^*a^*_{0
  e'}\an_{\alpha e'}\an_{\beta e} +a_{\alpha e}^*a^*_{\beta 
  e'}\an_{0e'}\an_{0e}),
\end{eqnarray*}
where
$$
t_{\alpha\beta}=\mu(u_\alpha,\cP_\z\upchi_\z(\x)F_{\ell
  s(\chi)}(-\Delta)\upchi_\z(\x)\cP_\z u_\beta)
$$
and 
$$
w_{\alpha\beta}=\ell^{-3}\iint 
u_\alpha(\x)\upchi_\z(\x)Y_{\omega/\ell}(\x-\y)\upchi_\z(\y)u_\beta(\x)d\x 
d\y.
$$
In $H_{\rm local}$ we have
ignored all error terms and hence also $\varepsilon(\upchi)\approx0$
and $\int\upchi^2\approx1$.

In the case of the one-component gas we get exactly the same local
Hamiltonian, except that we have only one type of particles, i.e, we
may set $\an_{\alpha-}=0$ above.

Let $\nu_\pm=\sum_{\alpha=0}^\infty a^*_{\alpha\pm}\an_{\alpha\pm}$
be the total number of particles in the box with charge $\pm1$.
For $\k\in\R^3$ we let 
$\upchi_{\k,\z}(\x)=\upchi_\z(\x)e^{i\k\cdot\x}$. We then introduce 
the operators 
$$
b_{\k\pm}=(\ell^3\nu_\pm)^{-1/2}a_\pm(\cP_\z\upchi_{\k,\z})a^*_{0\pm},
$$
where
$a_\pm(\cP_\z\upchi_{\k,\z})=
\sum_{\alpha=1}^\infty(\upchi_{\k,\z},u_\alpha)\an_{\alpha\pm}$
annihilates a particle in the state $\upchi_{\k,\z}$ with charge
$\pm1$.  It is then clear that the operators $b_{\k\pm}$ all commute
and a straightforward calculation shows that
$$
[\bn_{\k\pm},b_{\k\pm}^*]\leq(\ell^3\nu_\pm)^{-1}\|\cP_\z\upchi_\z\|^2
a^*_{0\pm}\an_{0\pm}\leq 1.
$$
If we observe that 
\begin{eqnarray*}
  \sum_{\alpha,\beta=1\atop e=\pm1}^\infty
  t_{\alpha\beta}a^*_{\alpha e}\an_{\beta e}
  &=&(2\pi)^{-3}\int \mu F_{\ell
    s(\chi)}(\k^2)\sum_{e=\pm}a_e(\cP_\z\upchi_{\k,\z})^*a_e(\cP_\z\upchi_{\k,\z})d\k\\
  &\geq&(2\pi)^{-3}\ell^3\int \mu F_{\ell
    s(\chi)}(\k^2)\sum_{e=\pm}b_{\k e}^*\bn_{\k e},
\end{eqnarray*}
we see that
\begin{eqnarray*}
  \lefteqn{H_{\rm local}\geq \mfr{1}/{2}(2\pi)^{-3}\int 
\mu\ell^3F_{\ell
    s(\chi)}(\k^2) \sum_{e=\pm}(b_{\k e}^*\bn_{\k e}+b_{-\k 
e}^*\bn_{-\k e})}&&\\
  &&{}+\sum_{ee'=\pm}\widehat{Y}_{\omega/\ell}(\k)\sqrt{\nu_e\nu_{e'}}ee'(b_{\k 
e}^*\bn_{\k,e'}
  +b_{-\k e}^*\bn_{-\k,e'}+b_{\k e}^*b_{-\k,e'}^*+\bn_{-\k 
e}\bn_{\k,e'})d\k\\
  &&-\sum_{\alpha\beta=1}w_{\alpha\beta}(a_{\alpha+}^*\an_{\beta+}+a_{\alpha-}^*\an_{\beta-})
\end{eqnarray*}
The last term comes from commuting $a^*_{0\pm}\an_{0\pm}$ to 
$\an_{0\pm}a^*_{0\pm}$.
It is easy to see that this last term is a bounded operator with norm 
bounded by 
$$
\const(\nu_++\nu_- 
)\ell^{-3}\|\widehat{Y}_{\omega/\ell}\|_\infty\leq\const\omega^{-2}(\nu_++\nu_- 
)\ell^{-1}.
$$
When summing over all boxes we see that the last term above gives a
contribution bounded by $\const
\omega^{-2}N\ell^{-1}=\omega^{-2}(N^{2/5}\ell)^{-1}N^{7/5}$ which is
lower order than the energy.

The integrand in the lower bound on $H_{\rm local}$ is precisely an
operator of the form treated in the Bogolubov method
Theorem~\ref{thm:FoldyBogolubov}.  Thus up to negligible errors we
see that the operator $H_{\rm local}$ is bounded below by
$$
  \mfr{1}/{2}(2\pi)^{-3}\int 
-(\cA(\k)+\cB(\k))+\sqrt{(\cA(\k)+\cB(\k))^2-\cB(\k)^2}\, d\k,
$$
where
$$
\cA(\k)=\mu\ell^3F_{\ell s(\chi)}(\k^2)\quad\hbox{and}\quad
\cB(\k)=\nu\widehat{Y}_{\omega/\ell}(\k)
$$
with $\nu=\nu_++\nu_-$ being the total number of particles in the
small box.  A fairly simple analysis of the above integral shows that
we may to leading order replace $\cA$ by $\mu\ell^3 \k^2$ and
$\cB(\k)$ by $4\pi\nu |\k|^{-2}$, i.e., we may ignore the cut-offs.
The final conclusion is that the local energy is given to leading 
order by
$$
  \frac {- 1}{2(2\pi)^{3}}\int 
 4\pi\nu|\k|^{-2}+\mu\ell^3|\k|^2
    -\sqrt{(4\pi\nu|\k|^{-2}+\mu\ell^3|\k|^2)^2
    -(4\pi\nu|\k|^{-2})^2} \, d\k
$$
$$=-2^{1/2}\pi^{-3/4}\nu\left(\frac{\nu}{\mu\ell^3}\right)^{1/4}
  \int_0^\infty 1+x^4-x^2(2+x^4)^{1/2}\, dx.
$$
If we finally use that 
$$
\int_0^\infty1+x^4-x^2(2+x^4)^{1/2}\, dx=\frac{2^{3/4}\sqrt{\pi}\Gamma(3/4)}{5\Gamma(5/4)}
$$
we see that the local energy to leading order is $ -I_0
\nu({\nu}/{\ell^3})^{1/4} $. For the one-component gas we should set
$\nu=\rho\ell^3$ and for the two-component gas we should set
$\nu=\phi^2\ell^3$ (see (\ref{eq:Dysonunscaled})).  After replacing
the sum over boxes by an integral and at the same time replace the
discrete Laplacian by a continuum Laplacian, as described above, we
arrive at asymptotic lower bounds as in Theorems~\ref{thm:Foldy} and
\ref{thm:Dyson}.

There is one issue that we have not discussed at all and which played
an important role in the treatment of the dilute gas.  How do we know
the number of particles in each of the small cubes?  For the dilute
gas a superadditivity argument was used to show that there was an
equipartition of particles among the smaller boxes. Such an argument
cannot be used for the charged gas.  For the one-component gas
one simply minimizes the energy over all possible particle numbers in
each little box.  It turns out that charge neutrality is essentially
required for the energy to be minimized. Since the background charge
in each box is fixed this fixes the particle number.

For the two-component there is a-priori nothing that fixes the 
particle
number in each box.  More precisely, if we ignored the kinetic energy
between the small boxes it would be energetically favorable to put all
particles in one small box.  It is the kinetic energy between boxes,
i.e., the discrete Laplacian term in Theorem~\ref{thm:kinloc}, that
prevents this from happening. Thus we could in principle again
minimize over all particle numbers and hope to prove the correct
particle number dependence (i.e., Foldy's law) in each small box. This
is essentially what is done except that boxes with very many or very
few particles must be treated somewhat differently from the ``good''
boxes. In the ``bad'' boxes we do not prove Foldy's law, but only 
weaker estimates that are adequate for the argument.

  \bigskip

\subsection{The Rigorous Upper Bounds}\label{sec:chargedupper}

\subsubsection{The upper bound for the two-component gas}

To prove an upper bound on the energy $E_0^{(2)}(N)$ of the form given
in Dyson's formula Theorem~\ref{thm:Dyson} we shall construct a trial
function from the prescription in the Bogolubov approximation.  We
shall use as an input a minimizer $\Phi$ for the variational problem
on the right side of (\ref{eq:Dyson}). That minimizers exist can be
easily seen using spherical decreasing rearrangements. It is however
not important that a minimizer exists. An approximate minimizer would
also do for the argument given here.  Define
$\phi_0(\x)=N^{3/10}\Phi(N^{1/5}\x)$. Then again $\int\phi_0^2=1$. In
terms of the unscaled function $\phi$ in (\ref{eq:Dysonunscaled}),
$\phi_0(\x)=N^{-1}\phi(\x)$.

Let $\phi_\alpha$, $\alpha=1,\ldots$ be an orthonormal family of real
functions all orthogonal to $\phi_0$.  We choose these functions
below.

We follow Dyson~\cite{D2} and choose a trial function which does not
have a specified particle number, i.e., a state in the bosonic Fock
space.

As our trial many-body wave function we now choose
\begin{eqnarray}\label{eq:trialstate} 
\Psi&=&\exp\left(-\lambda_0^2+\lambda_0a^*_{0+}+\lambda_0
a^*_{0-}\right)\nonumber\\&&\times\prod_{\alpha\ne0}(1-\lambda_\alpha^2)^{1/4}\exp\Bigl(-\sum_{e,e'=\pm1}\sum_{\alpha\ne
  0}\frac{\lambda_{\alpha}}{4} 
ee'a_{\alpha,e}^*a_{\alpha,e'}^*\Bigr)\left|0\right\rangle,
\end{eqnarray}
where $a_{\alpha,e}^*$ is the creation of a particle of charge
$e=\pm1$ in the state $\phi_\alpha$, $|0\rangle$ is the vacuum state,
and the coefficients $\lambda_0,\lambda_1,\ldots$ will be chosen
below satisfying $0<\lambda_\alpha<1$ for $\alpha\ne0$.

It is straightforward to check that $\Psi$ is a normalized function.

Dyson used a very similar trial state in \cite{D2}, but in his case
the exponent was a purely quadratic expression in creation operators,
whereas the one used here is only quadratic in the creation operators
$a^*_{\alpha e}$, with $\alpha\ne0$ and linear in $a^*_{0\pm}$.  As a
consequence our state will be more sharply localized around the mean
of the particle number.

In fact, the above trial state is precisely what is suggested by the 
Bogolubov approximation. 
To see this note that one has  
$$
(\an_{0\pm}-\lambda_0)\Psi=0,\quad
\hbox{ and }\quad
\left(a^*_{\alpha+}-a^*_{\alpha-}+\lambda_\alpha 
(\an_{\alpha+}-\an_{\alpha-})\right)\Psi=0
$$
for all $\alpha\ne0$. Thus the creation operators for the condensed
states can be replaced by their expectation values and an adequate
quadratic expression in the non-condensed creation and annihilation
operators is minimized.

Consider now the operator
\begin{equation}\label{eq:gamma}
  \gamma=\sum_{\alpha=1}^\infty
  \frac{\lambda_\alpha^2}{1-\lambda_\alpha^2}|\phi_\alpha\rangle\langle\phi_\alpha|.
\end{equation}
A straightforward calculation of the energy expectation in the state 
$\Psi$ gives that
\begin{eqnarray*}
  \left(\Psi,\sum_{N=0}^\infty H^{(2)}_N \Psi\right)&=&2\lambda_0^2 
\mu\int(\nabla \phi_0)^2
  +\hbox{Tr}\left(-\mu\Delta\gamma\right)\nonumber\\&&
  +2\lambda_0^2\hbox{Tr}\left(\cK\left(\gamma-\sqrt{\gamma(\gamma+1)}\right)\right),\label{eq:upper}
\end{eqnarray*}
where $\cK$ is the operator with integral kernel
\begin{equation}\label{eq:cK}
  \cK(\x,\y)=\phi_0(\x)|\x-\y|^{-1}\phi_0(\y).
\end{equation}
Moreover, the expected particle number in the state $\Psi$ is
$2\lambda_0^2+\hbox{Tr}(\gamma)$. In order for $\Psi$ to be well
defined by the formula (\ref{eq:trialstate}) we must require this
expectation to be finite.

Instead of making explicit choices for the individual functions
$\phi_\alpha$ and the coefficients $\lambda_\alpha$, $\alpha\ne0$ we
may equivalently choose the operator $\gamma$.  In defining $\gamma$
we use the method of coherent states.  Let $\upchi$ be a non-negative
real and smooth function supported in the unit ball in $\R^3$, with
$\int\upchi^2=1$.  Let as before $N^{-2/5}\ll\ell\ll N^{-1/5}$ and
define $\upchi_\ell(\x)=\ell^{-3/2}\upchi(\x/\ell)$. We choose
$$
\gamma=(2\pi)^{-3}\int_{\R^3\times\R^3}f(\u,|\p|)\cP_{\phi_0}^\perp|\theta_{\u,\p}\rangle\langle\theta_{\u,\p}|\cP_{\phi_0}^\perp
d\u d\p
$$
where $\cP_{\phi_0}^\perp$ is the projection orthogonal to
$\phi_0$,
$$
\theta_{\u,\p}(x)=\exp(i\p\cdot\x)\upchi_\ell(\x-\u),
$$
and
$$
f(\u,|\p|)=\frac{1}{2}\left(\frac{\p^4+16\pi\lambda_0^2\mu^{-1}\phi_0(\u)^2}{\p^2
    \left(\p^4+32\pi\lambda_0^2\mu^{-1}\phi_0(\u)^2\right)^{1/2}}-1\right).
$$
We note that $\gamma$ is a positive trace class operator, 
$\gamma\phi_0=0$, and
that all eigenfunctions of $\gamma$ may be chosen real. These are
precisely the requirements needed in order for $\gamma$ to define the
orthonormal family $\phi_\alpha$ and the coefficients $\lambda_\alpha$
for $\alpha\ne0$.

We use the following version of the Berezin-Lieb inequality
\cite{Berezin72,Lieb73}.  Assume that $\xi(t)$ is an {\it operator}  
concave function of $\R_+\cup\{0\}$ with $\xi(0)\geq0$.  Then if $Y$ 
is a
positive semi-definite operator we have
\begin{equation}\label{eq:berezinlieb}
  \hbox{Tr} \left(Y\xi(\gamma)\right)\geq (2\pi)^{-3}\int
  \xi(f(\u,|\p|))\left(\theta_{\u,\p},\cP_{\phi_0}^\perp
    Y\cP_{\phi_0}^\perp\theta_{\u,\p}\right) d\u d\p.
\end{equation}
We use this for the function $\xi(t)=\sqrt{t(t+1)}$.  Of course, if
$\xi$ is the identity function then (\ref{eq:berezinlieb}) is an
identity. If $Y=I$ then (\ref{eq:berezinlieb}) holds for all concave 
$\xi$ with
$\xi(0)\geq0$.

Proving an upper bound on the energy expectation
(\ref{eq:upper}) is thus reduced to the calculations of explicit 
integrals. After
estimating these integrals one arrives at the leading contribution
(for large $\lambda_0$)
\begin{eqnarray*}
  2\lambda_0^2 \mu\int(\nabla \phi_0)^2
  +&\displaystyle\iint&\left(\mu\p^2+2\lambda_0^2\phi_0(\u)^2\frac{4\pi}{\p^2}\right)f(\u,|\p|)\\
  &&-\frac{4\pi}{\p^2}2\lambda_0^2\phi_0(\u)^2\sqrt{f(\u,|\p|)(f(\u,|\p|)+1)}\ 
d\p d\u\\
&=&2\lambda_0^2\mu \int(\nabla \phi_0)^2-I_0\int 
(2\lambda_0^2)^{5/4}\phi_0^{5/2},
\end{eqnarray*}
where $I_0$ is as in Theorem~\ref{thm:Dyson}. 

If we choose $\lambda_0=\sqrt{N/2}$ we get after a simple rescaling
that the energy above is $N^{7/5}$ times the right side of
(\ref{eq:Dyson}) (recall that $\Phi$ was chosen as the minimizer).
We also note that the expected number of particles is 
$$
2\lambda_0^2+\hbox{Tr}(\gamma)=N+O(N^{3/5}),
$$
as $N\to\infty$.

The only remaining problem is to show how a similar energy could be
achieved with a wave function with a fixed number of particles $N$,
i.e., how to show that we really have an upper bound on 
$E^{(2)}_0(N)$.  We
indicate this fairly simple argument here.

We construct a trial function $\Psi'$ as above, but with an expected
particle number $N'$ chosen appropriately close to, but slightly
smaller than $N$. More precisely, $N'$ will be smaller than $N$ by an
appropriate lower order correction.  It is easy to see then that the
mean deviation of the particle number distribution in the state
$\Psi'$ is lower order than $N$. In fact, it is of order
$\sqrt{N'}\sim\sqrt{N}$. Using that we have a good lower bound on the
energy $E^{(2)}_0(n)$ for all $n$ and that $\Psi'$ is sharply
localized around its mean particle number, we may, without changing
the energy expectation significantly, replace $\Psi'$ by a normalized
wave function $\Psi$ that only has particle numbers less than $N$.
Since the function $n\mapsto E^{(2)}_0(n)$ is a decreasing function we
see that the energy expectation in the state $\Psi$ is, in fact, an
upper bound to $E^{(2)}_0(N)$.

\subsubsection{The upper bound for the one-component gas} 
The upper bound for the one-component gas is proved in a very similar
way as for the two-component gas.  We shall simply indicate the main
differences here.  We will again choose a trial state without a fixed
particle number, i.e., a grand canonical trial state. Since we know
that the one-component gas has a thermodynamic limit and that there is
equivalence of ensembles \cite{LN}, it makes no difference whether we
choose a canonical or grand-canonical trial state.

For the state $\phi_0$ we now choose a normalized function with
compact support in $\Lambda$, that is constant on the set
$\{x\in\Lambda\ |\ \hbox{dist}(x,\partial\Lambda)>r\}$. We shall
choose $r>0$ to go to zero as $L\to\infty$. Let us also choose the
constant $n$ such that $n\phi_0^2=\rho$ on the set where $\phi_0$ is
constant. Then $n\approx\uprho L^3$.

Let again $\phi_\alpha$, $\alpha=1,\ldots$ be an orthonormal family 
of real
functions  orthogonal to $\phi_0$.  
As our trial state we choose, this time,
\begin{eqnarray}\label{eq:trialstate1} 
\Psi&=&\prod_{\alpha\ne0}(1-\lambda_\alpha^2)^{1/4}\exp\Bigl(-\lambda_0^2/2+\lambda_0a^*_{0}
-\sum_{\alpha\ne
  0}\frac{\lambda_{\alpha}}{2}a_{\alpha}^*a_{\alpha}^*\Bigr)\left|0\right\rangle,
\end{eqnarray}
where $a^*_\alpha$ is the creation of a particle in the state
$\phi_\alpha$.  We will choose $\Psi$ implicitly by choosing the
operator $\gamma$ defined as in (\ref{eq:gamma}).

This time we obtain 
\begin{eqnarray}
  \lefteqn{\left(\Psi,\sum_{N=0}^\infty H^{(1)}_N 
\Psi\right)=\lambda_0^2 
    \mu\int(\nabla \phi_0)^2
  \nonumber }&&\\&&+\half\iint\frac{|\gamma(\x,\y)|^2}{|\x-\y|}d\x d\y
  +\half\iint\frac{|\sqrt{\gamma(\gamma+1)}(\x,\y)|^2}{|\x-\y|}
  d\x d\y\nonumber\\&&
  +\half\iint\limits_{\Lambda\times\Lambda}
  \left(\uprho-\uprho_\gamma(\x)-\lambda_0^2\phi_0(\x)^2\right)|\x-\y|^{-1}
  \left(\uprho-\uprho_\gamma(\y)-\lambda_0^2\phi_0(\y)^2\right)
  d\x d\y\nonumber\\&&+\hbox{Tr}\left(-\mu\Delta\gamma\right)
  +\lambda_0^2
  \hbox{Tr}
  \left(\cK\left(\gamma-\sqrt{\gamma(\gamma+1)}\right)\right),
  \label{eq:upperen}
\end{eqnarray}
where $\uprho_\gamma(\x)=\gamma(\x,\x)$ and $\cK$ is again given as in
(\ref{eq:cK}).  We must show that we can make choices such that the
first four terms on the right side above are lower order than the
energy, and can therefore be neglected.

We choose 
$$
\gamma=\gamma_\varepsilon=(2\pi)^{-3}\int_{|p|>\varepsilon\rho^{1/4}}f(|\p|)\cP_{\phi_0}^\perp
|\theta_{\p}\rangle\langle\theta_{\p}|\cP_{\phi_0}^\perp
d\p,
$$
where $\varepsilon>0$ is a parameter which we will let tend to 0 at 
the end of the calculation. Here
$\cP_{\phi_0}^\perp$ as before is the projection orthogonal to
$\phi_0$ and this time
$$
f(|\p|)=\frac{1}{2}\left(\frac{\p^4+8\pi\mu^{-1}\uprho}{\p^2
    \left(\p^4+16\pi\mu^{-1}\uprho\right)^{1/2}}-1\right)
$$
and
$$
\theta_{\p}(x)=\sqrt{n\uprho^{-1}}\exp(i\p\cdot\x)\phi_0(\x).
$$
Note that $n\uprho^{-1}\phi_0(\x)^2$ is 1
on most of $\Lambda$. We then again have the Berezin-Lieb inequality 
as before.
We also find that 
\begin{eqnarray*}
  \uprho_\gamma(\x)&=&(2\pi)^{-3}\int_{|p|>\varepsilon\rho^{1/4}} 
f(|\p|)d\p 
n\uprho^{-1}\phi_0(\x)^2\left(1+O(\varepsilon^{-1}\uprho^{-1/4}L^{-1})\right)\\
  &=&A_\varepsilon(\rho/\mu)^{3/4}n\uprho^{-1}\phi_0(\x)^2\left(1+O(\varepsilon^{-1}\uprho^{-1/4}L^{-1})\right),
\end{eqnarray*}
where $A_\varepsilon$ is an explicit function of $\varepsilon$.
We now choose $\lambda_0$ such that
$\lambda_0^2=n(1-A_\varepsilon\uprho^{-1/4}\mu^{-3/4})$, i.e., such 
that 
$$
\lambda_0^2\phi_0^2(\x)+\uprho_\gamma(\x)=
n\phi_0(\x)^2(1+O(\varepsilon^{-1}\uprho^{-1/2}L^{-1}))\approx\uprho.
$$
It is easy to see that the first term in (\ref{eq:upperen}) is of
order $\uprho L^3(rL)^{-1}$ and the fourth term in (\ref{eq:upperen})
is of order $\uprho L^3(\varepsilon^{-2}+\uprho r^2)$.  We may choose
$r$, depending on $L$, in such a way that after dividing by $\uprho 
L^3$ and letting
$L\to\infty$ only the error $\varepsilon^{-2}$ remains. This
allows choosing $\varepsilon\ll\rho^{-1/8}$.

To estimate the second term in (\ref{eq:upperen}) we use
Hardy's inequality to deduce
$$
\iint\frac{|\gamma(\x,\y)|^2}{|\x-\y|}d\x d\y\leq2(\Tr\gamma^2)^{1/2}
\Tr(-\Delta\gamma^2)^{1/2},
$$
and these terms can be easily estimated using the Berezin-Lieb
inequality in the direction opposite from before, since we are
interested now in an upper bound. The third term in (\ref{eq:upperen})
is controlled in exactly the same way as the second term.
We are then left with the last two terms in (\ref{eq:upperen}). They
are treated in exactly the same way as for the two-component gas again
using the Berezin-Lieb inequality.

\bibliographystyle{amsalpha}

\begin{thebibliography}{XXXXX}

\bibitem[AG]{astra} G.E. Astrakharchik and S. Giorgini, {\it Quantum 
Monte Carlo study of the three- to one-dimensional crossover for a 
trapped Bose gas}, Phys. Rev. A {\bf 66}, 053614-1--6 (2002). 

\bibitem[ABCG]{giorgini} G.E. Astrakharchik, J. Boronat, J. 
Casulleras, and 
S. Giorgini, {\it Superfluidity versus Bose-Einstein condensation in 
a Bose gas with disorder}, 
Phys. Rev. A {\bf 66}, 023603 (2002).

\bibitem[Ba]{BA} B. Baumgartner, \textit{The Existence of 
Many-particle Bound
States Despite a Pair Interaction with Positive Scattering
Length}, J. Phys. A \textbf{30} (1997), L741--L747.

\bibitem[Bm]{baym} G. Baym, in: {\it Math. Methods in Solid State and 
Superfluid Theory}, Scottish Univ. Summer School of Physics, Oliver 
and Boyd, Edinburgh (1969). 

\bibitem[Be]{Berezin72}
F.A. Berezin, Izv. Akad. Nauk, ser. mat., {\bf 36} (No. 5) (1972);
English translation: USSR Izv. {\bf 6} (No. 5) (1972). F.A. Berezin,
{\it General concept of quantization}, Commun. Math. Phys. {\bf 40}, 
153--174 (1975).

\bibitem[Bl]{blume} D. Blume, {\it Fermionization of a bosonic gas 
under 
highly elongated confinement: A diffusion quantum Monte Carlo study}, 
Phys. Rev. A {\bf 66}, 053613-1--8 (2002).

\bibitem[Bo]{BO} N.N. Bogolubov,  J.  Phys. (U.S.S.R.)  {\bf 11}, 23
(1947); N.N. Bogolubov, D.N.  Zubarev,  Sov. Phys.-JETP {\bf
1}, 83 (1955).

\bibitem[BBD]{bongs} K.\ Bongs, S.\ Burger, S.\ Dettmer, D.\ Hellweg, 
J.\ 
Artl, W. Ertmer, and K.\ Sengstok,  {\it Waveguides for Bose-Einstein 
condensates}, Phys. Rev. A, {\bf 63}, 031602 (2001)


\bibitem[B]{Bose} S.N. Bose, {\it Plancks Gesetz und 
Lichtquantenhypothese},
 Z. Phys. {\bf 26}, 178--181 (1924).

\bibitem[CS1]{CS01b} A.Y. Cherny, A.A. Shanenko,
{\it Dilute Bose gas in two dimensions: Density expansions and the
Gross-Pitaevskii equation}, Phys.\ Rev.\ E {\bf 64}, 027105 (2001)

\bibitem[CS2]{CS01a} A.Y. Cherny, A.A. Shanenko,
{\it The kinetic and interaction energies of a trapped Bose gas:
Beyond the mean field},
Phys.\ Lett.\ A {\bf 293}, 287 (2002).


\bibitem[CLY]{CLY} J. Conlon, E.H. Lieb, H.-T. Yau, \textit{The 
$N^{7/5}$ 
Law
for Charged Bosons}, Commun.\ Math.\ Phys.\ \textbf{ 116}, 417--448
(1988).


\bibitem[CCRCW]{Cornish}
S.L. Cornish, N.R. Claussen, J.L. Roberts, E.A. Cornell, C.E. Wieman,
{\it Stable $^{85}Rb$ Bose-Einstein Condensates with Widely Tunable
Interactions}, Phys. Rev. Lett. {\bf 85}, 1795--98 (2000).

\bibitem[DGPS]{DGPS}
F. Dalfovo, S.\ Giorgini, L.P.\ Pitaevskii, S.\ Stringari, {\it
Theory of Bose-Einstein condensation in trapped gases}, Rev. Mod.
Phys. \textbf{71}, 463--512 (1999).

\bibitem[DGW]{das2} K.K. Das, M.D. Girardeau, and E.M. Wright, {\it 
Crossover from One to Three Dimensions for a Gas of Hard-Core 
Bosons}, Phys. Rev. Lett. {\bf 89}, 110402-1--4 (2002).


\bibitem[DLO]{dunjko} V. Dunjko, V. Lorent, and M. Olshanii, {\it 
Bosons 
in Cigar-Shaped Traps: Thomas-Fermi Regime, Tonks-Girardeau Regime, 
and In Between}, Phys. Rev. Lett. {\bf 86}, 5413--5316 (2001).


\bibitem[D1]{dyson}
F.J. Dyson, \textit{Ground-State Energy of a Hard-Sphere Gas},
Phys. Rev. \textbf{106}, 20--26 (1957).

\bibitem[D2]{D2} F.J. Dyson, \textit{Ground State Energy of a Finite 
System
of Charged Particles}, J. Math. Phys. \textbf{8}, 1538--1545
(1967).

\bibitem[DLS]{DLS} F.J. Dyson, E.H. Lieb, B. Simon, {\it Phase 
Transitions
in Quantum Spin Systems with Isotropic and Nonisotropic
Interactions}, J. Stat. Phys. {\bf 18}, 335--383 (1978).


\bibitem[E]{Einstein} A. Einstein, {\it Quantentheorie des 
einatomigen idealen
Gases}, Sitzber. Kgl. Preuss. Akad. Wiss.,
261--267 (1924), and 3--14 (1925).

\bibitem[FS]{fetter} A.L. Fetter and A.A. Svidzinsky, {\it Vortices 
in a trapped dilute Bose-Einstein condensate}, 
J. Phys.: Condens. 
Matter {\bf 13}, R135 (2001).

\bibitem[FH]{fishho}
D.S. Fisher, P.C.\ Hohenberg, {\it Dilute Bose gas in two dimensions},
Phys. Rev. B {\bf 37}, 4936--4943 (1988).

\bibitem[F]{FO} L.L. Foldy, \textit{Charged Boson Gas}, Phys. Rev.
\textbf{124}, 649--651 (1961); Errata \textit{ibid} \textbf{125},
2208 (1962).

\bibitem[Gi]{gir} M.D.\ Girardeau, {\it Relationship between systems 
of impenetrable bosons and fermions in one dimension}, J. Math. Phys. 
{\bf 1}, 516 (1960).

\bibitem[GW]{girardeau2} M.D.\ Girardeau and E.M. Wright, {\it 
Bose-Fermi
variational Theory for the BEC-Tonks Crossover}, Phys. Rev. Lett. {\bf
87}, 210402-1--4 (2001).

\bibitem[GWT]{girardeau} M.D.\ Girardeau, E.M.\ Wright, and J.M.\ 
Triscari, 
{\it Ground-state properties of a one-dimensional system of hard-core 
bosons in a harmonic trap}, Phys. Rev. A {\bf 63}, 033601-1--6 (2001).

\bibitem[Go]{goerlitz} A. G{\"o}rlitz, {\it et al.}, {\it Realization 
of 
Bose-Einstein Condensates in Lower Dimension}, Phys. Rev. Lett. {\bf 
87}, 130402-1--4 (2001).  

\bibitem[GS]{GS} G.M.\ Graf and J.P.\ Solovej, \textit{A correlation 
estimate
with applications to quantum systems with Coulomb interactions},
Rev. Math. Phys. {\bf 6}, 977--997 (1994).

\bibitem[G]{greiner} M. Greiner, {\it et al.}, {\it Exploring Phase 
Coherence in a 2D Lattice of Bose-Einstein Condensates},
 Phys.  Rev.  Lett.  {\bf 87}, 160405 (2001).

\bibitem[Gr1]{G1961}E.P.\ Gross, {\it Structure of a Quantized Vortex 
in
    Boson Systems,} Nuovo Cimento {\bf 20}, 454--466 (1961).

\bibitem[Gr2]{G1963} E.P.\ Gross, {\it Hydrodynamics of a superfluid
condensate,} J. Math. Phys. {\bf 4}, 195--207 (1963).

\bibitem[HFM]{hines} D.F.\ Hines, N.E.\ Frankel, D.J.\ Mitchell, 
{\it Hard disc Bose gas}, Phys.\ Lett.\ {\bf 68A}, 12--14 (1978).

\bibitem[Ho]{Ho} P.C.\ Hohenberg, 
{\it Existence of Long-range Order in One and Two Dimensions}, Phys.\ 
Rev.\ {\bf 158},
383--386 (1966).

\bibitem[HoM]{HM} P.C.~Hohenberg and P.C.~Martin, {\it Microscopic
theory of helium}, Ann.\ Phys.\ (NY) {\bf 
34}, 291 (1965).

\bibitem[H]{huang} K. Huang, in: {\it Bose-Einstein Condensation}, A. 
Griffin, D.W. Stroke, S. Stringari, eds., Cambridge University Press, 
31--50 (1995).


\bibitem[HY]{Lee-Huang-YangEtc}K.~Huang, C.N.~Yang, Phys. Rev. {\bf
105}, 767--775 (1957); T.D.~Lee, K.~Huang, C.N.~Yang,  Phys.
Rev. {\bf 106}, 1135--1145 (1957); K.A. Brueckner, K. Sawada, Phys.
Rev. {\bf 106}, 1117--1127, 1128--1135 (1957); S.T. Beliaev, Sov.
Phys.-JETP {\bf 7}, 299--307 (1958); T.T. Wu, Phys. Rev. {\bf 115},
1390 (1959); N. Hugenholtz, D. Pines, Phys. Rev. {\bf 116}, 489
(1959); M. Girardeau, R. Arnowitt, Phys. Rev. {\bf 113}, 755
(1959); T.D. Lee,  C.N. Yang, Phys. Rev. {\bf 117}, 12 (1960).

\bibitem[JK]{jackson} 
A.D. Jackson and G.M. Kavoulakis, {\it Lieb Mode in a 
Quasi-One-Dimensional Bose-Einstein Condensate of Atoms}, Phys. Rev. 
Lett. {\bf 89}, 070403 (2002).


\bibitem[KLS]{KLS} T. Kennedy, E.H. Lieb, S. Shastry, {\it The $XY$ 
Model 
has
Long-Range Order for all Spins and all Dimensions Greater than
One}, Phys. Rev. Lett. \textbf{ 61}, 2582--2584 (1988).

\bibitem[KD]{TRAP} W. Ketterle, N. J. van Druten, {\it Evaporative 
Cooling of
Trapped Atoms}, in B. Bederson, H. Walther, eds.,  Advances in
Atomic, Molecular and Optical Physics, {\bf 37}, 181--236,
Academic Press (1996).

\bibitem[KNSQ]{KoSt2000}
E.B. Kolomeisky, T.J. Newman, J.P. Straley, X. Qi,
{\it Low-dimensional Bose liquids:
beyond the Gross-Pitaevskii approximation}, Phys. Rev. Lett.
{\bf 85}, 1146--1149 (2000).

\bibitem[KT]{KT} M.~Kobayashi and M.~Tsubota, {\it Bose-Einstein 
condensation and superfluidity of a 
dilute Bose gas in a random potential}, Phys. Rev. B {\bf 66}, 174516 
(2002).

\bibitem[KP]{komineas}
S. Komineas and N. Papanicolaou, {\it Vortex Rings and Lieb Modes in 
a Cylindrical Bose-Einstein Condensate}, Phys. Rev. Lett. {\bf 89}, 
070402 (2002).  

\bibitem[Le]{Lenard} A.~Lenard, {\it Momentum distribution in the 
ground state of the one-dimensional system of impenetrable bosons}, 
J. Math. Phys. {\bf 5}, 930--943 (1964). 

\bibitem[L1]{Lieb63} E.H. Lieb, \textit{Simplified Approach to the 
Ground
State Energy of an Imperfect Bose Gas}, Phys. Rev. \textbf{130}, 
2518--2528
(1963). See also Phys. Rev. \textbf{133} (1964),
A899--A906  (with A.Y. Sakakura) and Phys. Rev. \textbf{134}
(1964), A312--A315 (with W. Liniger).

\bibitem[L2]{EL2} E.H.~Lieb, \textit{The Bose fluid}, in 
W.E.~Brittin, ed.,
Lecture Notes in Theoretical Physics VIIC,  Univ. of Colorado
Press, pp.\ 175--224 (1964).

\bibitem[L3]{Lieb73}
E.H. Lieb, {\it The classical limit of quantum spin systems}, 
Commun.\ Math.\ Phys. {\bf 31}, 327--340 (1973).

\bibitem[L4]{L3} E.H.\ Lieb, {\it The Bose Gas: A Subtle Many-Body 
Problem}, 
in {\it Proceedings
of the XIII International Congress on Mathematical Physics, London},
A.~Fokas, et al. eds. International Press, pp.~91--111, 2001.

\bibitem[LL]{LL} E.H. Lieb, W. Liniger, \textit{Exact Analysis of an
Interacting Bose Gas. I.  The General Solution and the Ground
State}, Phys. Rev.\ \textbf{130}, 1605--1616 (1963); E.H. Lieb,
\textit{Exact Analysis of an Interacting Bose Gas. II. The
Excitation Spectrum}, Phys. Rev.\ \textbf{130}, 1616--1624 (1963).

\bibitem[LLo]{LL01}
E.H. Lieb, M. Loss, {\it Analysis}, 2nd ed., Amer. Math. Society,
Providence, R.I. (2001).

\bibitem[LN]{LN} E.H. Lieb, H. Narnhofer,  \textit{The Thermodynamic
Limit for Jellium}, J.  Stat. Phys. \textbf{12}, 291--310 (1975).
Errata J. Stat. Phys. \textbf{14}, 465 (1976).

\bibitem[LSe]{LS02} E.H.\ Lieb, R.\ Seiringer, {\it Proof of
Bose-Einstein Condensation for Dilute Trapped Gases},
Phys. Rev. Lett. {\bf 88}, 170409-1-4 (2002).

\bibitem[LSSY]{LSSY}
E.H.~Lieb, R.~Seiringer, J.P.~Solovej, and J.~Yngvason, {\it The
ground state of the Bose gas},  in: Current Developments in 
Mathematics, 2001, 131-178, 
International Press, Cambridge (2002).

\bibitem[LSeY1]{LSY1999} E.H.\ Lieb, R.\ Seiringer, J.\ Yngvason, 
\textit{
Bosons in a Trap: A Rigorous Derivation of the Gross-Pitaevskii
Energy Functional},  Phys.  Rev A \textbf{ 61}, 043602 (2000).

\bibitem[LSeY2]{LSY2d} E.H. Lieb, R. Seiringer, J. Yngvason, \textit{
A Rigorous Derivation of the Gross-Pitaevskii Energy Functional
for a Two-dimensional Bose Gas}, Commun. Math. Phys. {\bf 224}, 17
(2001).  

\bibitem[LSeY3]{LSYdoeb} E.H. Lieb, R. Seiringer, J. Yngvason, 
\textit{The
Ground State Energy and Density of Interacting Bosons in a Trap},
in \textit{ Quantum Theory and Symmetries}, Goslar, 1999,
H.-D.~Doebner, V.K.~Dobrev, J.-D.~Hennig and W. Luecke, eds., pp.
101--110, World Scientific (2000). 

\bibitem[LSeY4]{LSYnn} E.H.\ Lieb, R.\ Seiringer, J.\ Yngvason, 
\textit{Two-Dimensional Gross-Pitaevskii Theory}, in: Progress in 
Nonlinear Science, Proceedings of the International Conference 
Dedicated to the 100th Anniversary of A.A. Andronov, Volume II, A.G. 
Litvak, ed., 582-590, Nizhny Novgorod, Institute of Applied Physics, 
University of Nizhny Novgorod (2002).

\bibitem[LSeY5]{LSYsuper} E.H.~Lieb, R.~Seiringer, J.~Yngvason, {\it 
Superfluidity in Dilute Trapped
Bose Gases}, Phys.\ Rev.\ B {\bf 66}, 134529 (2002).

\bibitem[LSeY6]{LSY} E.H.~Lieb, R.~Seiringer, J.~Yngvason, {\it 
One-Dimensional Behavior of Dilute, Trapped Bose Gases}, 
Commun. Math. Phys. {\bf 244}, 347--393 (2004). See also: {\it 
One-Dimensional Bosons in Three-Dimensional Traps}, Phys. Rev. Lett. 
{\bf 91}, 150401-1-4 (2003).

\bibitem[LSeY7]{lsy02}
E.H.\ Lieb, R.\ Seiringer, and J.\ Yngvason, {\it Poincar\'e 
Inequalities in
Punctured Domains}, Ann. Math. {\bf 158}, 1067--1080 (2003).

\bibitem[LSo]{LS} E.H.\ Lieb, J.P.\ Solovej, \textit{ Ground State 
Energy
of the One-Component Charged Bose Gas}, Commun. Math. Phys.\
\textbf{217}, 127--163 (2001). Errata {\bf 225}, 219--221 (2002). 

\bibitem[LSo2]{LSo02} E.H.\ Lieb, J.P.\ Solovej, {\it Ground State 
Energy of the Two-Component Charged Bose Gas}, Commun. Math. Phys. 
(in press),
arxiv:math-ph/0311010, mp\_arc 03-490. 

\bibitem[LY1]{LY1998}
E.H. Lieb, J.\ Yngvason, \textit{ Ground State Energy of the low 
density
Bose Gas}, Phys. Rev. Lett. \textbf{80}, 2504--2507 (1998). 

\bibitem[LY2]{LY2d} E.H.\ Lieb, J.\ Yngvason, \textit{The Ground 
State Energy
of a Dilute Two-dimensional Bose Gas}, J. Stat. Phys. {\bf 103},
509 (2001). 

\bibitem[LY3]{LYbham} E.H. Lieb, J. Yngvason, \textit{ The
Ground State Energy of a Dilute Bose Gas}, in \textit{ Differential
Equations and Mathematical Physics, University of Alabama,
Birmingham, 1999}, R.~Weikard and G.~Weinstein, eds., 271--282
Amer. Math. Soc./Internat. Press (2000).  

\bibitem[MSKE]{esslinger} H. Moritz, T. St\"oferle, M. K\"ohl and
T. Esslinger, {\it Exciting Collective Oscillations in a Trapped 1D 
Gas},
Phys. Rev. Lett. {\bf 91}, 250402 (2003).

\bibitem[M]{M} W.J.\ Mullin, {\it Bose-Einstein Condensation in a 
Harmonic Potential}, 
J.\ Low Temp.\ Phys.\ {\bf 106}, 615--642 (1997).

\bibitem[Ol]{olshanii} M. Olshanii, {\it Atomic Scattering in the 
Presence of an External Confinement and a Gas of Impenetrable 
Bosons}, Phys. Rev. Lett. {\bf 81}, 938--941 (1998).

\bibitem[O]{Ovch}  A.A. Ovchinnikov, {\it On the description of a
two-dimensional Bose gas at low densities}, J. Phys. Condens. Matter
{\bf 5}, 8665--8676 (1993). See also JETP Letters {\bf 57}, 477
(1993); Mod. Phys. Lett. {\bf 7}, 1029 (1993).


\bibitem[PS]{PS} C.~Pethick, H.\ Smith, \textit{ Bose Einstein 
Condensation of 
Dilute Gases}, Cambridge University Press, 2001.

\bibitem[PSW]{petrov} D.S. Petrov, G.V. Shlyapnikov, and J.T.M. 
Walraven, 
{\it Regimes of Quantum Degeneracy in Trapped 1D Gases}, Phys. Rev. 
Lett. {\bf 85}, 3745--3749 (2000).

\bibitem[Pi]{P1961} L.P. Pitaevskii, {\it Vortex lines in an imperfect
    Bose gas}, Sov. Phys. JETP. {\bf 13}, 451--454 (1961).

\bibitem[PiSt]{PiSt} L. Pitaevskii, S. Stringari, {\it Uncertainty 
Principle, 
Quantum Fluctuations, and Broken
Symmetries}, J. Low Temp. Phys. {\bf 85}, 377 (1991).

\bibitem[Po]{popov} V.N. Popov, {\it On the theory of the 
superfluidity of
two- and one-dimensional Bose systems}, Theor. and Math. Phys. {\bf 
11},
565--573 (1977).

\bibitem[PrSv]{PrSv} N.V.~Prokof'ev and B.V.~Svistunov, {\it Two 
definitions of superfluid density}, 
Phys.\ Rev.\ B {\bf 61}, 11282 (2000).

\bibitem[S]{schick} M.\ Schick, \textit{ Two-Dimensional System of 
Hard Core Bosons}, Phys. Rev.\ A \textbf{ 3}, 1067--1073 (1971).

\bibitem[Sc]{schreck}  F.\ Schreck, {\it et al.}, {\it Quasipure 
Bose-Einstein Condensate Immersed in a Fermi Sea}, 
Phys. Rev. Lett. {\bf 87}, 080403 (2001).

\bibitem[Se1]{S1999}
R.\ Seiringer, Diplom thesis, University of Vienna, 1999.

\bibitem[Se2]{S4}
R.\ Seiringer, {\it Bosons in a Trap: Asymptotic Exactness of the
Gross-Pitaevskii Ground State Energy Formula}, in: {\it Partial
Differential Equations and Spectral Theory}, PDE2000 Conference in
Clausthal, Germany, M. Demuth and B.-W. Schulze, eds., 307--314,
Birkh\"auser (2001).

\bibitem[Se3]{rot1} 
R.\ Seiringer, {\it Gross-Pitaevskii Theory of the Rotating Bose Gas}, Commun. Math. Phys. {\bf 229}, 491--509 (2002);
{\it Ground state asymptotics of a dilute, rotating gas}, J. Phys. A: Math. Gen. {\bf 36}, 9755--9778 (2003).

\bibitem[Sh]{Shev}
S.I. Shevchenko, {\it On the theory of a Bose gas in a nonuniform 
field}, Sov.\ J.\ Low Temp.\ Phys.\ {\bf 18}, 223--230 (1992).


\bibitem[Si]{S79} B.\ Simon, {\it Trace ideals and their application},
Cambridge University Press (1979).

\bibitem[So]{So} J.P. Solovej, {\it Upper Bounds to the Ground
    State Energies of the One- and Two-Component Charged Bose gases},
  preprint, arxiv:math-ph/0406014.

\bibitem[T]{TE} G.\ Temple, \textit{The theory of Rayleigh's 
Principle as
Applied to Continuous Systems}, Proc.\ Roy.\ Soc.\ London A
\textbf{119},  276--293 (1928).

\bibitem[TT]{TT} D.R.~Tilley and J.~Tilley,  {\it Superfluidity and 
Superconductivity}, third edition,
Adam Hilger, Bristol and New York (1990).


\end{thebibliography}

\end{document}